\documentclass[a4paper,fleqn,usenatbib]{mn2e}
\usepackage[T1]{fontenc}
\usepackage{ae,aecompl}
\usepackage{graphicx}
\usepackage{amsmath}
\usepackage{amssymb}
\usepackage{subfigure}
\usepackage[linkcolor=blue,colorlinks=true,breaklinks,citecolor=blue,urlcolor=blue]{hyperref}

\newcommand{\aap}{A\&A}                   % "Astron. Astrophys."
\newcommand{\apjs}{ApJS}                  % "Astrophys. J. Suppl. Ser."
\newcommand{\apjl}{ApJ}                   % letter at ApJ

\newcommand{\s}{\rm\thinspace s}

\newcommand{\Ms}{\rm\thinspace Ms}
\newcommand{\Hz}{\rm\thinspace Hz}
\newcommand{\Msun}{\hbox{$\rm\thinspace M_{\odot}$}}

\newcommand{\keV}{\rm\thinspace keV}

\newcommand{\rg}{\rm\thinspace $r_\mathrm{g}$}

\title[Modelling X-ray reverberation]{Towards modelling X-ray reverberation in AGN: Piecing together the extended corona}
\author[D. R. Wilkins et al.]{D. R. Wilkins$^1$\thanks{E-mail: drw@ap.smu.ca}, E. M. Cackett$^2$, A. C. Fabian$^3$ and C. S. Reynolds$^{4,5}$\\
$^1$Department of Astronomy \& Physics, Saint Mary's University, Halifax, NS. B3H 3C3, Canada \\
$^2$Department of Physics \& Astronomy, Wayne State University, 666 W. Hancock St., Detroit, MI 48201, USA \\
$^3$Institute of Astronomy, University of Cambridge, Madingley Road, Cambridge. CB3 0HA, UK \\
$^4$Department of Astronomy, University of Maryland, College Park, MD 20742, USA \\
$^5$Joint Space-Science Institute (JSI), College Park, MD 20742, USA }
\begin{document}

\date{Accepted 2016 January 29. Received 2016 January 7; in original form 2015 November 5}

\pagerange{\pageref{firstpage}--\pageref{lastpage}} \pubyear{2016}

\maketitle

\label{firstpage}

\begin{abstract}
Models of X-ray reverberation from extended coron\ae\ are developed from general relativistic ray tracing simulations. Reverberation lags between correlated variability in the directly observed continuum emission and that reflected from the accretion disc arise due to the additional light travel time between the corona and reflecting disc. X-ray reverberation is detected from an increasing sample of Seyfert galaxies and a number of common properties are observed, including a transition from the characteristic reverberation signature at high frequencies to a hard lag within the continuum component at low frequencies, as well a pronounced dip in the reverberation lag at 3\keV. These features are not trivially explained by the reverberation of X-rays originating from simple point sources. We therefore model reverberation from coron\ae\ extended both over the surface of the disc and vertically. Causal propagation through its extent for both the simple case of constant velocity propagation and propagation linked to the viscous timescale in the underlying accretion disc is included as well as stochastic variability arising due to turbulence locally on the disc. We find that the observed features of X-ray reverberation in Seyfert galaxies can be explained if the long timescale variability is dominated by the viscous propagation of fluctuations through the corona. The corona extends radially at low height over the surface of the disc but with a bright central region in which fluctuations propagate up the black hole rotation axis driven by more rapid variability arising from the innermost regions of the accretion flow.
\end{abstract}


\section{Introduction}
The accretion of material onto supermassive black holes in active galactic nuclei (AGN) powers some of the most luminous objects in the Universe. An intense X-ray continuum is seen, originating from a corona of energetic particles associated with the innermost regions of the accretion flow. X-rays are produced when thermal ultraviolet photons from the accretion disc are Compton up-scattered by the particles of the corona \citep{sunyaev_trumper}, thought to be accelerated and confined by magnetic fields associated with the disc \citep{galeev+79,haardt+91,merloni_fabian}. Much remains unknown, however, about the exact structure of the corona, the process by which it is formed and how it evolves to give rise to the extreme variability that is seen in the X-ray emission. A deep understanding of the corona is a vital component to understanding how supermassive black holes can power some of the most luminous objects in the Universe.

In addition to being observed directly, the X-ray continuum illuminates the accretion disc, from which it is `reflected.' More precisely, it is back-scattered through the processes of Compton scattering, photoelectric absorption and subsequent fluorescent line emission, and bremsstrahlung \citep{ross_fabian}. The spectrum of the reflected X-rays is shifted in energy and blurred by Doppler shifting and associated relativistic beaming due to the orbital motion of the reflecting material in the disc as well as the gravitational redshift associated with the spacetime curvature in such close proximity to the black hole \citep{fabian+89,laor-91}. In particular, the relativistically blurred reflection from the accretion disc results in a `soft excess' in emission above the power law continuum between around 0.3 and 1\keV\ composed of a number of emission lines from the ionised disc material. These become blurred and smoothed. The prominent K$\alpha$ fluorescence line of iron is seen at 6.4\keV\ with a broad wing extending as low as 3\keV\ where photons emitted in this line from the inner regions of the disc are shifted to lower energy.

The X-ray emission from accreting black holes, in particular the narrow line Seyfert 1 (NLS1) galaxies, is seen to be highly variable \citep{leighly-99_2,turner+99,ponti+2012} and in recent years studies of this variability have added a further dimension to our understanding of the corona and the inner regions of the accretion flow. The reflected X-rays vary in luminosity according to variations in the flux received from the corona.Where there are energy bands in which the emission is predominantly composed of reflection from the accretion disc (\textit{i.e.} in the 0.3-1\keV\ soft excess and the broadened iron K$\alpha$ line between around $3\sim 4$ and 7\keV), the variability is expected to lag behind correlated variations in energy bands dominated by the directly observed continuum emission (1-4\keV). The time delay corresponds to the additional light travel time from the coronal source to the reflecting accretion disc \citep{reverb_review}.

Since their initial detection \citep{fabian+09}, X-ray reverberation lags between the continuum and soft excess have been detected in an ever-growing sample of Seyfert galaxies \citep{emmanoul+2011,demarco+2011,zoghbi+2011,demarco+2012,iras_fix,cackett+2013,alston+2013} and also between the continuum and broad iron K line \citep{zoghbi+2012,zoghbi+2013,kara_1h0707,kara+13}. The advent of the hard X-ray imaging telescope, \textit{NuSTAR}, has enabled the detection of reverberation lags between the continuum and the broad Compton hump that appears around 20\keV\ in the reflection spectrum \citep{zoghbi+2014,kara+2015}.

X-ray timing analysis is typically conducted as a function of Fourier frequency; that is the different frequency components that make up the X-ray light curves. X-ray reverberation is detected in the higher frequency components representing the variability occurring on the shortest timescales. Measuring the time lag (with respect to some reference band) of successive X-ray energy bands over these high frequency components reveals a profile reminiscent of the reflection spectrum. Emission in the soft excess and the iron K$\alpha$ line is seen to arrive later than the continuum emission \citep{kara_1h0707,kara_pg1244} while the redshifted wing of the iron K line responds more rapidly than the core of the line as might be expected if the redshifted wing is originating from the inner parts of the disc, closer to a relatively compact coronal X-ray source \citep{zoghbi+2012}. Remarkably, the high frequency energy dependence over the iron K line of the lag shows a common profile across many objects, once scaled for the absolute magnitude of the lag, suggesting that there is a common reverberation process producing the lag in these objects \citep{kara+13}.

The magnitude of the reverberation lag between both the soft excess \citep{demarco+2012} and iron K line \citep{kara+13} is found to scale approximately linearly with the black hole mass suggesting that the process inducing the lag is occurring on a similar size scale across the sample of objects in terms of the characteristic length scale in the gravitational field, represented by the gravitational radius (1\rg\,$=GM/c^2$). Reverberation lags have even been detected in Galactic stellar mass X-ray binaries where the equivalent lags are on the order milliseconds \citep{uttley+2011}. The measured lags correspond to just a few gravitational radii showing that reverberation measurements are probing the innermost regions of the accretion flow in the immediate vicinity of the black hole event horizon.

While the signature of reverberation from the disc (the `soft lag' or `reverberation lag') is seen in the high frequency components of the X-ray variability, at lower frequencies, the measured time lags undergo a transition to a different behaviour. At low frequencies, variability in harder X-rays are seen to systematically lag behind that in lower energy `(softer) X-rays. Measuring the frequency dependence of the lag between the soft 0.3-1\keV\ reflection-dominated band (the soft excess) and the hard 1-4\keV\ continuum-dominated band, a clear transition is seen with the hard band leading the soft at high frequencies (\textit{i.e.} reverberation) but the soft band leading the hard. The lag time increases smoothly, following an approximately log-linear dependence on energy. Before its discovery in AGN, this `hard lag' was a well known feature in X-ray binaries \citep{miyamoto+88,miyamoto+89,nowak+99} and is often attributed to the propagation of fluctuations in mass accretion rate through the disc, energising the less energetic outer parts of the corona before reaching the more energetic inner parts \citep{kotov+2001,arevalo+2006}.

Following the almost ubiquitous detection of X-ray reverberation across the AGN in Seyfert galaxies, attention has turned to modelling the phenomenon in order to translate the frequency and energy dependence of reverberation lags into accurate measurements of the geometry of the inner accretion flow and the processes forming and injecting energy into the corona. The modelling of X-ray reverberation has been largely motivated by \citet{reynolds+99} who computed `2D transfer functions' giving the observed flux as a function of observed photon energy and time from a single, localised, X-ray flare with\citet{young_reynolds} relating this work to observable signatures that may be detected by future generation X-ray observatories, comparing the observed reverberation of X-ray flares with a library of computed transfer functions to determine the mass of the black hole as well as the location of the flare. \citet{lag_spectra_paper} develop general relativistic ray tracing simulations of the propagation of X-rays from coron\ae, both point-like and extended, to model the frequency dependence of the lag and show how the reverberation lag provides a measure of the height of a point like corona or average vertical extent of a corona extended above the accretion disc, while being relatively insensitive to the radial extent.

\citet{cackett_ngc4151} extend this analysis to the frequency and energy dependence of lags arising from the reverberation of X-rays originating from a point source. As well as considering the source height, the dependence of the lags on a number of parameters commonly associated with spectral modelling of X-ray reflection is computed and the frequency and energy dependence of the lags in the Seyfert galaxy NGC\,4151 are fitted with this model. Further attempts to fit point source models of X-ray reverberation to lag measurements have reinforced the requirement for compact coron\ae\ confined to a small region around the black hole and inner regions of the accretion flow \citep{kara_pg1244,emman+2014,chainakun+2015}. Point source models of the X-ray emission have been found, however, to have limited applicability to reproducing the full details of X-ray timing measurements \citep{chainakun+2012}. In particular the self-consistent combination of hard lags produced by propagation through an extended emitting region and reverberation from a compact, point-like corona as well as the reproduction of the fine features of the energy dependence is challenging. Moreover, measurements of the illumination pattern of the accretion disc by the coronal X-ray source, exploiting the energy shifts imparted on photons reflected at different radii, has suggested that accretion discs in a number of NLS1 galaxies are illuminated by an extended corona \citep{1h0707_emis_paper,understanding_emis_paper,iras_fix,mrk335_corona_paper}.

\begin{figure*}
\centering
\subfigure[Lag-frequency spectrum] {
\includegraphics[width=56mm]{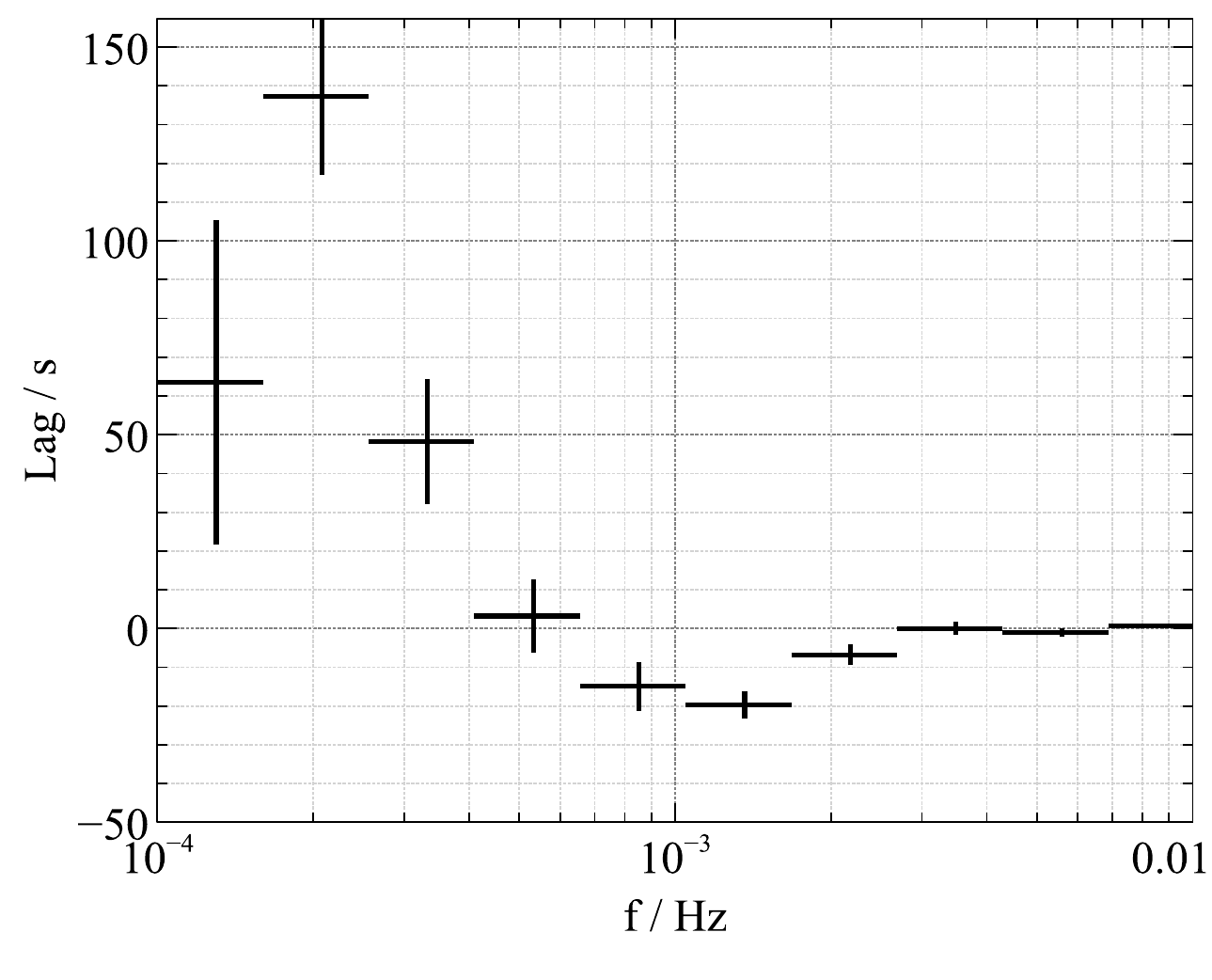}
\label{1h0707_lagspec.fig:lagfreq}
}
\subfigure[Lag-energy, low frequency] {
\includegraphics[width=56mm]{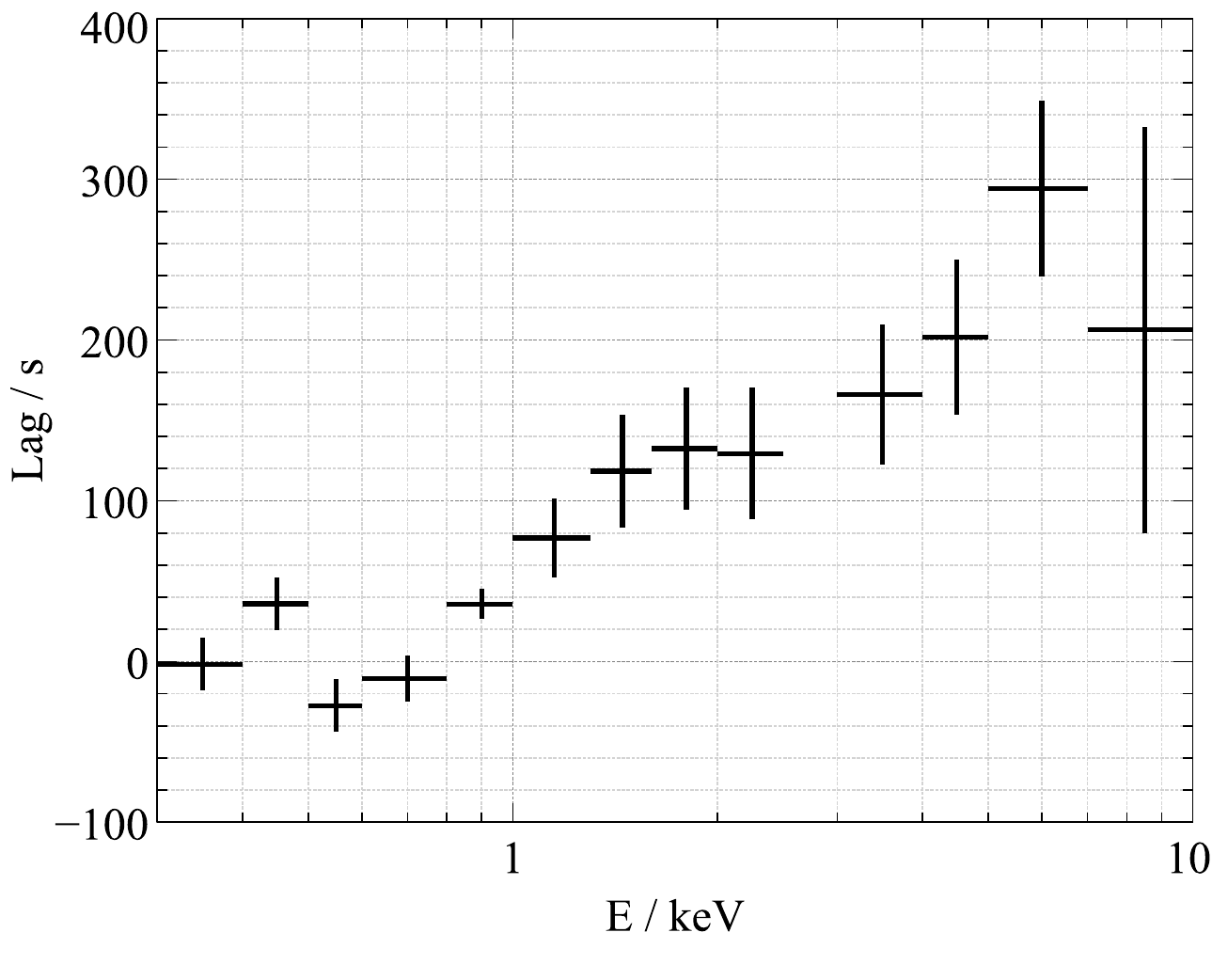}
\label{1h0707_lagspec.fig:lagen_low}
}
\subfigure[Lag-energy, high frequency] {
\includegraphics[width=56mm]{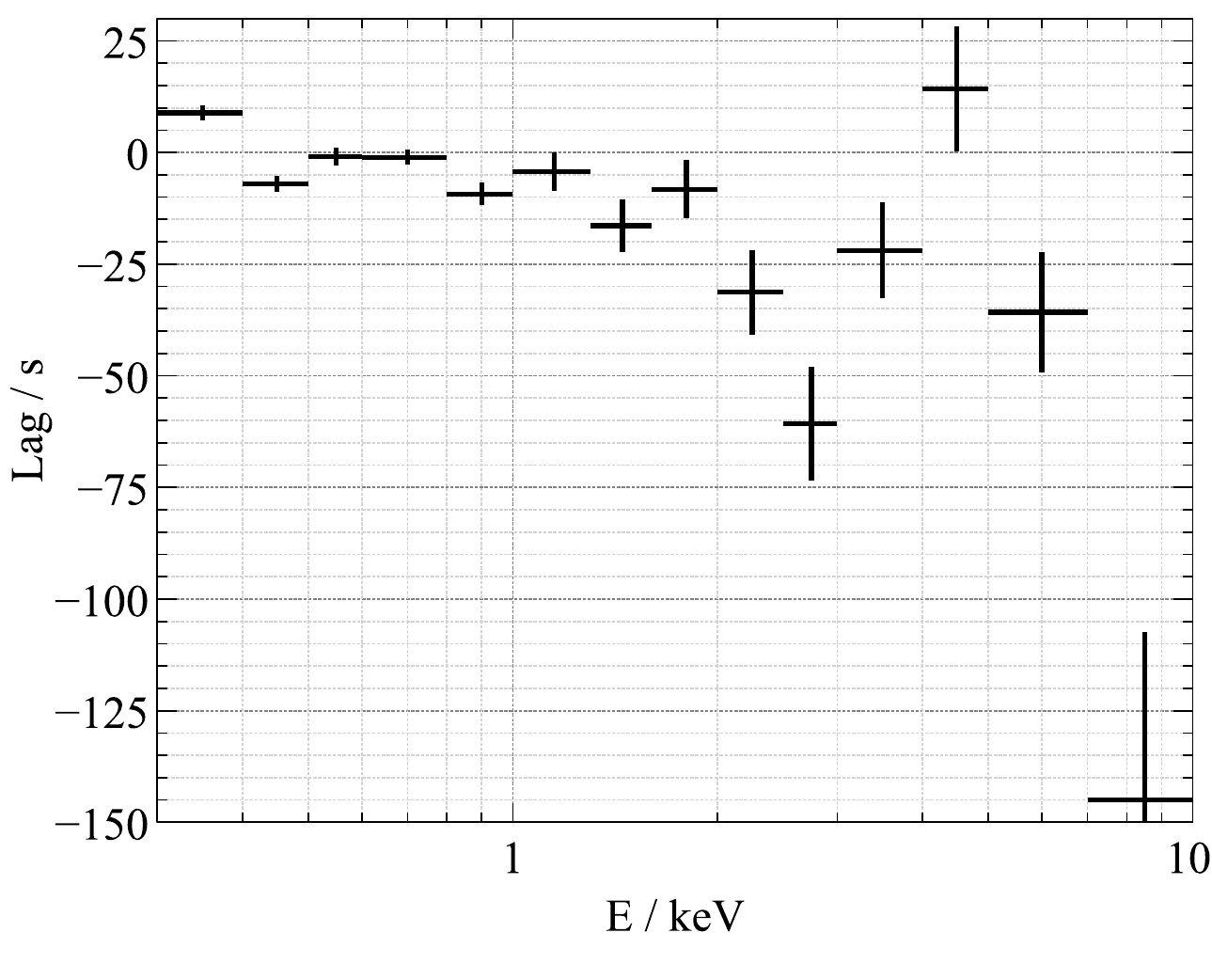}
\label{1h0707_lagspec.fig:lagen_high}
}
\caption[]{\subref{1h0707_lagspec.fig:lagfreq} The broadband lag-frequency spectrum calculated from 1.3\Ms\ of observations of the NLS1 galaxy 1H\,0707$-$495 with \textit{XMM-Newton} as in \citet{kara_1h0707}. The lag is shown as a function of Fourier frequency (\textit{i.e.} for long and short timescale components of the observed variability) between the reflection-dominated 0.3-1\keV\ energy band and the continuum-dominated 1-4\keV\ band. A positive lag indicates that the harder, 1-4\keV\ band is lagging behind the softer, thus X-ray reverberation is seen at higher frequencies where the lag is negative and the reflection-dominated band is lagging behind the continuum. Lag-energy spectra are computed, averaged over frequency ranges that show positive and negative lags. \subref{1h0707_lagspec.fig:lagen_low} The lag-energy spectrum of 1H\,0707$-$495 over the low frequency range $(1.65-5.44)\times 10^{-4}$\Hz\ showing the hard lag, where the lag tie increases systematically for higher photon energies. \subref{1h0707_lagspec.fig:lagen_high} The high frequency lag-energy spectrum, averaged over the high frequency range $(0.96-2.98)\times 10^{-3}$\Hz\ where reverberation is seen. Variability in the soft excess between 0.3 and 1\keV\ and in the iron K$\alpha$ fluorescence line from the disc between 4 and 7\keV\ is seen to lag behind the continuum, while there is a characteristic dip in the lag-energy spectrum at 3\keV. Many Seyfert galaxies show a common lag-energy spectrum shape between 2 and 10\keV.}
\label{1h0707_lagspec.fig}
\end{figure*}

We here develop a self-consistent model of the reverberation of X-rays illuminating the disc from an extended corona, accounting for the finite speed at which luminosity fluctuations can propagate through its extent. The techniques for X-ray timing and measuring X-ray reverberation as well as the common results are reviewed in Section~\ref{measurement.sec} before techniques for simulating X-ray reverberation through general relativistic ray tracing simulations are discussed in Section~\ref{sim.sec}. The features of reverberation from point-like X-ray sources are revisited in Section~\ref{point.sec} before models of extended coron\ae\ are developed in Section~\ref{ext.sec}. These begin with simplified prescriptions for the propagation of fluctuations before introducing a model for the hard lag and linking propagation to the underlying accretion disc. The results of these simulations are discussed in the context of explaining the lags measured in the NLS1 galaxy 1H\,0707$-$495, displaying X-ray timing properties representative of the findings in Seyfert galaxies and on which high quality data are available obtained from more than 1\Ms\ of observations with \textit{XMM-Newton}.

\section{Measurement of X-ray Reverberation}
\label{measurement.sec}
Simulations of X-ray reverberation are developed in the context of the commonly employed techniques for its measurement from light curves in distinct energy bands obtained from X-ray telescopes, reviewed by \citet{reverb_review}. In this section, these techniques are outlined as relevant to the simulation of reverberation and developing models of the process and typical features of the measurements are reviewed.

Following \citet{nowak+99}, time lags between X-ray energy bands, used as proxies for different spectral components, are computed from the Fourier transforms of the light curves; the variation over time of the arrival rate of photons falling within those specific energy bands. The light curve, $F(t)$, is considered to be the sum over Fourier components of all frequencies, $\omega$, describing the contributions of long and short time scale variations to the overall variability:
\[ F(t) = \frac{1}{\sqrt{2\pi}}\int \tilde{F}(\omega)e^{-i\omega t}\,d\omega \]

The amplitude and phase of each frequency component is given by the Fourier transform of the light curve and can be written $\tilde{F}(\omega)=\left|\tilde{F}(\omega)\right|e^{i\varphi}$ and is computed by
\[ \tilde{F}(\omega) = \frac{1}{\sqrt{2\pi}}\int F(t)e^{i\omega t}\,dt \]
The phase lag between two time series, say the hard and soft light curves, $H(t)$ and $S(t)$ respectively, can be found by considering the complex form of their Fourier transforms $\tilde{H}(\omega)=\left|\tilde{H}(\omega)\right|e^{-i\varphi}$ and $\tilde{S}(\omega)=\left|\tilde{S}(\omega)\right|e^{-i\theta}$ and computing the cross spectrum
\begin{equation}
	\label{crossspec.equ}
	\tilde{C}(\omega) = \tilde{S}^*(\omega) \tilde{H}(\omega) = \left|\tilde{S}(\omega)\right| \left|\tilde{H}(\omega)\right| e^{i(\theta - \varphi)}
\end{equation}
The argument of which gives the time lag, $\tau$, since $\varphi = \omega t$ (and converting from angular to linear frequency, $f$)
\begin{equation}
	\label{lag.equ}
	\tau(f) = \frac{1}{2\pi f}\arg\left(\tilde{C}(f)\right)
\end{equation}
Following this sign convention, a positive time lag indicates that the variability in the hard band, $H(t)$, is lagging behind that in the soft band, $S(t)$.

The lag time is a function of both the Fourier frequency, that is whether the lag is seen in rapid components of the variability or in more slowly varying components, and the photon energy, via the energy channels that are included when accumulating the light curves from which the cross spectrum is computed. In order to maximise signal to noise, the lag is measured as a function of Fourier frequency between light curves in two broad energy bands henceforth referred to as the (\textit{broadband}) \textit{lag-frequency spectrum}. When investigating reverberation from the accretion disc, energy bands will be selected such that one is expected to be dominated by the directly observed continuum emission and the other is expected to be dominated by X-rays reflected from the disc. The cross spectrum is averaged over a range of Fourier frequencies to produce higher signal to noise in a lag spectrum binned by frequency.

Alternatively, the lag can be computed as a function of photon energy to produce a \textit{lag-energy spectrum}. In this case, cross spectra are computed between each (narrow) energy band of interest and a common reference band. To reduce uncertainty due to counting statistics, a reference band spanning a broad energy range is taken, commonly over the full instrument bandpass (0.3-10\keV\ in the case of \textit{XMM-Newton}) or over the reflected soft excess (0.3-0.8\keV) where the effective area of the detector is high. The energy channel of interest is subtracted from the reference band during the calculation of each cross spectrum such that errors are not correlated between the bands \citep{zoghbi+2012}. Finally, the lag of each energy band with respect to the reference band is computed from the cross spectrum averaged over a broad range of frequencies (to minimise the errors). Frequency ranges are selected from the broadband lag-frequency spectrum to encompass a single region that appears to be dominated by the same process (for instance whether the lag is positive or negative).

Reverberation measurements are illustrated in Fig.~\ref{1h0707_lagspec.fig}, showing the broadband lag-frequency spectrum as well as the lag-energy spectrum at both low and high frequencies for the NLS1 galaxy 1H\,0707$-$495. The features displayed are common to many NLS1 galaxies. The lag-frequency spectrum is computed between the 0.3 to 1\keV\ energy band, expected from the time-averaged X-ray spectrum to be dominated by reflection from the accretion disc \citep{zoghbi+09}, and the 1 to 4\keV\ band, dominated by directly observed continuum emission. A positive lag indicates that the harder band is lagging behind the softer, hence reverberation can be seen at higher frequencies (between around $5\times 10^{-4}$ and $3\times 10^{-3}$\Hz). Lower frequencies show the hard lag, thought to arise within the continuum and characteristic of both AGN and accreting X-ray binaries. Frequency ranges for the lag-energy spectrum are selected from this plot, taking the hard-lag dominated range showing a positive lag for the low frequency lag-energy spectrum and the reverberation-dominated range where the lag is negative for the high frequency lag-energy spectrum. Zero time lag corresponds to the average arrival time of the reference band, here taken to be 0.3-0.8\keV, with more negative times indicating an earlier response. The low frequency lag-energy spectrum shows a steady rise in lag as a function of photon energy, while the high frequency spectrum is reminiscent of the X-ray reflection spectrum. The soft excess (below 1\keV) and iron K$\alpha$ line (5-7\keV) are seen to respond later than the X-ray continuum (1-2\keV). The earliest response is seen from photons around 3\keV\ producing a dip in the lag-energy spectrum. The origin of this dip has not previously been well understood but it is a common feature seen across Seyfert galaxies. Once scaled for the absolute value of the lag, the shape of the 2-10\keV\ high frequency lag-energy spectrum is almost identical across the sampled objects \citep{kara+13}.

\section{Simulation of X-ray Reverberation}
\label{sim.sec}
The reverberation of fluctuations in luminosity propagating through an extended source region or corona is considered in the above formalism of a cross spectrum computed from extended light curves. In this formulation, the underlying variability in the luminosity of the X-ray source is described by some time series $L(t)$. These fluctuations will then propagate through the source causing variation in the X-ray flux from different regions.

Once the fluctuation reaches a given region of the source, the emanating rays will propagate in the curved spacetime around the black hole to reach the observer directly to be observed as part of the continuum. The propagation of the fluctuation through the source region and subsequent propagation of the emitted rays to the observer from each location within the source is encoded in the \textit{impulse response function} $T_\mathrm{C}(E,t)$. This describes the response seen by the observer; the number of photons received as a function of photon energy (\textit{i.e.} the shape of the continuum spectrum) and time after an instantaneous ($\delta$-function) flash propagating through the corona from the site of energy injection. The light curve in a given energy band is given simply by the convolution of this response function with the time series describing the underlying variability.
\begin{equation}
L_E(t) = L(t) \otimes T_\mathrm{C}(E,t)
\end{equation}

X-rays emitted from the coronal X-ray source will also reach the accretion disc whereupon they will produce the reflection spectrum. The reflected X-rays will then propagate from the location of their emission upon the accretion disc to the observer. The response seen by the observer in the reflection from the accretion disc of the fluctuation in the X-ray continuum is again described by an impulse response function, $T_\mathrm{R}(E,t)$. The photon count is expressed as a function of energy (the spectrum of the reflected X-rays in the rest frame of the material in the accretion disc shifted according to the gravitational redshift and Doppler shift determined by the location and orbital velocity of the material) and total propagation and ray travel time from the patch of the X-ray source to different sites of reflection on the disc and then to the observer.

The total response seen by the observer is obtained simply by summing the responses of the directly observed continuum and the reflection from the accretion disc.
\begin{equation}
T(E,t) = T_\mathrm{C}(E,t) + T_\mathrm{R}(E,t)
\end{equation}

It is important to note that in this formalism, as is necessary when X-ray reverberation is measured from the cross spectra computed from extended light curves, the propagation and spectral response is assumed to be constant across all fluctuations. If there is any variation (\textit{e.g.} in the origin of the fluctuation, how it propagates and its spectral response), this will be implicitly averaged by the Fourier transforms of the measured time series.

Given this definition of the impulse response function and noting that the convolution theorem allows the Fourier transform of the light curve in a given energy band to be expressed as the product of the Fourier transforms of the underlying variability and of the response function, $\tilde{L}_E = \tilde{L}\tilde{T}_E$, the cross spectrum of one energy band with some reference band can be written
\begin{equation}
\label{crossspec.equ}
\tilde{C} = \tilde{L}^*_E \tilde{L}_\mathrm{ref} = \left|\tilde{L}\right|^2 \tilde{T}^*_E \tilde{T}_\mathrm{ref}
\end{equation}
The time lag is again calculated from the argument of this and is entirely encoded in the term $\tilde{T}^*_E \tilde{T}_\mathrm{ref}$. $\left|\tilde{L}\right|^2$ is the (frequency-dependent) power spectral density of the time series and is typically a power law in frequency, $\left|\tilde{L}\right|^2\propto f^{-\alpha}$ with $\alpha\sim 2$. Being a real quantity, this does not affect the lag in narrow frequency bins. It becomes important, however, when the cross spectrum is averaged over broad frequency bins since this term acts as a low-pass filter, increasing the influence of the lower frequency components where the lag varies significantly over a bin. It is also important in lag measurements of real data since it will decrease the signal to noise at high frequency.

\subsection{The Reverberation Response Function}
Impulse response functions are calculated through general relativistic ray tracing simulations of the propagation of rays in the Kerr spacetime around a rotating black hole with spin parameter $a = J/Mc$ following the procedure of \citet{understanding_emis_paper,lag_spectra_paper}.

In order to simulate X-rays originating from an extended corona, rays are started at random locations within a cylindrical source region, defined by an outer radius ($R$) and lower and upper vertical height above the equatorial plane in which the accretion disc is taken to lie. The probability distributions are uniform in vertical position $z$, azimuthal angle $\varphi$, and radial distance $\rho$ from the black hole rotation axis. Each ray is weighted by the volume element $2\pi\rho\,d\rho\,d z$ to produce uniform luminosity per unit volume of the corona. Rays propagate in random directions, with uniform probability distributions in $\cos\alpha$ and $\beta$ which are respectively the polar and azimuthal angle of the rays measured in the rest frame of the emitting material. 

Propagation of the luminosity fluctuation through the corona is accounted for by setting the starting time co-ordinate, $t$ of the ray at its origin. The ray is emitted once the propagation reaches that part of the corona.

Rays are propagated through numerical integration of the geodesic equations. In the Kerr geometry, the null geodesic equations describing the propagation of photons can be written in terms of the first derivatives of the Boyer-Lindquist co-ordinates with respect to some affine parameter.
\begin{align}
	\label{tdot.equ}	
	\dot{t} &= \frac{
		\left[(r^2 + a^2\cos^2\theta)(r^2+a^2) + 2a^2 r\sin^2\theta\right]k - 2arh
	}
	{ 
		r^2\left(1+\frac{a^2\cos^2\theta}{r^2} - \frac{2}{r}\right)\left(r^2+a^2\right) + 2a^2r\sin^2\theta
	} \\
	\label{phidot.equ}
	\dot{\varphi} &= \frac{
		2ar k \sin^2\theta + (r^2+a^2\cos^2\theta - 2r)h
	}
	{
		(r^2+a^2)(r^2+a^2\cos^2\theta-2r)\sin^2\theta + 2a^2 r\sin^4\theta
	}
\\
	\label{thetadot.equ}
	\dot{\theta^2} &= \frac{Q + (ka\cos\theta - h\cot\theta)(ka\cos\theta + h\cot\theta)}{\rho^4}
\\
	\label{rdot.equ}
	\dot{r}^2 &= \frac{\Delta}{\rho^2}\left[ k\dot{t} - h\dot{\varphi} - \rho^2\dot\theta^2  \right]
\end{align}

With the starting co-ordinates of the ray set and the constants of motion $h$ and $Q$ determined from the initial direction of propagation ($k$ is set such that each photon as unit energy; this does not affect the propagation), the affine parameter is advanced to propagate the ray. Clearly, the equations in $\dot{r}$ and $\dot{\theta}$ allow for both positive and negative solutions, corresponding to propagation in the direction of increasing and decreasing values of these co-ordinates. The initial signs of these steps are set to produce the correct initial direction of motion, then the sign is switched during propagation whenever $\dot{r}^2$ or $\dot{\theta}^2$ passes through zero between integration steps.

The geodesic equations are integrated until the ray reaches either the equatorial plane, where it is taken to have hit the accretion disc so long as it hits outside the innermost stable orbit for the value of the spin parameter in question, until it reaches a limiting outer radius of 1000\rg\ or it is lost through the black hole event horizon. The rays that hit the disc are binned in $r$ and $\varphi$ and their redshifts, which represent both the shift in photon energy and photon arrival rate are calculated (for a source emitting photons at a given energy and constant rate along each ray in its own rest frame as measured by some observer on the disc). For an observer with 4-velocity $\mathbf{v}_\mathrm{O}$ receiving photons (4-momentum $\mathbf{p}$) from an emitter with 4-velocity $\mathbf{v}_\mathrm{E}$, the total redshift (which includes both the gravitational redshift and Doppler shift) is given by
\begin{equation}
	\label{redshift.equ}
 g^{-1} \equiv \frac{\nu_\mathrm{O}}{\nu_\mathrm{E}} = 
 \frac{\mathbf{v}_\mathrm{O}\cdot\mathbf{p}(\mathrm{O})}{\mathbf{v}_\mathrm{E}\cdot\mathbf{p}(\mathrm{E})} = \frac{g_{\mu\nu}v_O^\mu p^\nu(O)}{g_{\rho\sigma}v_E^\rho p^\sigma(E)}
\end{equation}

Once the illumination of the disc by the source has been computed, it is necessary to calculate the appearance of this disc to the distant observer carrying out the reverberation measurements. In order to do this, the observer will measure the rays travelling parallel to one another arriving at the telescope. Hence to visualise the accretion disc, an `image plane' is constructed a large distance (in this case 10,000\rg\ to be free from the gravitational influence of the black hole) centred on a line of sight inclined at some angle $i$ to the rotation axis. The image plane is a regular grid of parallel rays (travelling perpendicular to its face and parallel to the line of sight to the black hole from the centre of the image plane) whose dimensions are taken to be $500\times 500$\rg.

The Kerr metric is time-reversible under the transformation $a\rightarrow -a$, hence rather than tracing all possible rays from the illuminated accretion disc and finding those which reach the plane, the grid of parallel rays can be traced backwards in time to find where they originated on the disc. As for the illumination of the disc by the corona, the rays are traced from the plane until they reach the equatorial plane where they are again binned and the average travel time (the time taken for a ray from part of the disc to reach the observer) and redshift are calculated for each bin. The number of rays reaching each bin, divided by the proper area of the bin, represents the projected area of that bin as seen by the observer and accounts for the magnification of parts of the disc by gravitational lensing. It is implicitly assumed that each ray hitting the disc illuminates equal proper area of the disc surface.

Hence, for each ray that hits the disc from the source, it is possible to look up from a pre-calculated table the total travel time from the source to the disc to the observer and the perceived brightness of reflection from that part of the disc. When a ray hits the disc, it is assumed to produce the rest frame reflection spectrum according to the \textsc{reflionx} model of \citet{ross_fabian}, which for each ray, is shifted by the appropriate redshift from the disc to the observer. The numbers of photons arriving at the observer (with the arrival rate along each ray appropriately shifted by the redshifts from the source to the disc and the disc to the observer) are then binned by energy and time, representing the impulse response function, $T_\mathrm{R}(E,t)$.

\subsection{The Continuum Response Function}
When considering a spatially extended X-ray source, it is necessary to consider the response to the luminosity fluctuation seen in the continuum as well as in its reverberation from the accretion disc, owing to the range of ray paths from an extended source to the observer, not to mention the additional delay caused by propagation of the fluctuation through the source on a finite time scale. On the contrary, from a point source, there is precisely one ray path that reaches the observer.

The impulse response function is calculated in the same way as the observation of the accretion disc by the observer. For a corona extended radially over the accretion disc, all rays are taken to originate from the same height above the disc (\textit{i.e.} $dz \ll R$). Rays are again traced from an image plane, a regular grid of parallel rays 10,000\rg\ centred on a line of sight at inclination $i$ to the black hole rotation axis, backward in time, until they intercept the plane taken to represent the X-ray source. The ray travel times, redshifts and numbers of rays (representing the projected area and gravitational lensing) are recorded for bins on the source plane. Propagation of the source fluctuation is accounted for by incrementing the travel time of the ray according to its radius of origin in the source (\textit{i.e.} the radius at which the backward-traced ray lands on the source plane) and, again, the impulse response function, $T_\mathrm{C}(E,t)$ is computed as the number of rays arriving at the observer from the continuum source as a function of photon energy and time.

The reflection and continuum response functions are summed with coefficients appropriate to produce the desired reflection fraction, here taken to be a ratio of 2.5 between the reflection and continuum spectrum photon count rates over the $0.1-10$\keV\ energy band in line with the best-fitting model to the X-ray spectrum of 1H\,0707$-$495.

\begin{figure}
\centering
\includegraphics[width=85mm]{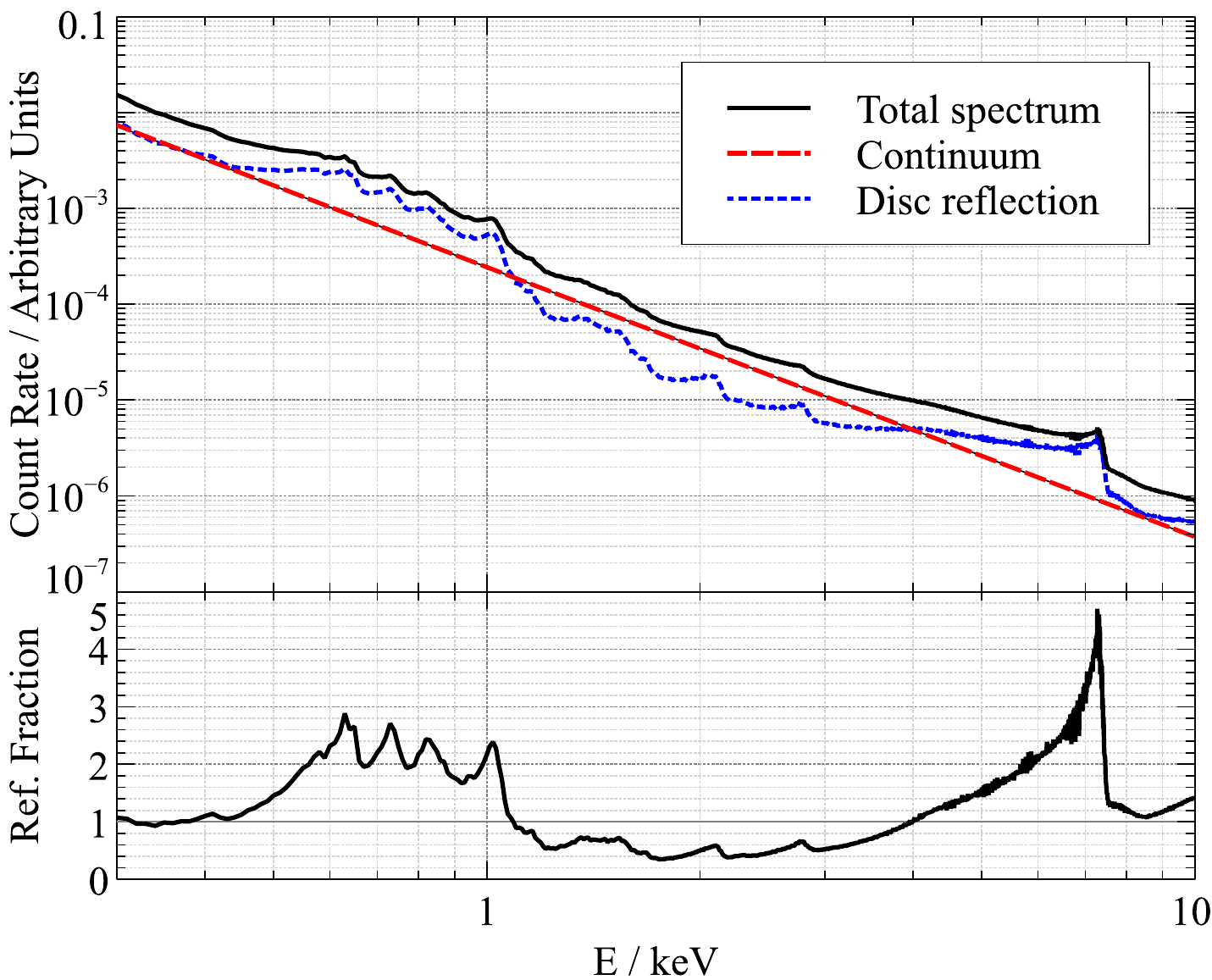}
\caption[]{The time-average spectrum of the model showing the contributions from the directly observed continuum emission and that reflected from the accretion disc, along with the fractional contribution of the reflection from the accretion disc \textit{vs.} the continuum at each energy.}
\label{spectrum.fig}
\end{figure}

We adopt the best-fitting spectral parameters of the NLS1 galaxy, 1H\,0707$-$495, shown in Table~\ref{par.tab} for the photon index of the power law continuum emitted by each point in the corona and for the parameters of the black hole, accretion disc and rest frame reflection spectrum. The time-average spectrum of the model is shown in Fig.~\ref{spectrum.fig}.

\begin{table}
\caption{Parameters of the best-fitting model reflection spectrum to observations of the NLS1 galaxy, 1H\,0707$-$495, used as the basis for simulations of X-ray reverberation.}
\begin{tabular}{lll}
  	\hline
   	\textbf{Component} & \textbf{Parameter} & \textbf{Value} \\
	\hline
	Continuum & Photon index, $\Gamma$ & 3.0 \\
	\hline
	Disc reflection & Inclination, $i$ / deg & 53 \\
	& Spin parameter, $a$ / $GMc^{-2}$ & 0.998 \\
	& Inner radius / \rg & 1.235 \\
	& Outer radius / \rg & 1000 \\
	&  Iron abundance / Solar & 8 \\
	& Ionisation parameter, $\xi$ & 50 \\
	\hline
	& Reflection fraction (0.1-10\keV) & 2.5 \\
	\hline
\end{tabular}
\label{par.tab}
\end{table}

Ray tracing is conducted in spatial units of gravitational radii (1\rg$=GM/c^2$) and temporal units of $GM/c^3$ in order to exploit the scale invariance of the gravitational field with respect to the black hole mass, $M$. When expressed in gravitational radii, the effect of light bending at the equivalent distance from the singularity is identical regardless of the black hole mass. So too is the (relativistic) Keplerian orbital velocity and the  associated Doppler shift and gravitational redshift. The results of ray tracing calculations can trivially be applied to the reverberation of X-rays around any black hole simply by converting the natural gravitational units into physical units. The reverberation lags scale simply as the black hole mass, $M$, and the frequencies at which they are observed, measured in the equivalent units of $c^3 (GM)^{-1}$, scale inversely with the black hole mass. The predicted profiles of the lag-frequency and lag-energy spectra are unchanged. Taking the mass of the black hole in 1H\,0707$-$496 to be $2\times 10^{6}$\Msun\ \citep{zhou_wang}, $GM/c^3 \sim 10$\\s and $c^3 (GM)^{-1} \sim 0.1$\Hz.

\section{Lag Spectra for a Point Source}
\label{point.sec}
\begin{figure}
\centering
\includegraphics[width=85mm]{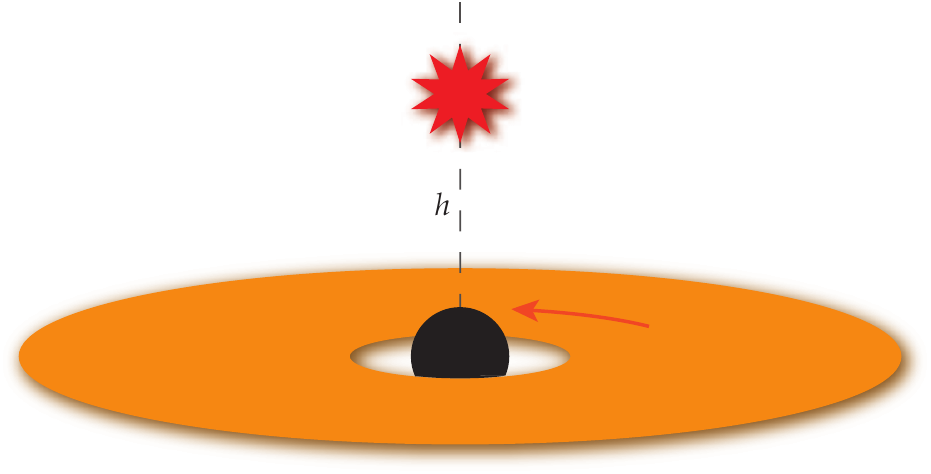}
\caption[]{In the simplest reverberation model, X-rays are emitted from an isotropic point source that is stationary on the black hole rotation axis above the plane of the accretion disc.}
\label{pointsource_setup.fig}
\end{figure}

\begin{figure*}
\centering
\subfigure[$h = 5$\rg] {
\includegraphics[width=85mm]{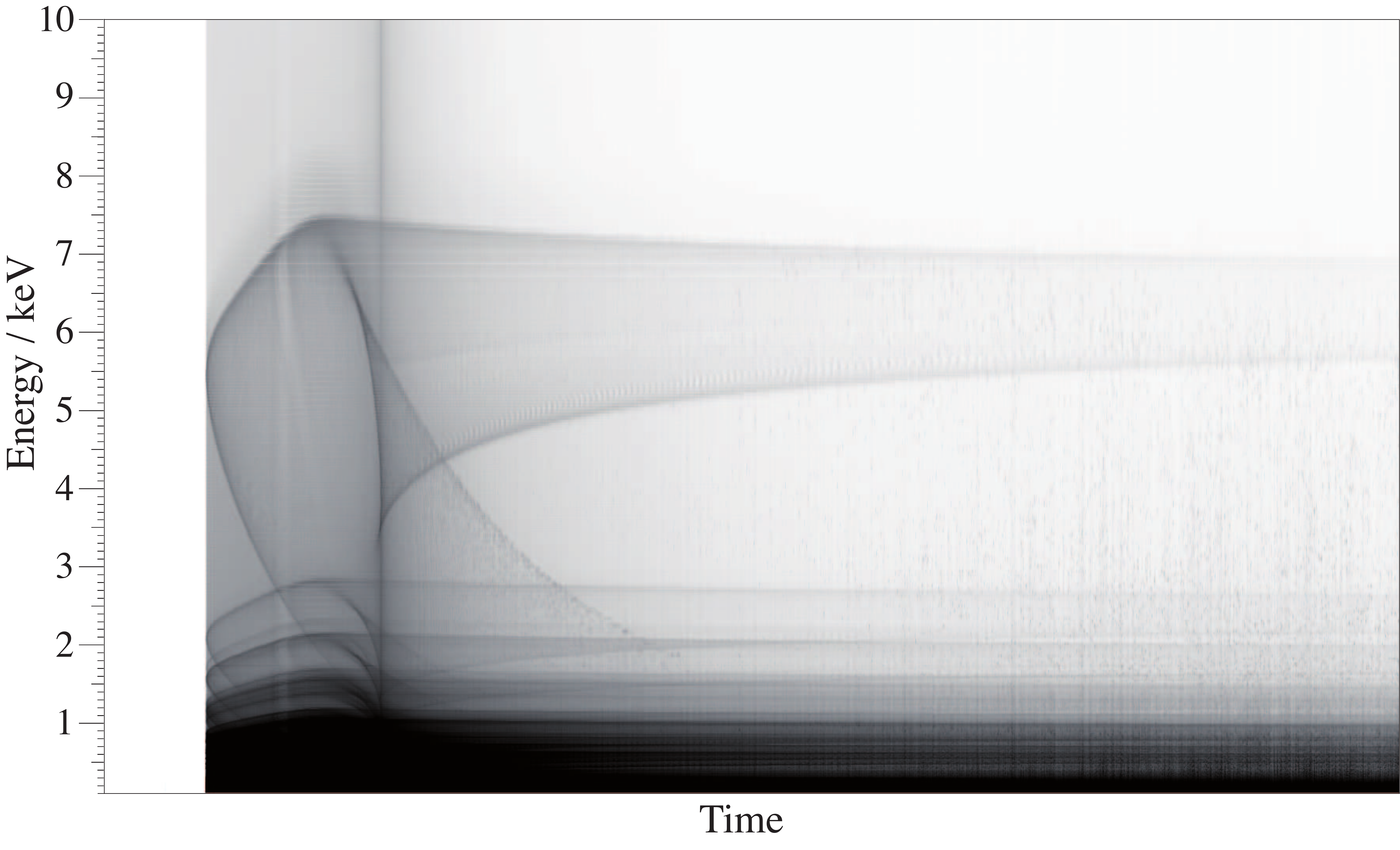}
\label{ent_point.fig:h5}
}
\subfigure[$h = 1.5$\rg] {
\includegraphics[width=85mm]{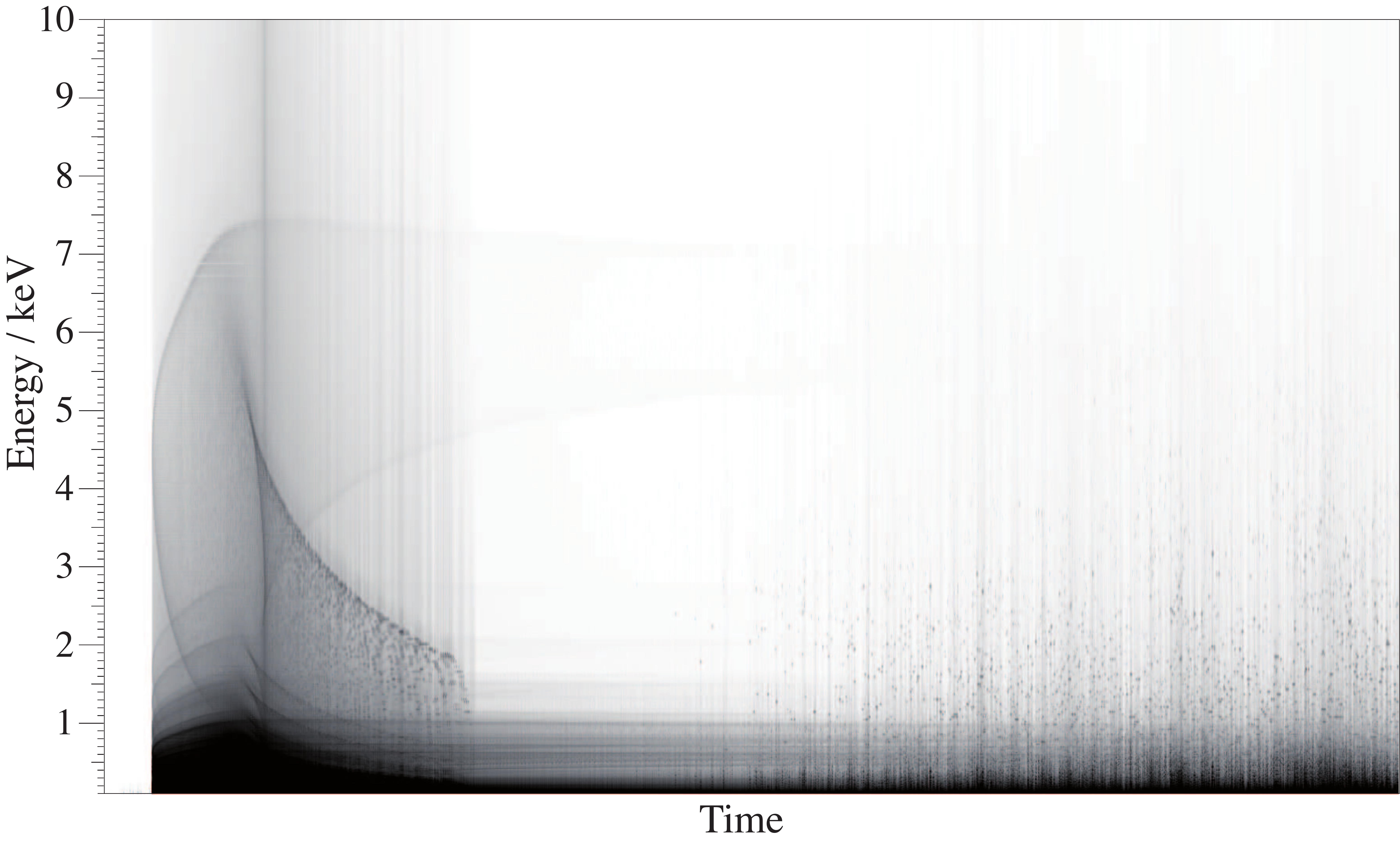}
\label{ent_point.fig:h2}
}
\centering
\caption[]{Time and energy resolved response functions for an accretion disc illuminated by an instantaneous flash from a point source at a height of \subref{ent_point.fig:h5} 5\rg\ and \subref{ent_point.fig:h2} 2\rg\ above the singularity located on the spin axis of the black hole. The rate of photon arrival from the accretion disc at a distant observer is shown as a function of time and for distinct photon energies (as measured by the observer, accounting for Doppler shifts and gravitational redshift) with darker shading indicating greater photon flux.}
\label{ent_point.fig}
\end{figure*}

Before considering luminosity fluctuations propagating through extended coronae, we look first at the simple case of a point source shown in Fig.~\ref{pointsource_setup.fig} in order to illustrate the basic features of lag spectra. X-ray reverberation and lag spectra from point sources is considered at length by \citet{lag_spectra_paper} and \citet{cackett_ngc4151}. Fig.~\ref{ent_point.fig} shows the time and energy resolved response functions of rays reflected from the accretion disc. Shading corresponds to the photon flux received by an observer as a function of time and photon energy after the continuum rays from the point source are received.

The simplest model of the lag-energy spectrum is the average arrival time of photons at a given energy in the impulse response function, calculated as the mean of the arrival time from the response function;
\begin{equation}
\bar\tau(E) = \frac{1}{\int N(E,t)\,dt}\int t\,N(E,t)\,dt
\end{equation}

\begin{figure*}
\centering
\subfigure[] {
\includegraphics[width=85mm]{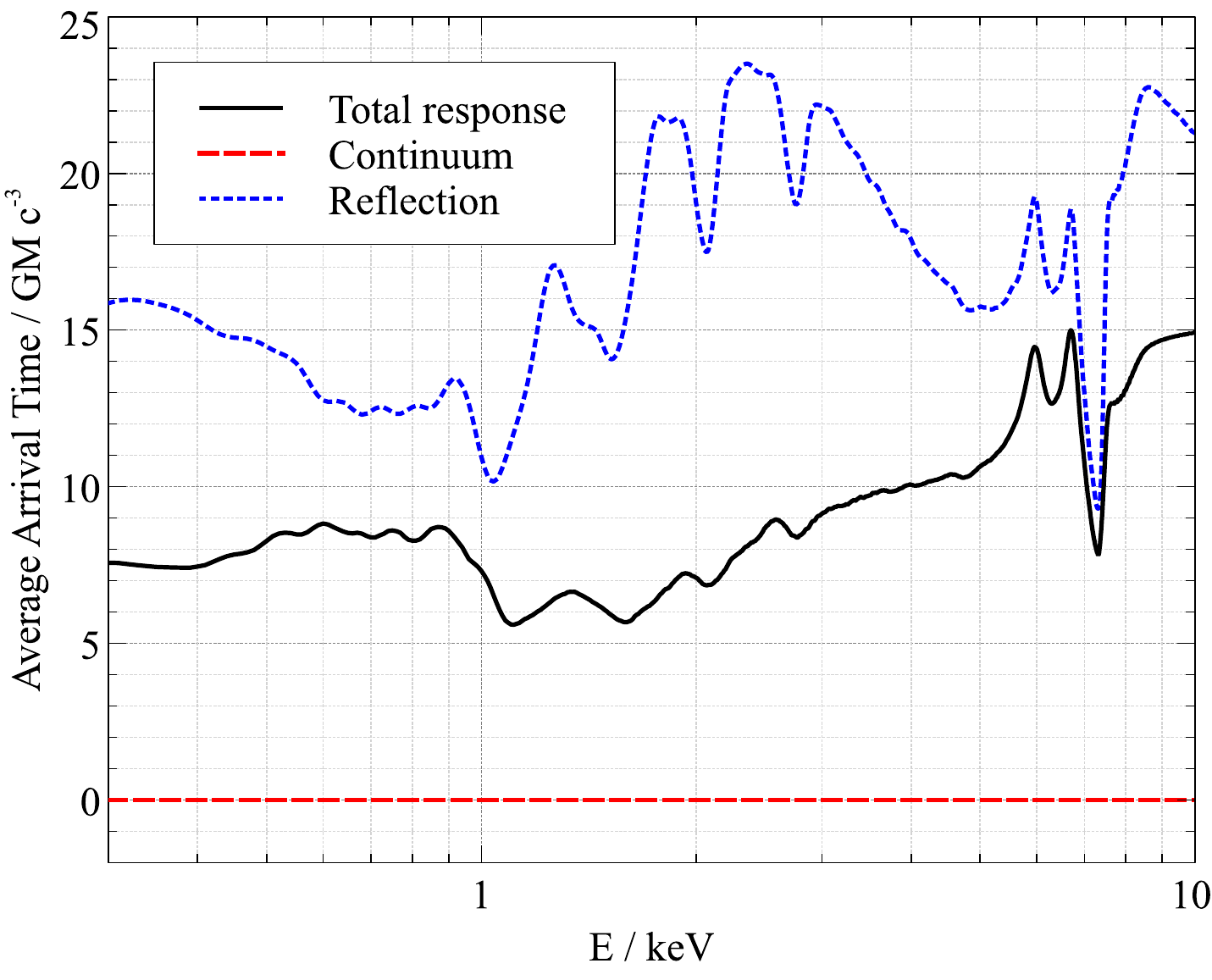}
\label{pointsource.fig:lagen}
}
\subfigure[] {
\includegraphics[width=85mm]{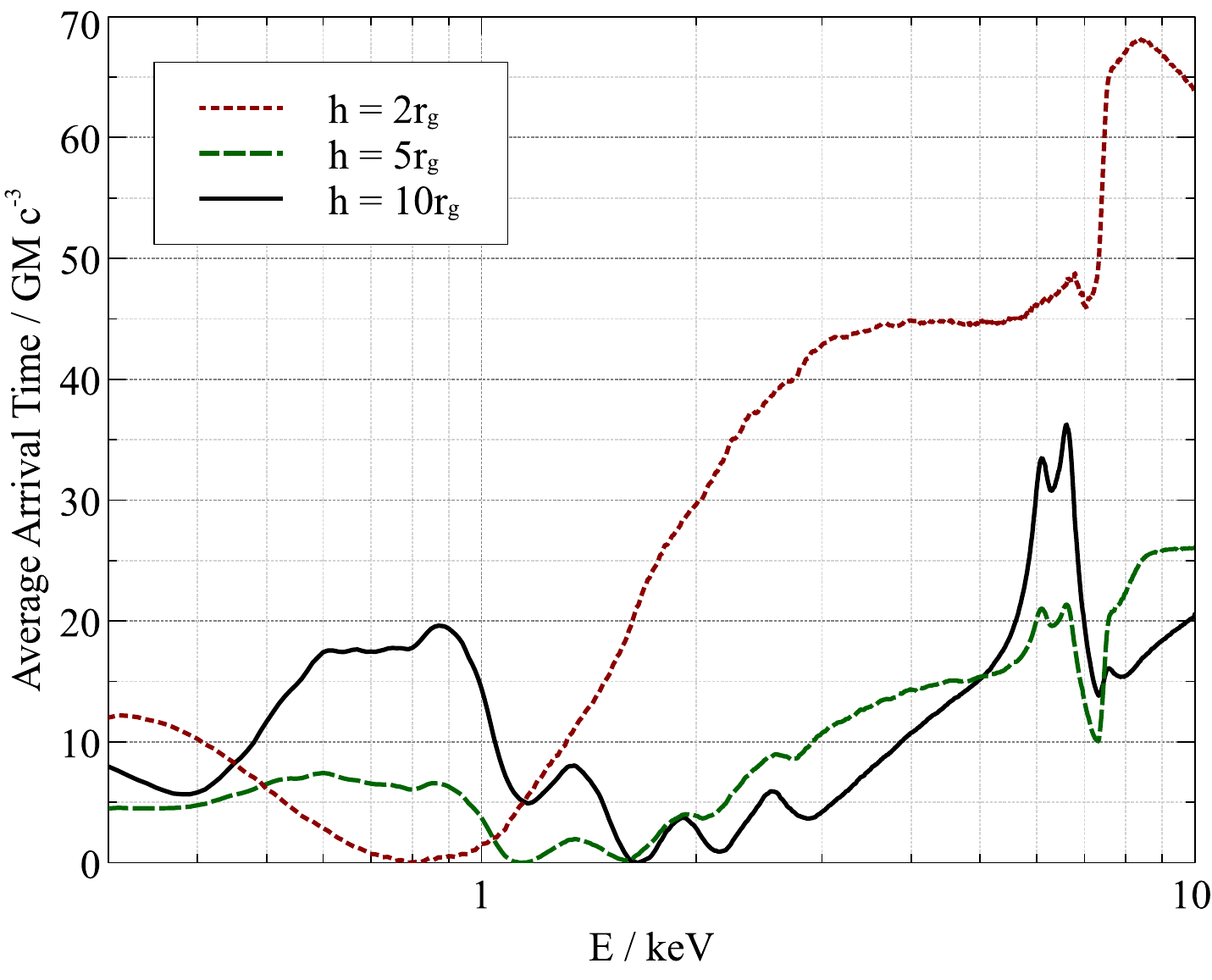}
\label{pointsource.fig:height}
}
\caption[]{\subref{pointsource.fig:lagen} The average arrival times (relative to the earliest arrival of photons in each case) of photons as a function of energy after an instantaneous flash of emission from a point source located 5\rg\ above the plane of the accretion disc on the black hole's rotation axis. Also shown are the arrival times of the photons that make up the continuum and reflected components. \subref{pointsource.fig:height} The variation in the overall average photon arrival times as the point source is moved in height above the singularity. }
\label{pointsource.fig}
\end{figure*}

The average arrival time of photons originating from an isotropic point source located at a height of 5\rg\ above the singularity is shown in Fig.~\ref{pointsource.fig:lagen}. The average arrival time of all photons is shown as well as that for photons in just the directly-observed continuum and reflected components. The form of the lag-energy spectrum can be simply interpreted, with the earliest arriving photons in the 1-2\keV\ energy band, which is dominated by the directly-observed continuum photons, as can be seen from the reflection fraction shown in Fig.~\ref{spectrum.fig}. These continuum photons travel directly from the point source to the observer and, therefore, will always travel the shortest path.

The energy bands 0.5-1\keV\ and 4-7\keV\ are dominated by the reflection from the accretion disc, corresponding to the soft excess and the iron K fluorescence line, respectively. Due to the extra distance travelled by these photons from the primary source to the disc, they are delayed with respect to the continuum-dominated 1-2\keV\ band. The time lag corresponds to the average light travel time over all paths from the source to the disc in the curved spacetime around the black hole (with light travelling more slowly the closer it passes to a massive object, \citealt{shapiro}). The time lag is then diluted according to the reflection fraction in each of the energy bands as both the continuum and reflected components will contribute in varying proportions to each band, delaying the arrival of a `continuum-dominated' band and advancing the arrival of one `reflection dominated.' Varying the overall reflection fraction simply introduces a linear scaling of the time lag, while preserving the overall shape of the lag spectrum \citep{lag_spectra_paper,cackett_ngc4151}.

A deeper understanding of the light travel times associated with X-ray reverberation can be gleaned from the average arrival time of photons in just the reflected component shown in Fig.~\ref{pointsource.fig:lagen} and also in the time- and energy-resolved response function (Fig.~\ref{ent_point.fig}), which will be instructive when interpreting the lag-energy spectra obtained from luminosity fluctuations that propagate through extended coron\ae. Focusing particularly on the iron K$\alpha$ fluorescence line between 3 and 10\keV\ in which photons are shifted from their rest frame energy of 6.4\keV, the earliest photons to be observed are those at 5\keV. This is expected considering the light travel time of photons in flat, Euclidean space from a point source, to the disc and then to the observer, neglecting the effect of spacetime curvature. The classical reflected path length from a point source 5\rg\ above the disc plane to an observer at an inclination of $53\deg$ is minimised for reflections approximately 8\rg\ from the centre and this annulus on a relativistic accretion disc around a black hole will shift the 6.4\keV\ line emission to between 4 and 7.5\keV, with a weighted average at 5\keV. The light travel time is longer to the outer parts of the disc, producing the delayed double-peaked lag profile centred on the rest frame energy of 6.4\keV. The double peaks correspond to the Doppler shifting of reflection from the more slowly approaching and receding material orbiting in the outer disc.

What is not expected classically, however, is the long delay experienced by line photons between 2 and 4\keV. These photons are delayed from their classically predicted arrival time to a greater extent by the Shapiro delay they experience passing close to the black hole (both towards and away from the disc). This results in the leading edge of the response function curving to later times for the most redshifted and blueshifted emission received from the inner edge of the disc on the nearside to the observer.

A distinctive `loop back' pattern is produced in the emission line response, visible between around 2 and 7\keV\ where the line tracing the delayed reception of the most blueshifted photons appears to curve back in energy forming a second peak at later times in the redshifted wing of the line. This consists of photons reflected from the back side of the disc (that would classically be hidden behind the black hole's shadow). These photons are gravitationally lensed into the observer's line of sight, starting with those blueshifted from the approaching side of the back of the disc (just out of sight of the observer) and shifting to those that have travelled round the black hole from the receding side of the disc, leading to a re-emergence of the red wing of the line. An intense response is seen from the part of the disc immediately behind the black hole shadow. This produces the flash represented by the dark band and the following triangular structure between 4 and 5\keV\ (redshifted from the rest frame energy of the line as these photons are reflected from the innermost disc radii). The region of the accretion disc immediately behind the singularity is most magnified as its emission is lensed into the line of sight \citep{reynolds+99}. Since the full reflection spectrum is considered here, including not just line emission but the reflected continuum, this re-emergence results in a flash of emission across all X-ray energies, not just the redshifted energies in the line, represented by the vertical band seen to coincide with the re-emergence of lien photons down the response function.

Photons reflected from the innermost parts of the disc become `trapped' close to the photon orbit at 2\rg\ around a maximally spinning black hole. These extremely redshifted photons can spiral around the black hole several times before being turned in the direction of the observer and being observed at a much later time. The combination of these effects leads to the extended low-energy tail of the response function and the 2-4\keV\ hump in the average arrival time of the reflected photons, however very few of the line photons are reflected down to this energy band so they become lost in the continuum (that arrives much earlier) and are not detected in the overall lag-energy spectrum.

If the radius at which photons were reflected could be determined by the distant observer, in addition to the outward propagating ellipsoid representing the response from successively larger radii in the disc at later times, the response from the inner disc would appear to move inwards at later times. The two components to the response would split at the radius from which the earliest response is seen and the delay experienced by photons passing close to the black hole would lead to the inner disc response appearing as an inward propagating ring towards the innermost stable orbit. 

A similar effect is seen in the soft X-ray excess. Once shifted by Doppler and gravitational redshifts, the responses of the component emission lines sit atop one another in energy so the specific features of each line cannot be distinguished. In the overall lag-energy spectrum, a broad hump is seen from the delayed arrival of these reflected photons with respect to the continuum and in the average arrival of just the reflected photons, slightly earlier arrival is seen above 0.5\keV\, with later arrival below this of the redshifted wings of these lines.

Fig.~\ref{pointsource.fig:height} shows the variation in the lag-energy spectrum as a function of the source height above the singularity. Comparing the average arrival times of photons originating from sources at 5 and 10\rg\ above the black hole, it is clear that the most significant effect of increasing the source height is increasing the average lag time between the arrival of continuum photons, dominating the 1-4\keV\ energy band and both the reflection seen in the soft excess (0.3-1\keV) and the iron K line (4-7\keV). Indeed, \citet{lag_spectra_paper} show that the source height is the most significant factor in determining the reverberation lag time, this defining the additional light travel time between source and reflector. The lag-energy profile remains similar between the sources at 5 and 10\rg, however changes significantly when the source is extremely close to the black hole, as can be seen in the lag-energy spectrum for a source at 2\rg\ above the singularity. When the source is so close to the black hole, the majority of X-rays that are emitted are focused towards the black hole and hence onto the inner regions of the accretion disc. This results in a significant response from the redshifted wing of the line that is delayed as the passage of photons propagating in the strong gravitational field is slowed. By contrast, few photons are able to reach the outer parts of the accretion disc, resulting in little response from the line core at late times. Moreover, few photons are able to escape to be observed directly as part of the continuum, explaining the skewed shape of the profile with respect to those for sources at greater heights.

\citet{cackett_ngc4151} consider the dependence of reverberation time lags on the parameters generally associated with the spectral modelling of relativistically blurred reflection from accretion discs. In addition to the height of the X-ray source above the disc, they find that the lag spectra depend upon the inclination of the line of sight to the accretion disc. Observing the disc at a higher inclination (closer to edge-on) increases the component of the orbital velocity projected along the line of sight, thus increases the range of Doppler shifts. For greater inclination, the lagged emission (relative to the continuum) extends up to greater energy with a broader line profile observed in the lag-energy spectrum, while there is little effect on the lag-frequency spectrum between two broad energy bands. The black hole spin was found to influence lag measurements through the location of the innermost stable orbit. For more rapidly spinning black holes, the accretion disc extends closer in and enables the reflection of more highly redshifted photons in a more extended wing of the line, lagging behind the continuum. Truncation of the disc at larger radius for more slowly spinning (and for retrograde spin) also places the inner edge of the disc further from the point source above the black hole, increasing the average lag. These dependences will remain in lag spectra for extended coron\ae\ and the propagation of fluctuations through these and these parameters will not be considered further in the present work.

\begin{figure*}
\centering
\subfigure[Lag-frequency spectrum] {
\includegraphics[width=85mm]{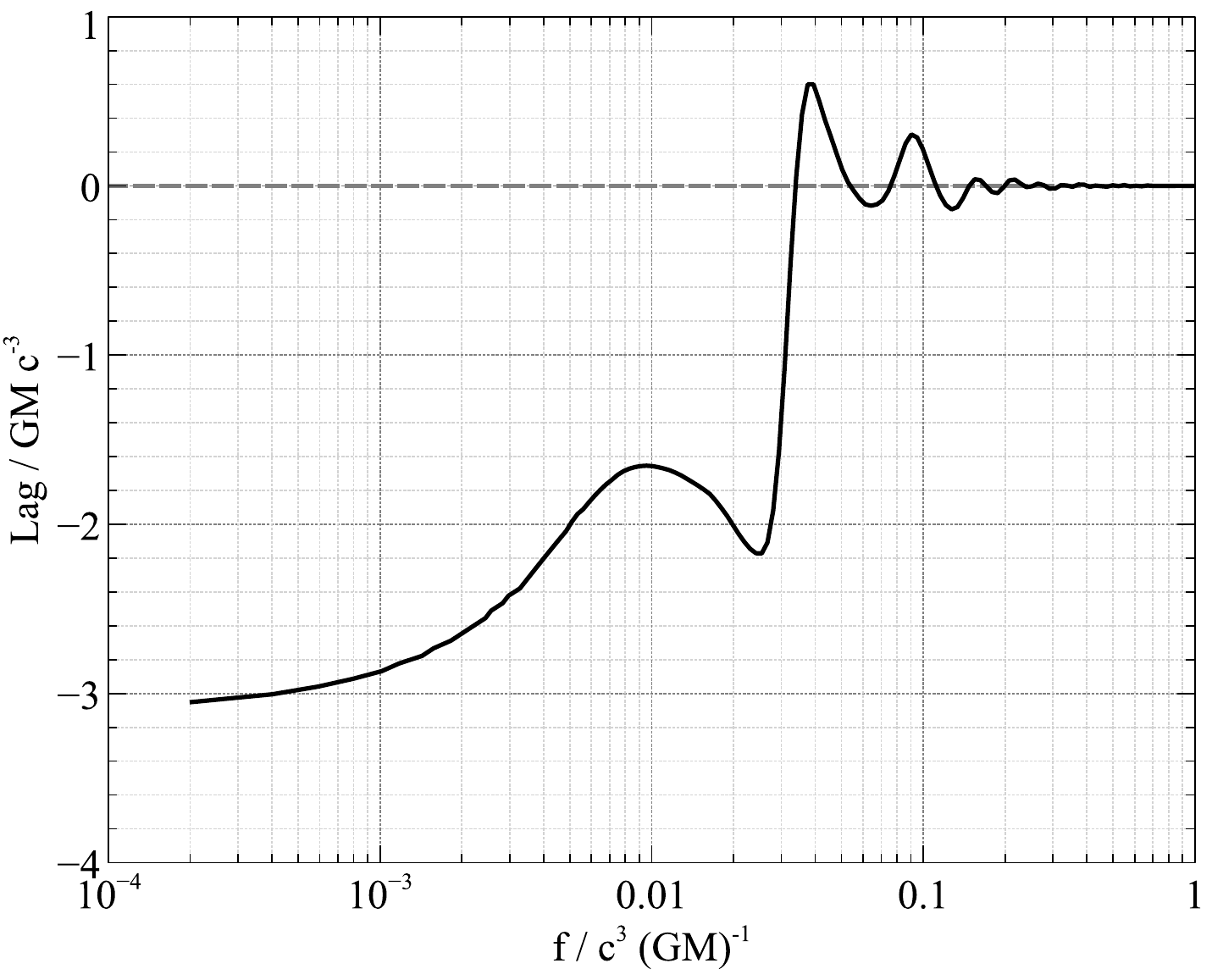}
\label{freqdependence.fig:lagfreq}
}
\subfigure[Frequency dependence of lag-energy spectrum] {
\includegraphics[width=85mm]{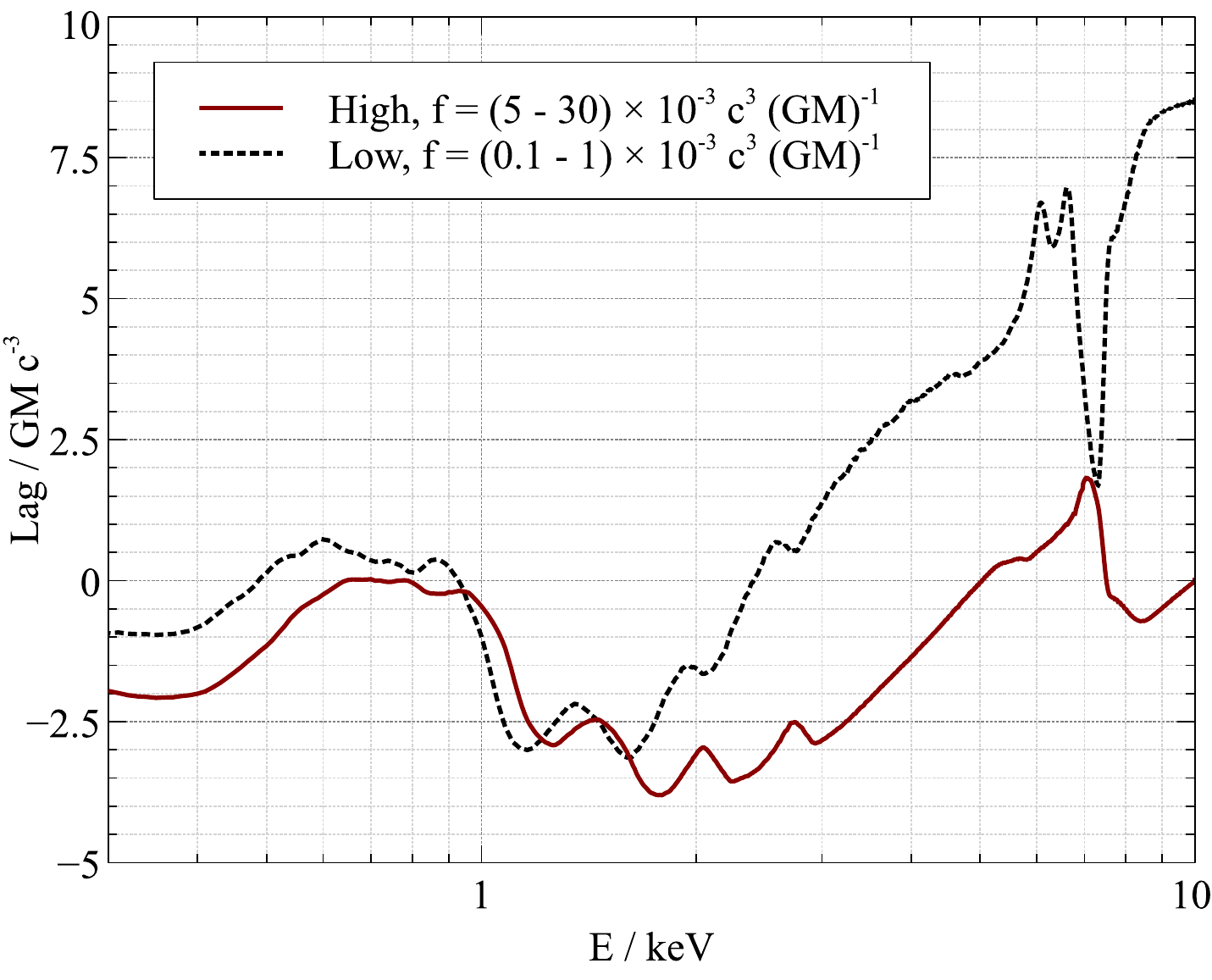}
\label{freqdependence.fig:lagen}
}
\caption[]{\subref{freqdependence.fig:lagfreq} The lag between the 0.3-1\keV\ reflection-dominated band and 1-4\keV\ continuum-dominated band for reverberation from an accretion disc illuminated by point source 5\rg\ above the singularity. A positive lag indicates the harder band lags behind the soft. A negative lag hence represents X-ray reverberation where the softer reflection-dominated band is lagging behind the continuum. \subref{freqdependence.fig:lagen} The variation in the lag-energy spectrum (the lag between variability in the 0.1-10\keV\ reference band and that in narrow energy bands) averaged over high and low frequency intervals, illustrating the effect of measuring the lag-energy spectrum at different frequencies.}
\label{freqdependence.fig}
\end{figure*}

\subsection{Frequency dependence of the lag}

The lag-frequency spectrum, showing the time lag as a function of Fourier frequency (of the temporal components making up the light curve) of the continuum-dominated 1-4\keV\ band with respect to the reflection-dominated 0.3-1\keV\ band is shown in Fig.~\ref{freqdependence.fig:lagfreq}. Following the usual convention, a positive lag indicates that the harder band lags behind the soft, hence the negative lag apparent here indicates that at all frequencies, variability in the reflection-dominated soft band is lagging behind that in the continuum-dominated hard band, as expected due to the extra light travel time from the source to the disc.

As discussed by \citet{lag_spectra_paper}, the lag for the lowest frequency components is equal to the mean of the lag from the response function. A time lag $\tau$ greater than $\frac{1}{2f}$ cannot be detected since at this point the (sinusoidal) Fourier component has been shifted by half of its wavelength. It is therefore impossible to tell (given the definition of the phase angle between the two components in the range $-\pi$ to $\pi$) whether the component has been shifted forward or back by half a wavelength. At $f = \frac{1}{2\tau}$ the phase wraps around from $-\pi$ to $\pi$ and the sign of the measured lag flips. Moving to higher and higher frequency, the sign continues to flip every $2\pi$ shift in phase and the lag, averaged across the time series, decays to zero.

In the lag-frequency spectrum for smoothly decaying response functions (as is expected for reverberation from the disc), the time lag contributed by the longest photon paths for which $\tau > \frac{1}{2f}$ are averaged to zero, leaving only the shorter paths to contribute to the measured lag. This results in the gradual decrease in the measured lag towards high frequencies.

When measuring the lag-energy spectrum at a given frequency, as for real data, the overall shape of the spectrum is retained from the profile of the average photon arrival time as a function of energy when only the reverberation process is considered. The lag times are scaled to lower values when the measurement is taken at higher frequency and the sharp features in the spectrum become smoothed out, as demonstrated in Fig.~\ref{freqdependence.fig:lagen}. We therefore consider the average photon arrival times as a proxy for the lag-energy spectrum to understand the form and dependence on different parameters such that observed features can be identified but note that if these models were to be used to fit real data, the lag-energy spectrum should be computed from the modelled response function for the frequency range considered.

\subsection{The requirement for more complex models}
It is apparent, comparing the frequency and energy dependence of the lag measured in 1H\,0707$-$495 (shown in Fig.~\ref{1h0707_lagspec.fig}) and other Seyfert galaxies with the predictions of models in which X-rays originate from a point source (Figs.~\ref{pointsource.fig} and \ref{freqdependence.fig}), that this simplified model of X-ray reverberation is inadequate to explain the observed features. The dip observed in many objects in the high frequency lag-energy spectrum at 3\keV, apparent in Fig.~\ref{1h0707_lagspec.fig:lagen_high}, is not explained by X-ray reverberation from a point source. There is precisely one ray path from an infinitesimal point source to the observer and this path will always be shorter than those passing via the accretion disc. Hence the earliest response should always be seen in the continuum-dominated 1-2\keV\ band rather than a sharp dip at 3\keV.

Point source models do not account for the transition to the `hard lag,' denoted by not only the softer reflection-dominated band leading the harder continuum-dominated band in the broadband lag-frequency spectrum but by the change in energy dependence from the characteristic profile of X-ray reflection to the systematic increase in lag time with X-ray energy. Models in which luminosity fluctuations propagate inwards from the less energetic outer regions to more energetic inner regions of the corona or from outer to inner regions of the accretion disc to explain this behaviour implicitly assume a finite spatial extent to the corona, thus cannot trivially be reconciled with a point source origin of the X-ray continuum.

\section{Extended Coronae and Propagating Source Fluctuations}
\label{ext.sec}
Measurements of the emissivity profiles of AGN accretion discs, that is the pattern of illumination of the disc by the X-rays emitted from the corona, suggest that in a growing number of cases, the coron\ae\ extend over the surface of the accretion disc. They have been found to extend up to a few tens of gravitational radii and around 30\rg\ in the case of 1H\,0707$-$495 rather than being compact, point-like sources of emission located above the black hole \citep{1h0707_emis_paper,understanding_emis_paper,iras_fix,mrk335_corona_paper}. It is therefore a natural extension to point source models to consider the lag spectra that would be expected from these extended coronae to understand how such models can be further constrained by X-ray timing analyses and to test whether these can explain the observed features of reverberation.

In this section, models of X-ray reverberation from extended coron\ae\ are developed, both for the inward and outward propagation of fluctuations through coron\ae\ extending over the accretion disc at a low height and upward through vertically collimated coron\ae. We begin with simplified prescriptions of propagation at a constant velocity through the corona in Sections~\ref{propout.sec} and \ref{propin.sec}, before considering propagation linked to viscous propagation through the underlying accretion disc in Section~\ref{viscous.sec}. Once the propagation is established, the hard lag is self-consistently included as a gradient in the photon index of the continuum produced by the corona in Section~\ref{hardlag.sec} and the seeding of the corona by stochastic variations across its extent are considered in Section~\ref{propfluc.sec}.

Cylindrical coron\ae\ are considered with the X-ray emitting region in each case defined by a maximum radius over the disc and by a lower and upper vertical bound above the plane of the disc and the coron\ae\ are taken to have uniform luminosity throughout their extent. While this is not a completely realistic model (for instance the corona may be closer to spheroidal and brighter in the central regions where more energy is injected from the accretion flow) these simplified models will allow the basic properties of extended coron\ae\ to be explored through X-ray timing. Where good agreement is found to observed data, the uniform cylindrical model will represent the approximate geometry and extent of the corona from which the bulk of the X-ray emission is originating.

When considering extended coron\ae\ that are governed by processes on the underlying accretion disc, it is unphysical to take an instantaneous flash from all regions of the corona since for a corona extending 10\rg\ over the surface of the accretion disc but situated only 1\rg\ above the disc, the light travel time through the corona is ten times longer than the reverberation time scale to the disc. It is therefore necessary to account for the propagation of the flash or the fluctuation in luminosity from the site at which energy is injected into the corona. The simplest case of radial propagation at constant velocity, $v$, originating at some radius $\rho_0$ corresponding to either the centre or the edge is considered along with propagation up vertically extended coron\ae. The start time of each ray is calculated from its originating radius, $\rho$, within the corona (measured in the plane parallel to the accretion disc), with $t_0 = (\rho - \rho_0) / v$. 

In these models, the geometry and extent of the corona are put in by hand along with the propagation of luminosity fluctuations through its extent. No assumption is made about the means by which the corona is energised by the underlying accretion flow, although it is commonly speculated that this is by the reconnection of magnetic fields associated with the accretion disc. Magnetic flux loops can arise buoyantly from a dynamo action within the disc. These may reconnect with one another above the disc energising the surrounding plasma, forming a corona that extends over the surface of the disc \citep{liu+03}.

While most likely not accounting for the full complexities of the system, the models herein allow the characteristics of energy injection and propagation mechanisms to be explored through X-ray timing and enable some constraints to be placed on these as-yet unknown phenomena from observational data. They enable constraints to be placed on the geometry of the corona and the propagation of fluctuations which will in turn enable constraints to be placed upon models of underlying emission mechanisms. Reverberation measurements are simulated for a range of coronal parameters, detailed in Table~\ref{model_params.tab} (in addition to the frozen parameters of the continuum and reflection spectrum shown in Table~\ref{par.tab}), to enable the manifestation of these parameters to be explored in X-ray observations.

\begin{table}
\caption{Coronal parameters in the ray tracing simulations. Unless otherwise specified, reverberation measurements are simulated for various combinations of these parameters to reproduce radially extended and vertically collimated coron\ae. Measurements are simulated taking the fixed values of the parameters for the continuum and reflection spectrum specified in Table~\ref{par.tab}.}
\begin{tabular}{lll}
  	\hline
   	\textbf{Property} & \textbf{Parameter} \\
	\hline
	Coronal extent & Inner cylindrical radius (fixed at 0) \\
	& Outer cylindrical radius, $R$ \\
	& Lower vertical extent, $z_0$ (fixed at 1.5\rg) \\
	& Upper vertical extent, $z_0 + \Delta z$ \\
	\hline
	Propagation & Direction (outward, inward, upward) \\
	& Velocity, $v$ \\
	\hline
	Luminosity profile & Radial and vertical variation in \\
	& luminosity (here taken to be uniform) \\
	\hline
\end{tabular}
\label{model_params.tab}
\end{table}

\subsection{Outward Propagation through Radially Extended Coronae}
\label{propout.sec}

\begin{figure}
\centering
\includegraphics[width=85mm]{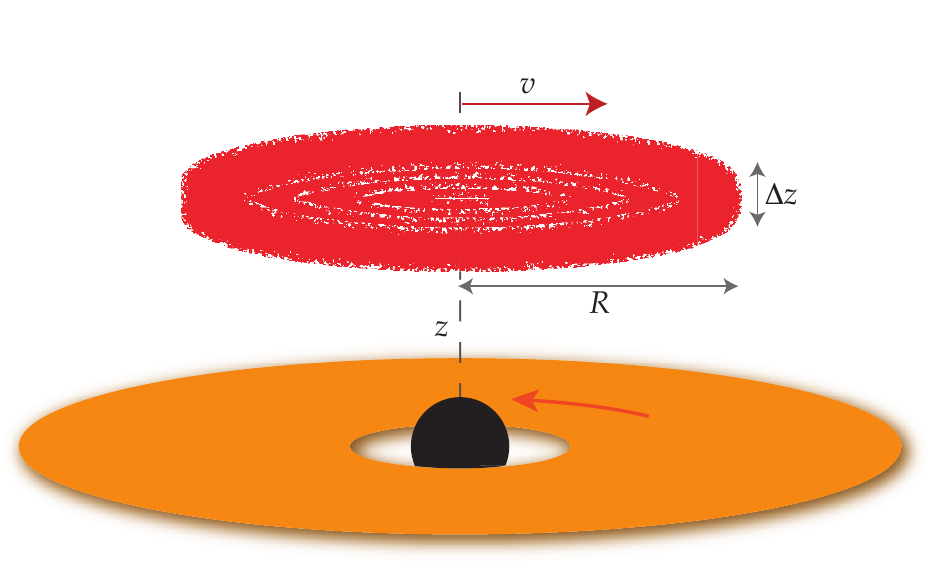}
\caption[]{Schematic of the model of an extended corona. X-rays originate in a cylindrical region spanning vertically between 1.5 and 2\rg\ above the plane of the disc and with varying radius, $R$. The start time of each ray is set to reproduce the outward propagation of a single flash at speed $v$ from the centre.}
\label{propout.fig}
\end{figure}
We first consider the case of luminosity fluctuations propagating outward through a corona extended radially over the surface of the accretion disc, illustrated in Fig.~\ref{propout.fig}. This could represent a scenario in which the majority of the energy from the accretion flow is not injected into the corona until the innermost radii on the accretion disc from which point this energy flows outwards through its volume, whether through the energetic particles themselves that are accelerated in the inner regions or through the accumulation of magnetic flux loops outwards that then accelerate particles in the outer part of the corona.

\begin{figure*}
\centering
\subfigure[$v = c$] {
\includegraphics[width=85mm]{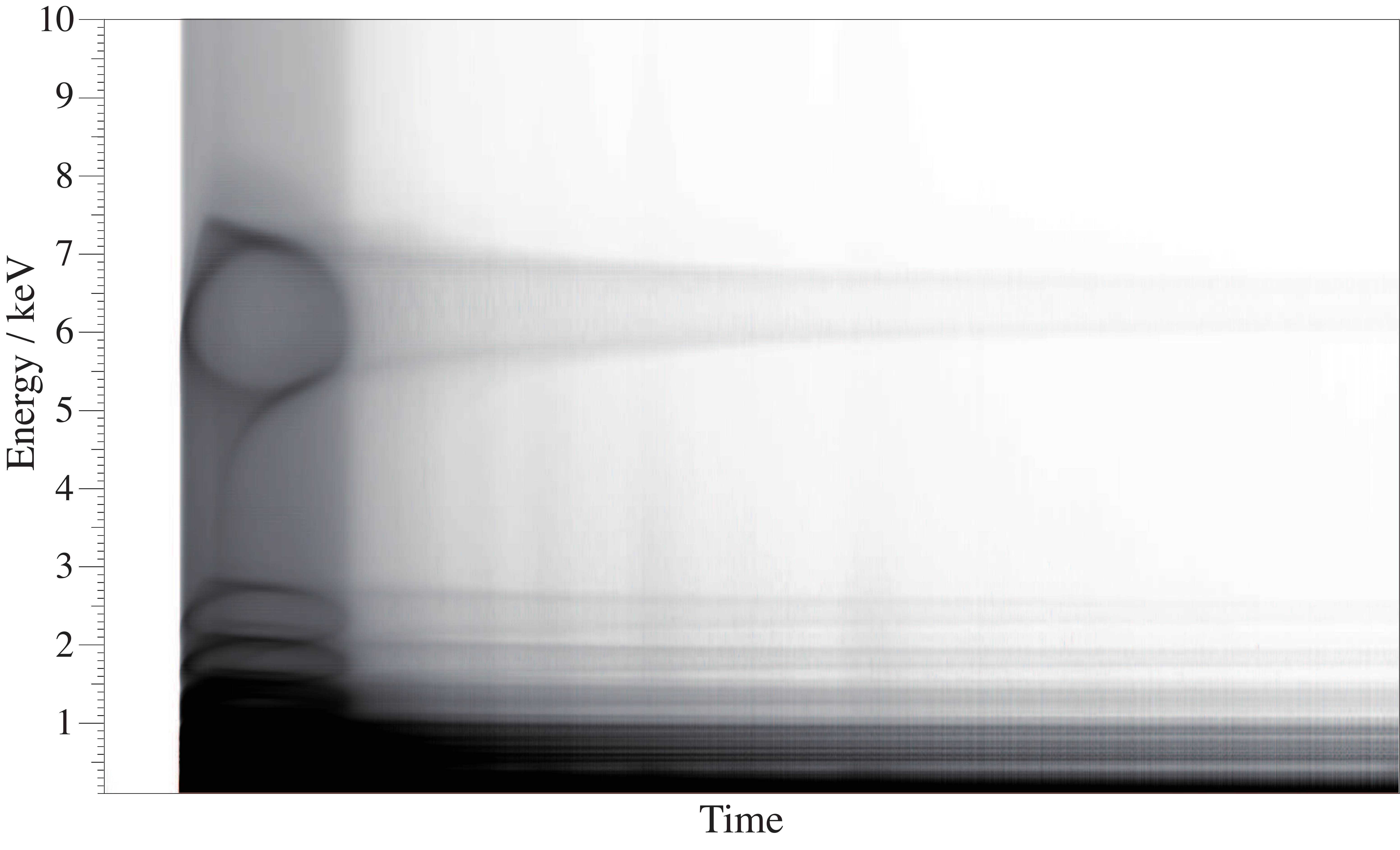}
\label{ent_propout.fig:c}
}
\subfigure[$v = 0.1c$] {
\includegraphics[width=85mm]{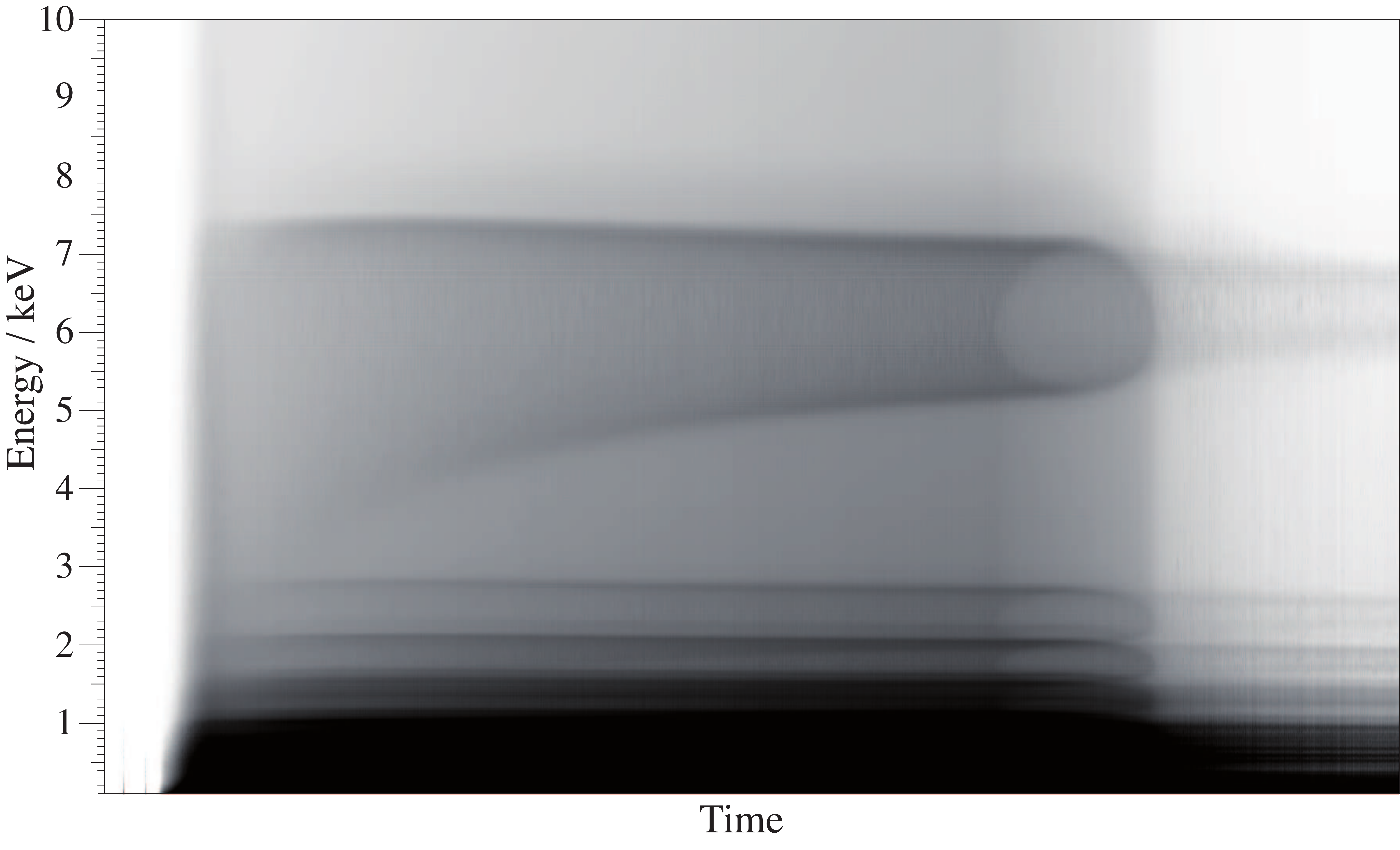}
\label{ent_propout.fig:0.1c}
}
\caption[]{Time and energy resolved functions for the reverberation from the accretion disc of a single flash propagating outward through a corona extending over the surface of the disc to 30\rg\ at \subref{ent_propout.fig:c} the speed of light and \subref{ent_propout.fig:0.1c} one-tenth the speed of light. This response function represents continuous luminosity fluctuations that propagate outwards by convolving this response function with the underlying time series describing the variability.}
\label{ent_propout.fig}
\end{figure*}

While the radial extent of the corona over the disc surface is varied, the corona is fixed vertically. The X-ray emitting region extends between 1.5 and 2\rg\ above the disc plane, as found by \citet{lag_spectra_paper} in simpler models (not accounting for propagation) to reproduce the measured reverberation lag as a function of frequency of 1H\,0707$-$495. \citet{lag_spectra_paper} show that varying the height of a radially extended corona simply scales the measured lag time. The lag profile as a function of energy should not be altered in the regime where the radial extent remains much greater than the vertical, since the radial extension provides consistent illumination over the surface of the disc even while the height is varied. This fixed height will act as a starting point to explore the effect of the radial extent and propagation mechanism on the measured lag profiles and allow the analysis to be extended to the energy dependence of the lag.

The time- and energy-resolved response functions of an accretion disc illuminated by such a corona, extending to a radius of 30\rg\ over the plane of the disc, through which a single flash propagates outwards are shown in Fig.~\ref{ent_propout.fig}. The case of light speed propagation is shown alongside propagation at $0.1c$. When the flash or luminosity fluctuations propagate through the corona at the speed of light, the response of the accretion disc is qualitatively similar to that in the case of illumination by a point source. Early response is seen from the inner parts of the disc with the double-peaked line profile from the approaching and receding sides of the outer disc responding at later times. The tail of the emission line response is seen to low energies along with the loop-back of delayed emission to progressively lower energies as delayed, redshifted photons arrive from the back side of the disc. The loop in the response function spans a shorter range in energy due to the illumination of the disc by the extended corona irradiating not only the innermost radii but providing prompt illumination of the disc out to 30\rg\ wherefrom a smaller range of Doppler shifting and gravitational redshifting is measured. The propagation of the flash from the inner to the outer region of the corona delays the arrival of the double-peaked line from the outer disc with respect to the redshifted tail of the response and stretches the loop in the response function to later times.

When the flash propagates outward significantly slower than the speed of light, the response function changes shape dramatically. The flash propagates outward sufficiently slowly that at early times, illumination is provided to only the inner regions of the disc. The line response is broad early on as highly redshifted and blueshifted emission is seen from the innermost parts of the disc, while the response becomes narrower at later times. The flash is delayed reaching the outer part of the corona which provides the predominant source of illumination to the outer disc, since rays emitted from the inner regions are focused onto the inner parts of the disc. The loop back is much more subtle as photons emitted throughout the corona (over a wide range of times) can be bent towards the inner part of the disc to re-emerge delayed and redshifted from the inner part of the back side of the disc.

\begin{figure*}
\centering
\subfigure[$v = c$] {
\includegraphics[width=56mm]{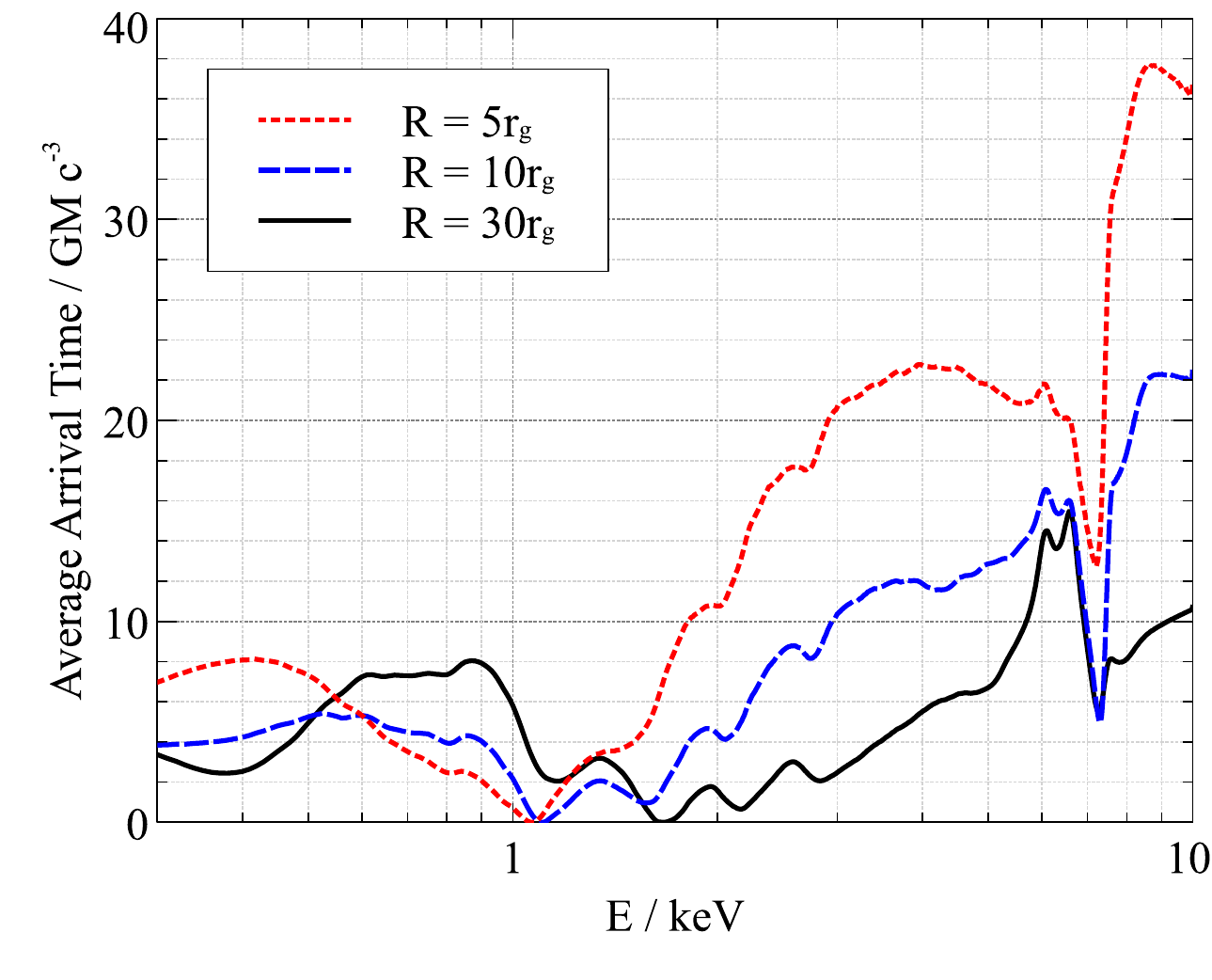}
\label{avg_arrival_propout.fig:c}
}
\subfigure[$v = 0.1c$] {
\includegraphics[width=56mm]{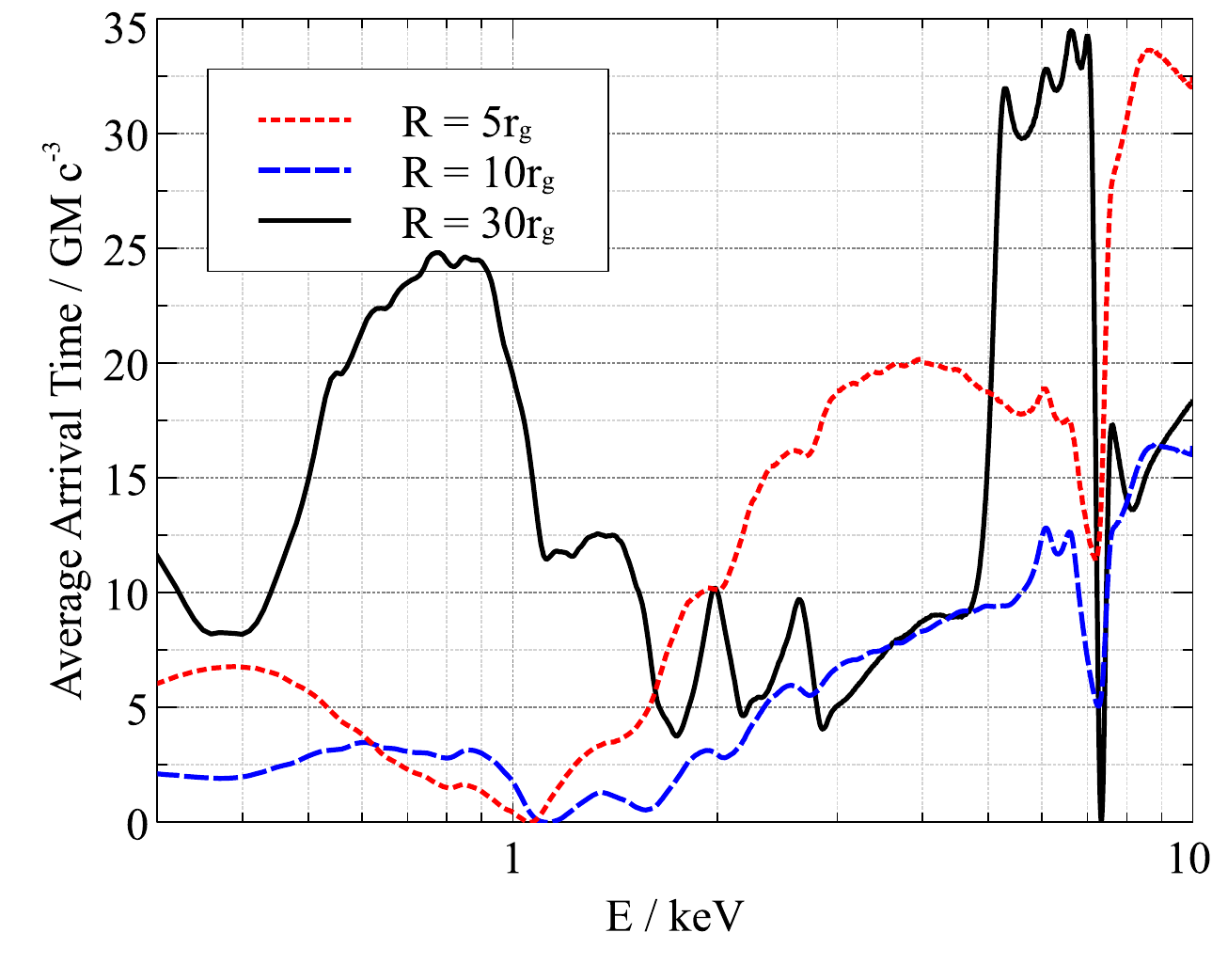}
\label{avg_arrival_propout.fig:0.1c}
}
\subfigure[$v = 0.01c$] {
\includegraphics[width=56mm]{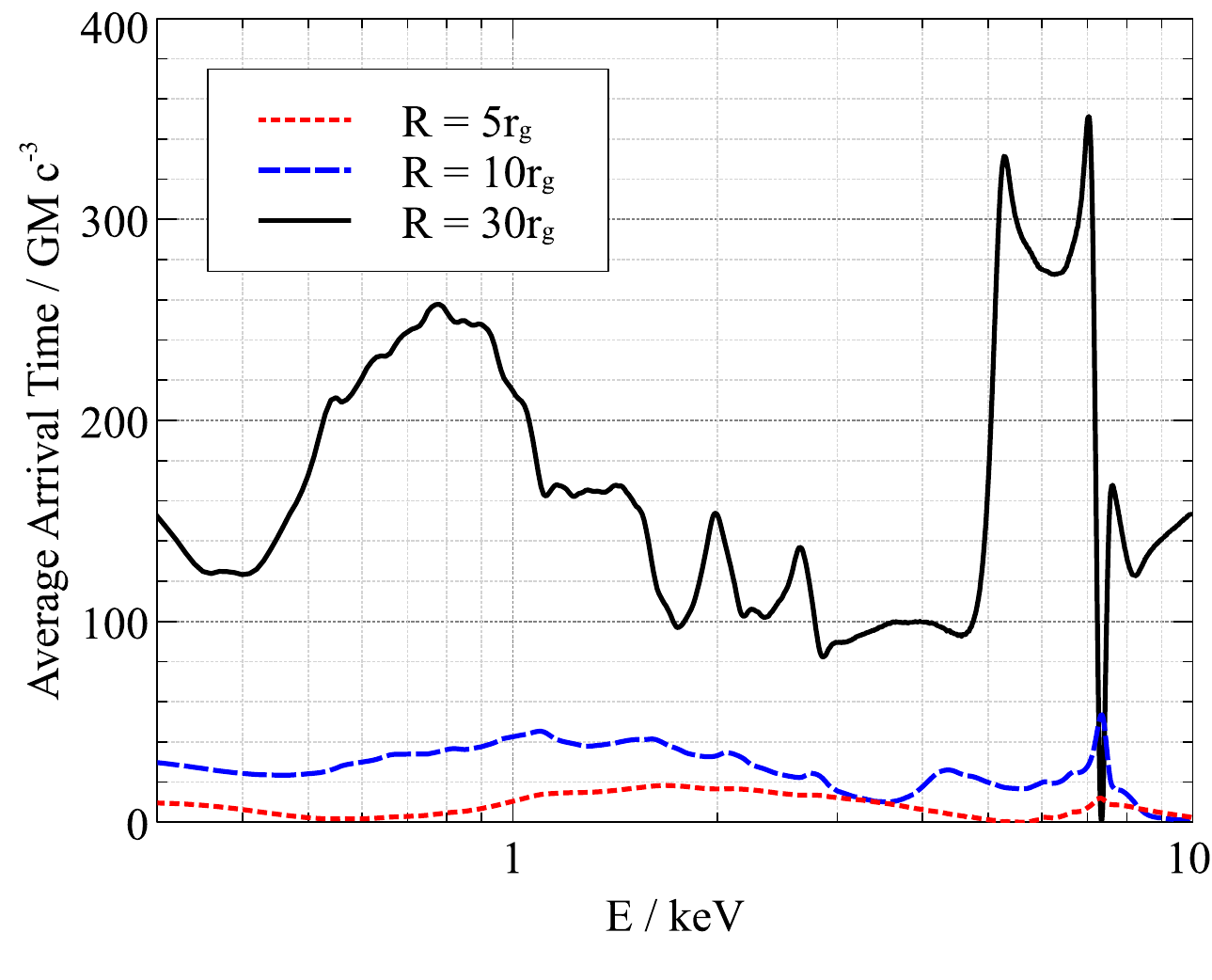}
\label{avg_arrival_propout.fig:0.01c}
}
\caption[]{The average arrival times (relative to the earliest arrival of photons in each case) of photons as a function of energy after a luminosity fluctuation propagating radially outwards through the corona at varying radial velocity $v$. Coron\ae\ are considered of different radial extent, $R$, over the surface of the accretion disc and are located between 1.5 and 2\rg\ above the disc plane. }
\label{lagspec_propout.fig}
\end{figure*}

The effects of propagation through the extended corona are readily seen in the average arrival time as a function of photon energy (proxy to the lag-energy spectrum) shown in Fig.~\ref{lagspec_propout.fig}. Shown are the lag-energy functions for coronae of varying radius over the surface of the disc and for outward propagation at speeds of $c$, $0.1c$ and $0.01c$.

Considering the case of light-speed propagation of fluctuations for coron\ae\ extending to different radii, we see that the propagation has little effect on the lag as a function of energy, with the response functions looking very similar to those for reverberation of X-rays originating from a point source. The earliest arrival is seen in the continuum-dominated $1-3$\keV\ band, with delayed emission in the soft excess at $0.5-1$\keV. The redshifted wing of the 6.4\keV\ iron K$\alpha$ emission line reflected from the inner disc is seen to respond earlier than the core of the emission line from the outer disc and the core of the line is again split into two peaks, blueshifted and redshifted from the approaching and receding sides of the disc.

For more compact coron\ae, more of the rays are focused towards the black hole and hence onto the inner parts of the accretion disc. This changes the shape of the lag-energy spectrum now that the broad iron K$\alpha$ line is dominated by emission from the inner part of the disc. This enhances the redshifted wing of the line which now dominates the lag-energy spectrum, with little contribution from the outer disc, producing the core of the line. A similar effect is seen in the soft excess emission. As reported by \citet{lag_spectra_paper}, the reverberation lag time (the delay between the arrival of photons in continuum- and reflection-dominated bands) becomes shorter for more radially extended coronae since the rays travelling from the outer parts of the corona to the accretion disc do not pass so close to the black hole, meaning they experience less delay from passing through the strong gravitational field.

Once the propagation time becomes much longer than the light travel time from the corona to the disc, the lag-energy spectrum starts to change shape. The propagation through the corona dominates the long timescale variability with the response changing over time as the disc appears to be illuminated by an X-ray source moving, from the inner to the outer regions. Most notably, for fluctuations propagating outward, the emission lines take a distinctive square shape. The inner disc is illuminated initially while the fluctuation is still on the inner parts of the corona and the narrow response around the rest-frame energy of the line is delayed until the fluctuation reaches the outer parts of the corona. 

Each of the lag-energy spectra show that the emission in the 1-4\keV\ band, on average, leads that in the 0.3-1\keV\ band, hence the broadband lag-frequency spectra between continuum- and reflection-dominated bands will show the latter lagging behind the former over all frequencies (\textit{i.e.} the lag remains negative, following the convention of a positive lag indicating the harder band leading the softer). At the lowest frequencies, the lag is found to be approximately $2\,GM/c^3$ whatever the radial extent of the corona when the fluctuations propagate at light speed. This then decreases towards zero as higher and higher frequency Fourier components are considered.

When the fluctuation propagates more slowly through the corona, outward propagation can cause the continuum-dominated 1-4\keV\ band to appear delayed with respect to the 0.3-1\keV\ reflection-dominated band. Rays from the innermost parts of the corona, which leads the fluctuation, are bent towards the black hole and, hence, onto the inner regions of the disc. Comparatively few of these early rays are able to escape to be observed as part of the continuum which is predominantly made up of rays from the outer part of the corona that are emitted later. Emission in the redshifted wing of the line at 3-4\keV\ can arrive, on average,  earlier than the continuum-dominated band between 1 and 3\keV\ for the slowest propagation through the more compact coron\ae\ causing the lag-frequency spectrum to turn over to become positive at the lowest frequencies as discussed by \citet{lag_spectra_paper}. This phenomenon does not, however, explain the observed turn-over in lag-frequency spectra since the energy-dependence of the lag at low frequency is simply the average arrival time of the rays shown in Fig.~\ref{lagspec_propout.fig}. The low frequency lag-energy spectrum in this instance follows the characteristic shape of the reflection spectrum with the delayed soft excess emission and iron K$\alpha$ line, and not the smooth increase in arrival time as a function of X-ray energy that is observed.

For slow propagation through the largest coron\ae, however, this turn-over is not seen. In this case of propagation at $0.1c$ and $0.01c$ through coron\ae\ extending 30\rg\ over the plane of the disc, the time delay introduced by the propagation between the illumination of the inner and outer parts of the disc leads to a peak and trough appearing in the soft excess. The redshifted wings of these emission lines seen between 0.2 and 0.5\keV, illuminated by the inner parts of the corona, appear early while the cores of these lines from the outer disc over the energy range 0.5-1\keV\ are delayed. The cores of these lines are now delayed along with the directly-observed continuum emission, counteracting the turn-over in the lag at low frequency, but coron\ae\ larger than around 15\rg\ are required.

\subsection{Inward Propagation}
\label{propin.sec}

\begin{figure}
\centering
\includegraphics[width=85mm]{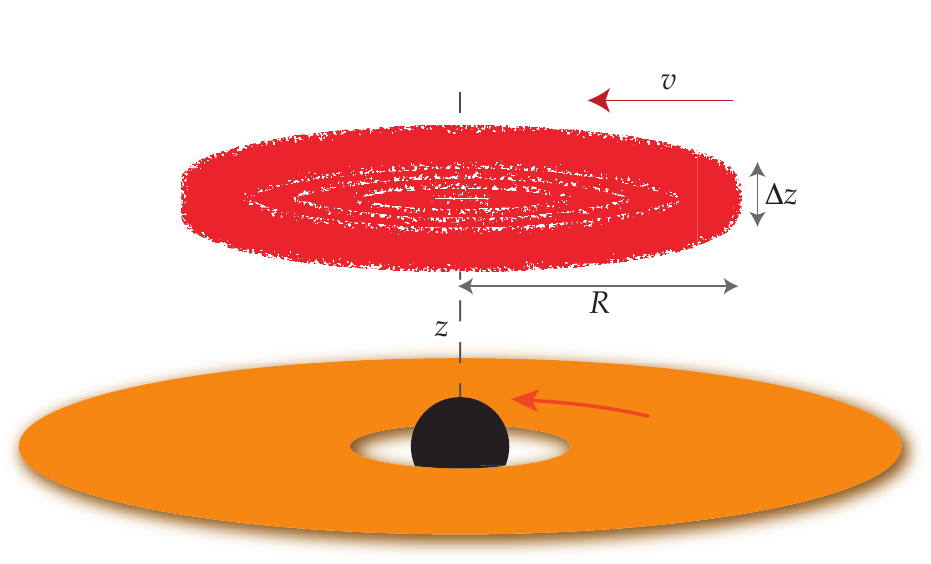}
\caption[]{Schematic of the model of an extended corona filling a cylindrical region spanning vertically between 1.5 and 2\rg\ above the plane of the disc through which fluctuations in luminosity propagate inwards from the outer edge at speed $v$.}
\label{propin.fig}
\end{figure}

\begin{figure*}
\centering
\subfigure[$v = c$] {
\includegraphics[width=85mm]{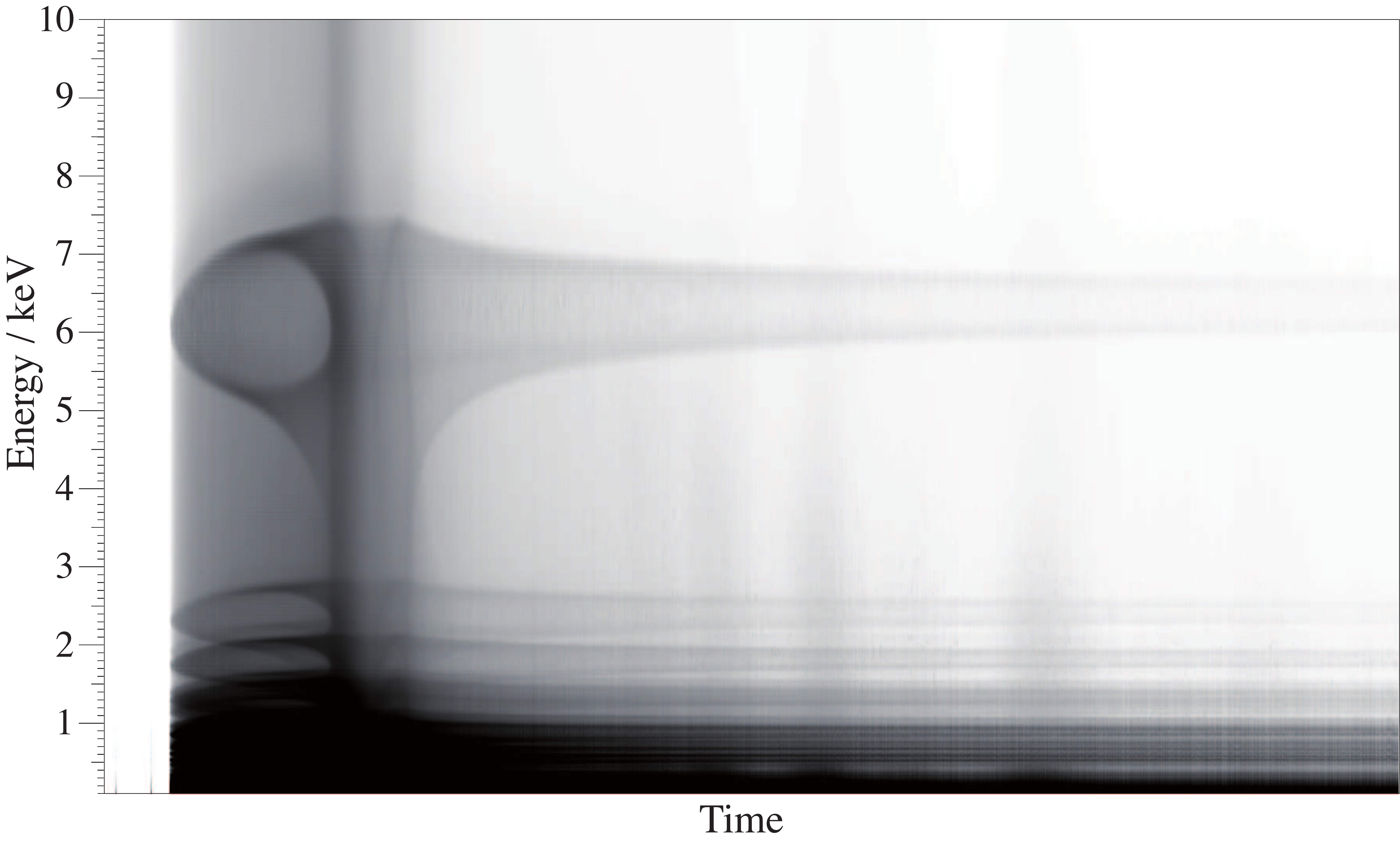}
\label{ent_propin.fig:c}
}
\subfigure[$v = 0.1c$] {
\includegraphics[width=85mm]{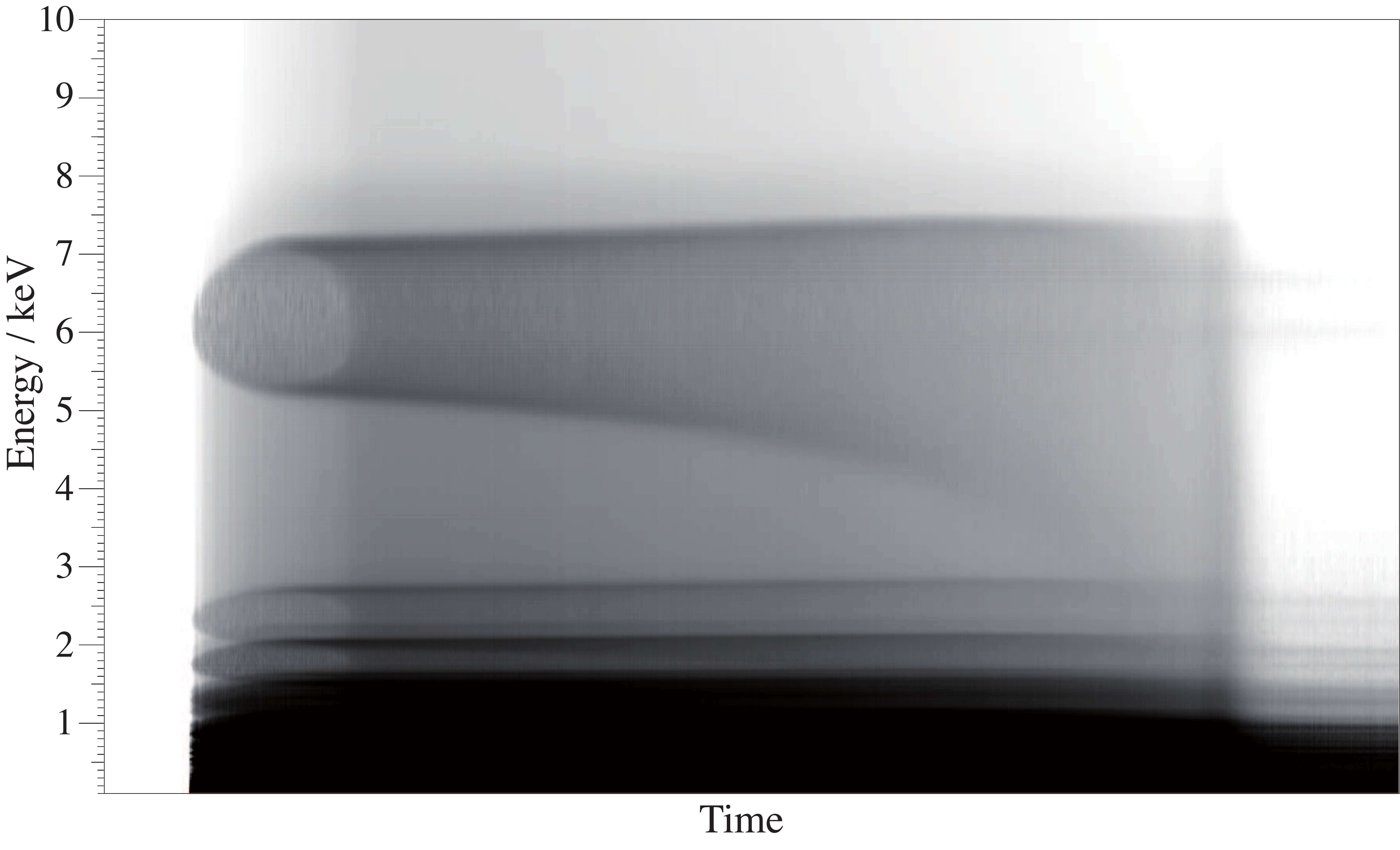}
\label{ent_propin.fig:0.1c}
}
\caption[]{Time and energy resolved functions for the reverberation from the accretion disc of a single flash propagating inward through a corona extending over the surface of the disc to 30\rg\ at \subref{ent_propin.fig:c} the speed of light and \subref{ent_propin.fig:0.1c} one-tenth the speed of light.}
\label{ent_propin.fig}
\end{figure*}

Inward propagation of fluctuations in luminosity that originate at the edge of a radially extended corona, illustrated in Fig.~\ref{propin.fig} could reflect a scenario in which energy is injected directly into the corona from the underlying accretion flow. The X-ray emission could originate in a hot atmosphere atop the disc or from energetic particles that are accelerated by magnetic fields arising from the disc. The fluctuation then propagates inwards, either with the underlying accretion flow or directly through the body of the corona (for instance through the motion of the accelerated particles towards the black hole or the propagation of the magnetic field). We first consider the generalised case of propagation at an (arbitrary) constant velocity to understand the basic properties of inwardly propagating fluctuations and will return to propagation on viscous time scales through the disc later.

The energy- and time-resolved response function of the accretion disc to illumination by a flash propagating inwards through a radially extended corona is shown in Fig.~\ref{ent_propin.fig}. In this case, the emission line response begins narrow with the illumination of the outer disc seen first. The earliest response is the portion of the disc directly in front of the observer beneath the outer edge of the corona, moving transverse to the line of sight in its orbit and, hence, producing emission close to the rest-frame energy of the line. The response then spreads to the blue- and redshifted emission from the approaching and receding sides of the outer disc. It is not until later times that the bulk of the response is seen from the inner disc with the greatest fraction of photons illuminating these regions originating from the inner parts of the corona, illuminated later.

The tail of the response function is seen as redshifted emission is detected from the inner disc at later times and the photons reflected from the back side of the disc is again seen as the loop-back in the response function as well as the band across all energies that is seen when the full reflection spectrum is considered; this is the re-emergence of the reflected continuum alongside the emission lines. When the propagation is much slower than the speed of light, the tail of the response function is drawn out over a long time and the re-emergence is less pronounced, smeared over an extended time interval as photons from a range of radii in the corona (and hence a range of start times) hit the inner parts of the back side of the disc.

\begin{figure*}
\centering
\subfigure[$v = c$] {
\includegraphics[width=56mm]{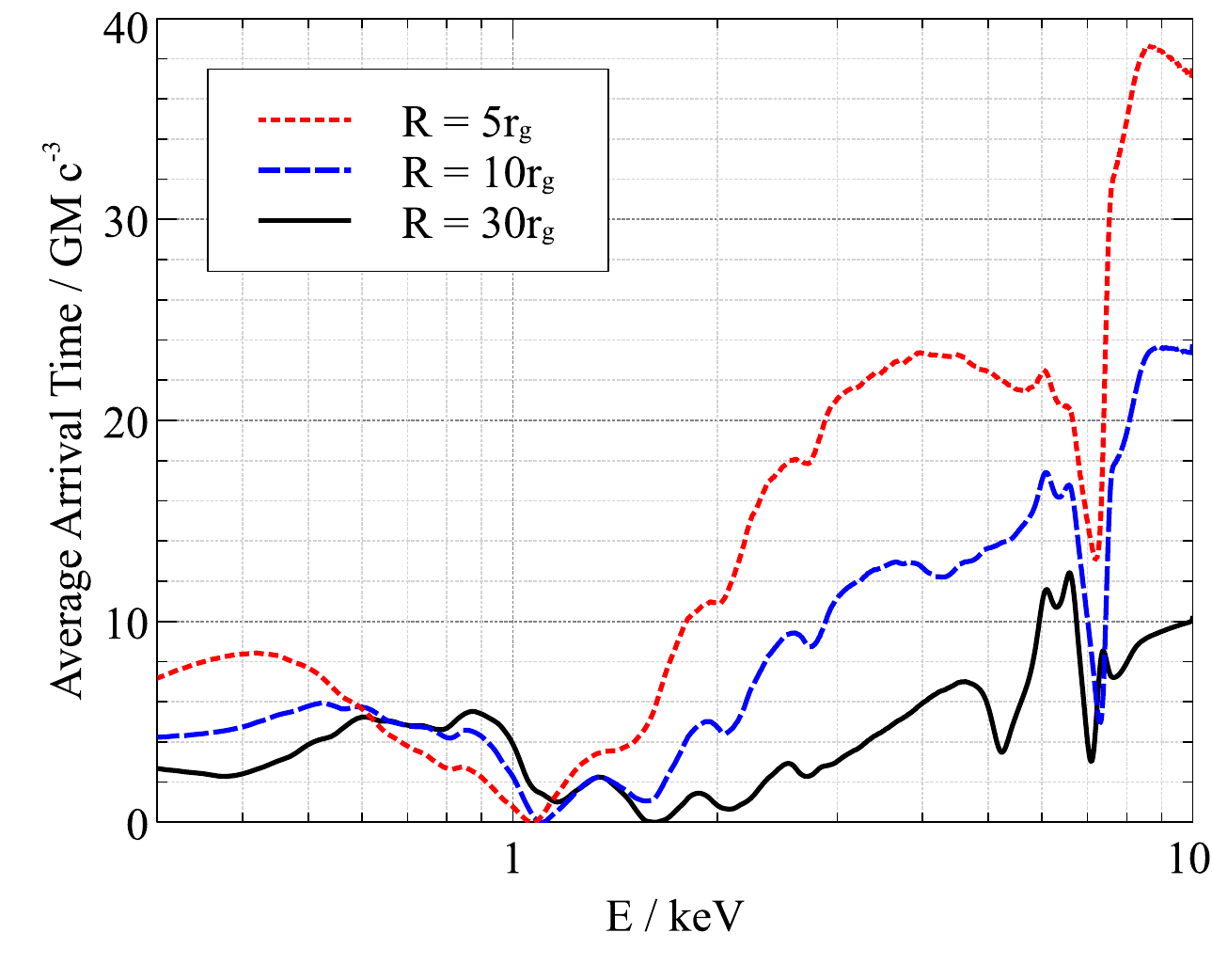}
\label{avg_arrival_propin.fig:c}
}
\subfigure[$v = 0.1c$] {
\includegraphics[width=56mm]{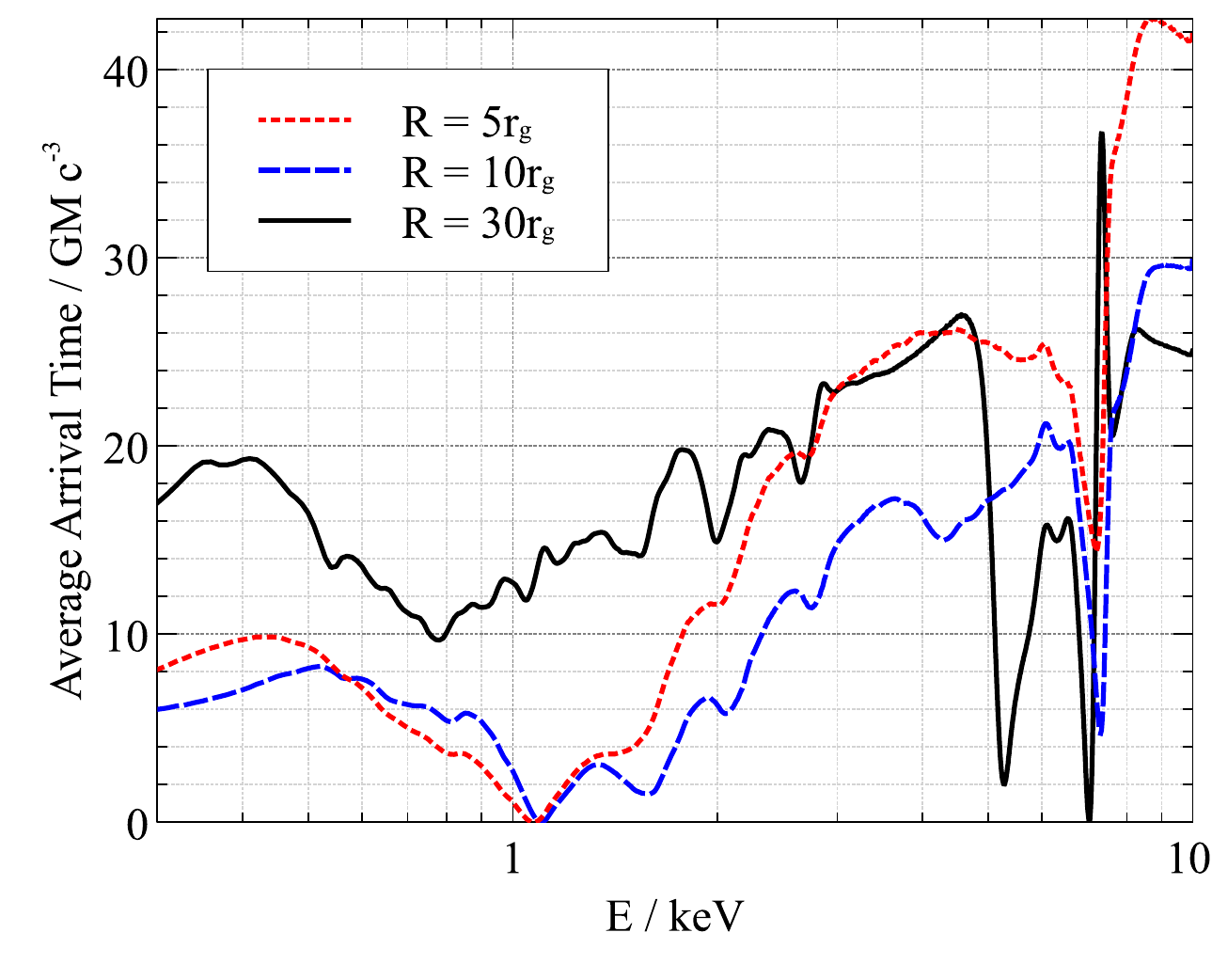}
\label{avg_arrival_propin.fig:0.1c}
}
\subfigure[$v = 0.01c$] {
\includegraphics[width=56mm]{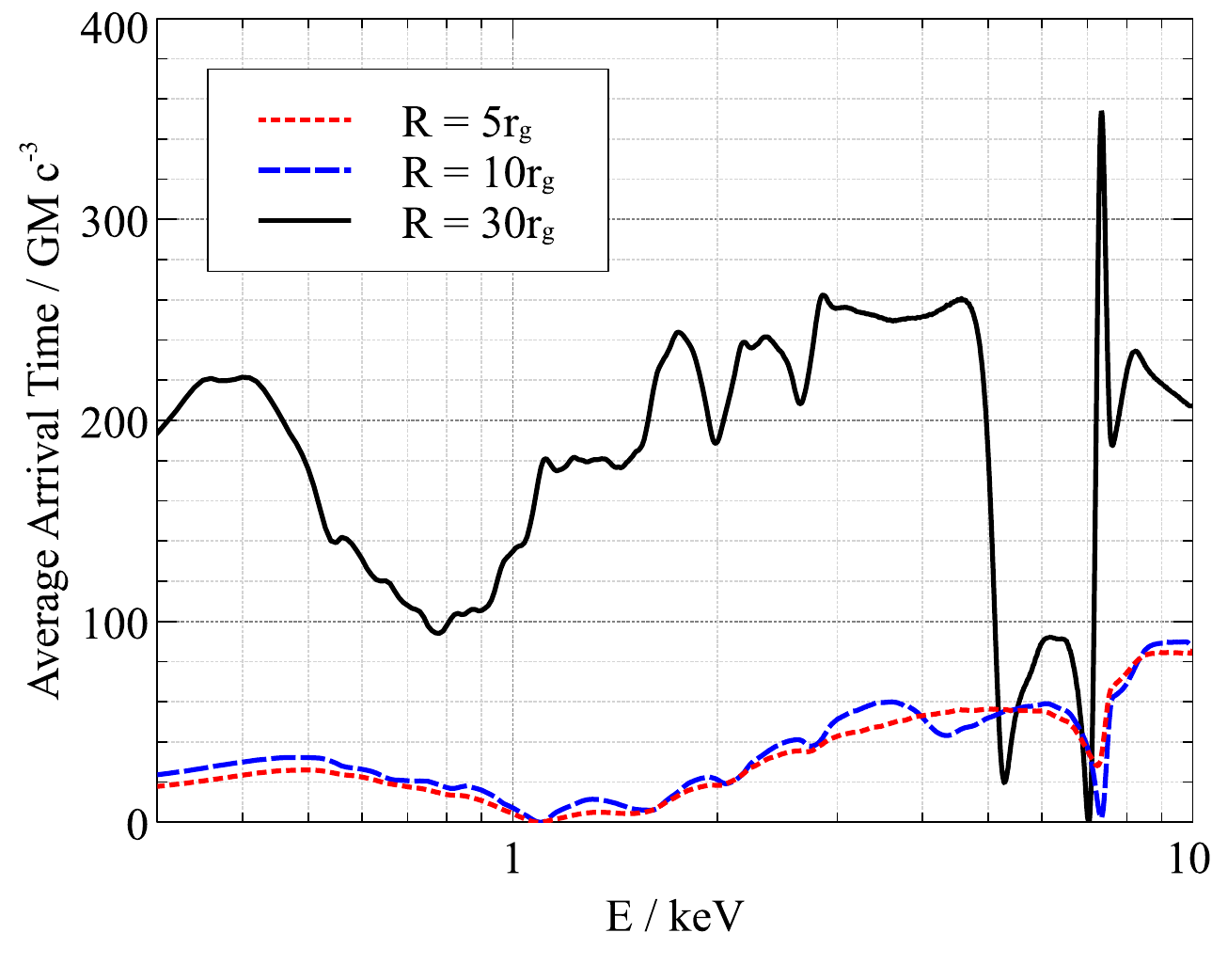}
\label{avg_arrival_propin.fig:0.01c}
}
\caption[]{The average arrival times (relative to the earliest arrival of photons in each case) of photons as a function of energy after a luminosity fluctuation propagating radially inwards through the corona at varying radial velocity $v$ through coronae of different radial extent, $R$ over the surface of the accretion disc.}
\label{lagspec_propin.fig}
\end{figure*}

The average arrival time as a function of photon energy from the reverberation of continuum fluctuations propagating inwards is shown in Fig.~\ref{lagspec_propin.fig}. Again shown are the lag-energy functions for coronae of varying radius over the surface of the disc and for propagation at speed $c$, $0.1c$ and $0.01c$.

While the effect of the propagation of the luminosity fluctuation through the corona is again subtle for light speed propagation, the most obvious effect for more slowly propagating fluctuations is the apparent inversion of the iron K$\alpha$ emission line. Two dips are seen in the lag profile of the emission line with the earliest arrival from photons just below the rest frame energy at $4\sim 5$\keV\ and just above at $7\sim 8$\keV. The dips move further apart (the lowest to lower energy and the highest to higher energy) for more compact coronae.

These energies correspond to the double-peaked emission line profiles from thin annuli on the disc just below the outer edge of the corona. Now the fluctuation starts on the outer edge of the corona, the first part of the disc to be illuminated is that directly beneath the outer edge. At later times, the core of the line arises from the outer disc, once the rays from the outer edge of the corona have had time to propagate, then the slow-moving fluctuation gradually illuminates more of the inner disc producing late emission from the redshifted wing of the line.

The broadband lag-frequency spectra will again be dominated by reverberation from the accretion disc. It is clear from the lag-energy spectra that the  reflection-dominated 0.3-1\keV\ band lags behind the continuum-dominated 1-4\keV\ band so, following the usual convention, a negative reverberation lag would be detected down to the lowest frequencies, again corresponding to the average height of the corona above the disc and being fairly insensitive to the radial extent. For the coron\ae\ considered here, extending vertically between 1.5 and 2\rg\ above the plane of the accretion disc, the lag in each case is measured to be between 2 and $2.5\,GM/c^3$ for fluctuations propagating at the speed of light, increasing for slower propagation through larger coron\ae. For coron\ae\ spanning 10\rg\ over the disc, the reverberation lag is measured to be around $4\,GM/c^3$ when fluctuations propagate at $0.1c$ and around $12\,GM/c^3$ when propagating at $0.01c$. The apparent increase in the lag time is due to the slower propagation of the fluctuation inwards through the larger corona. The fluctuation begins in the outer regions which contribute a greater fraction of the continuum emission, since photons can escape more easily from these regions. A greater portion of the reflected photons is then contributed by the inner regions of the corona since the strong gravitational field close to the black hole focuses more photons emitted from the inner regions towards the black hole and, hence, onto the inner regions of the disc. The overall effect is to further delay the arrival of the reflected photons with respect to the continuum.

The exception is the case of the slowest propagation through the largest coron\ae. In this case, the slow propagation from the outer to the inner parts of the corona delays even further the extremely redshifted emission below 4\keV\ re-emerging from the innermost parts of the back side of the disc. These photons are delayed with respect to the cores of the 0.5-1\keV\ lines in the soft excess that arise from the outer disc, illuminated early by the outer parts of the corona and early compared to the average arrival of the continuum over the whole corona. The skewed shape of the lag-energy profile between 0.3 and around 4\keV\ can result in a turn-over in the broadband lag-frequency spectrum to a positive lag at the lowest frequencies for slow inward propagation through the largest coron\ae. As in the case of slow outward propagation, however, the lag-energy spectrum at these frequencies would follow the average arrival time of the photons as a function of energy which would again show distinctive features of the reflection spectrum rather than the smooth increase in arrival time with photon energy that is observed.

\subsection{The low frequency hard lag}
\label{hardlag.sec}

\begin{figure}
\centering
\includegraphics[width=85mm]{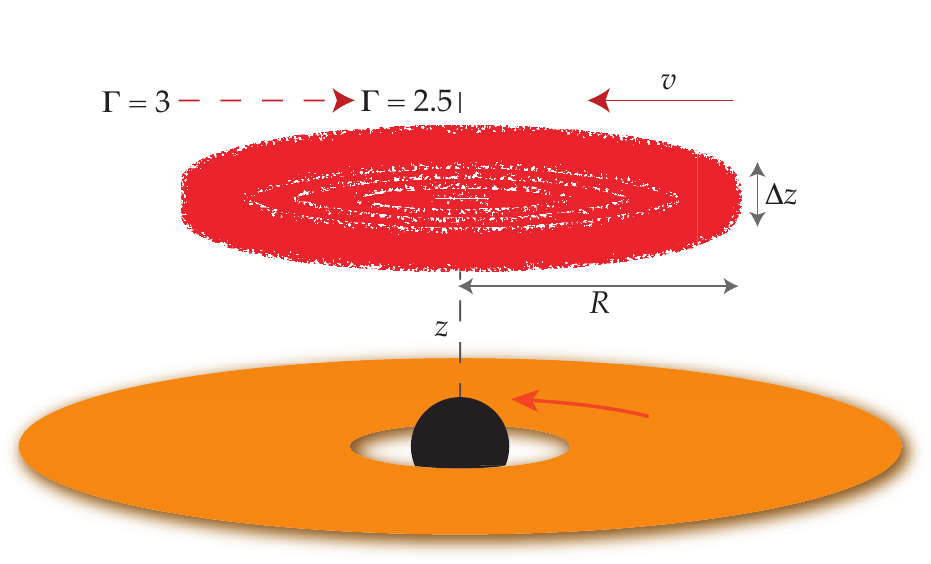}
\caption[]{The hard lag is introduced into the continuum emission through a linear gradient in photon index produced as a function of radius in the corona. The corona is less energetic in the outer regions, producing a softer continuum spectrum that hardens towards the centre.}
\label{propin_gammagrad.fig}
\end{figure}
It is clear that reverberation alone from the accretion disc cannot reproduce the shape of the lag spectra commonly observed in AGN, with the hard (continuum-dominated) band lagging behind the soft (reflection-dominated) band at low frequencies and the reverberation lag (the soft band lagging behind the hard band) appearing at higher frequencies with distinct lag-energy spectra at low and high frequencies. The observed energy dependence of the lags change in their behaviour, with a steady increase in lag as a function of energy at low frequency.

\begin{figure*}
\centering
\subfigure[$v = c$] {
\includegraphics[width=56mm]{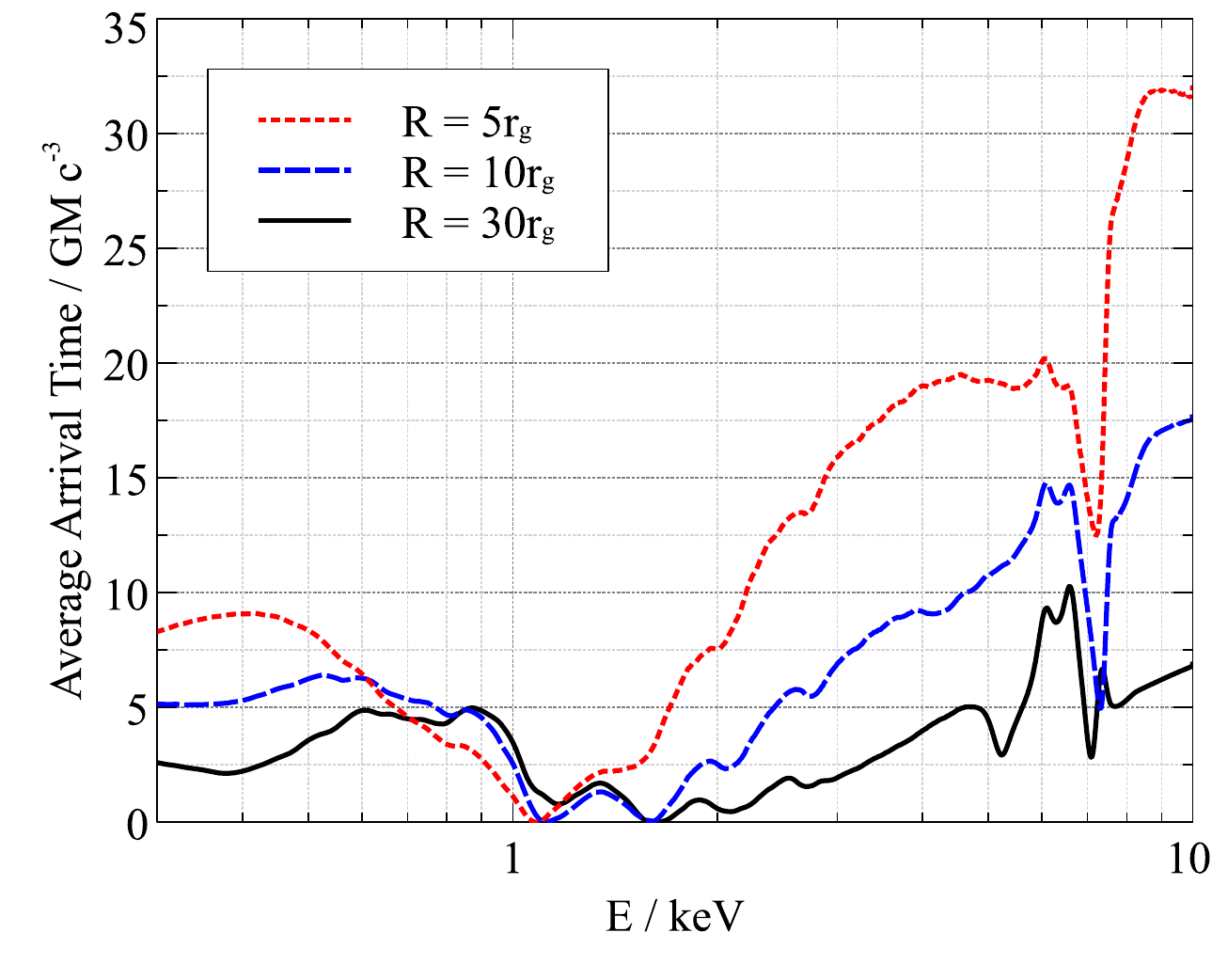}
\label{avg_arrival_propin.fig:c}
}
\subfigure[$v = 0.1c$] {
\includegraphics[width=56mm]{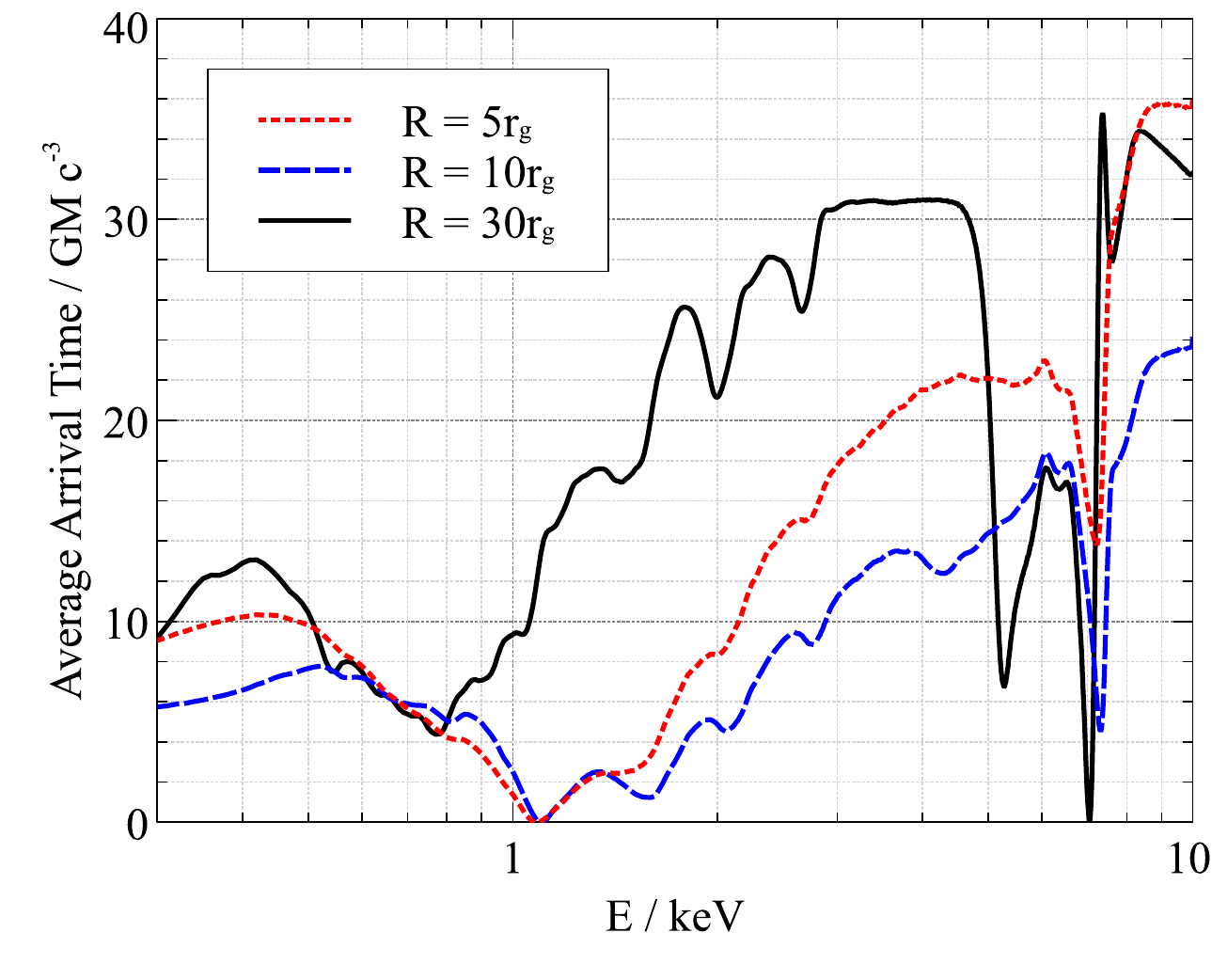}
\label{avg_arrival_propin.fig:0.1c}
}
\subfigure[$v = 0.01c$] {
\includegraphics[width=56mm]{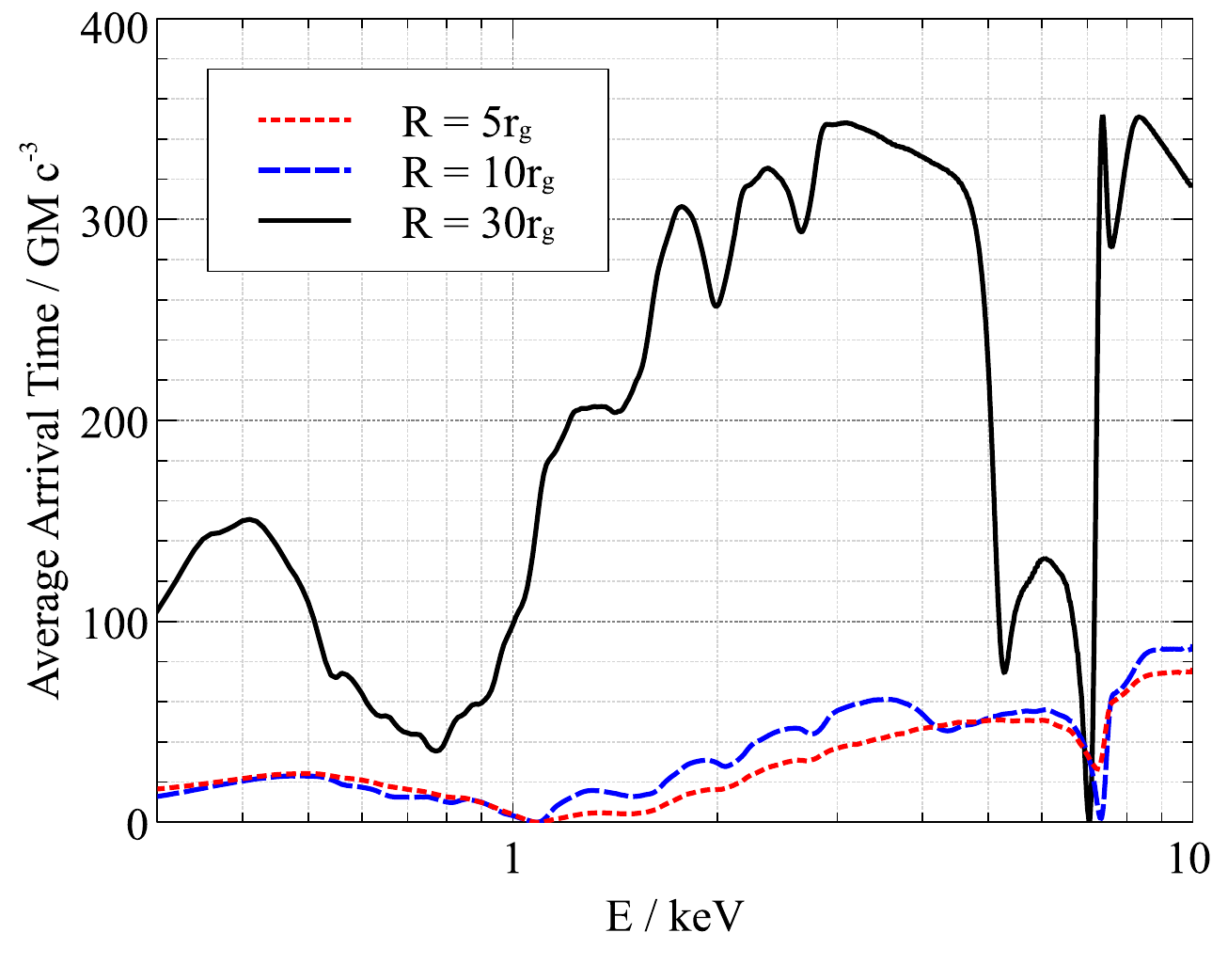}
\label{avg_arrival_propin.fig:0.01c}
}
\subfigure {
\includegraphics[width=56mm]{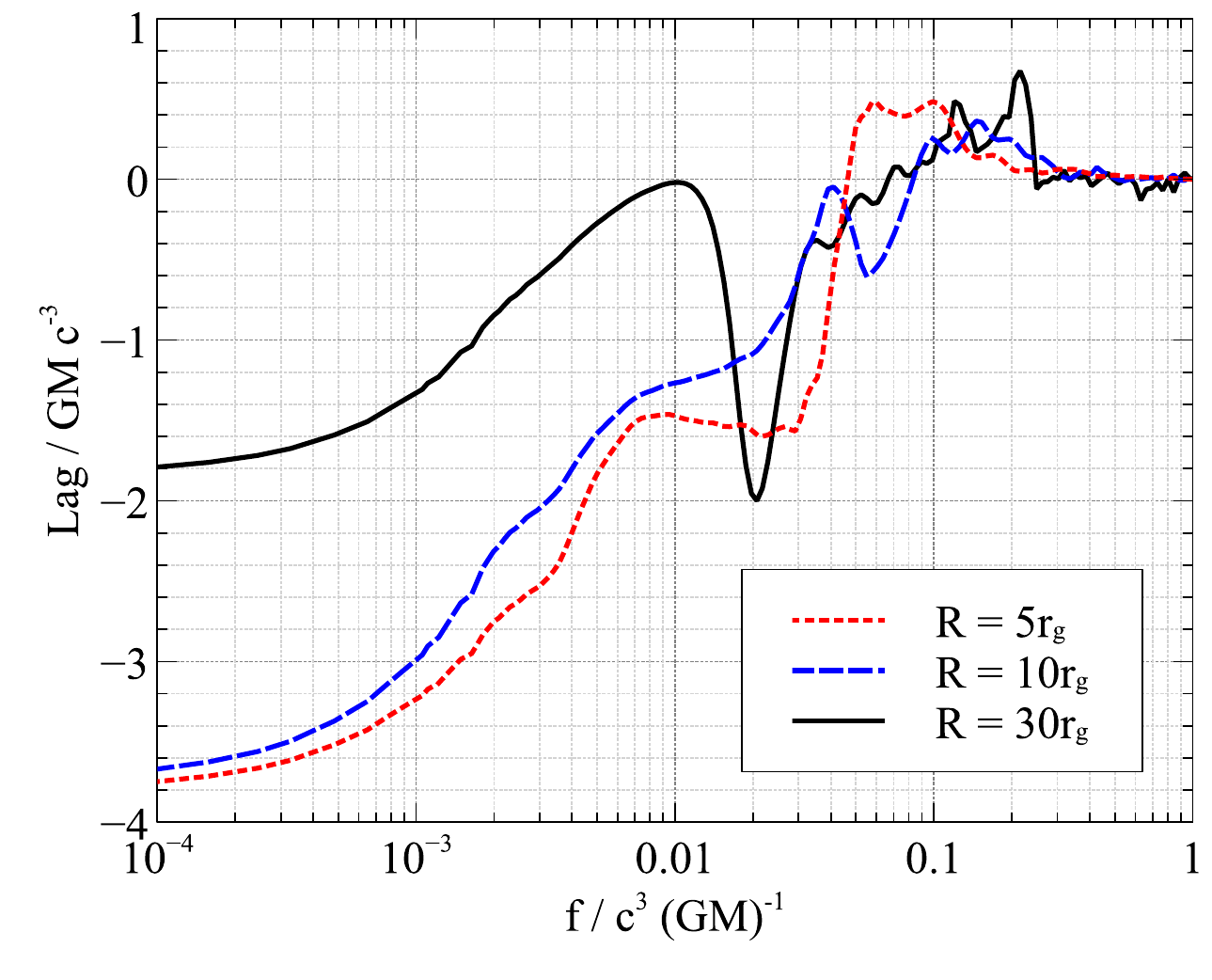}
}
\subfigure {
\includegraphics[width=56mm]{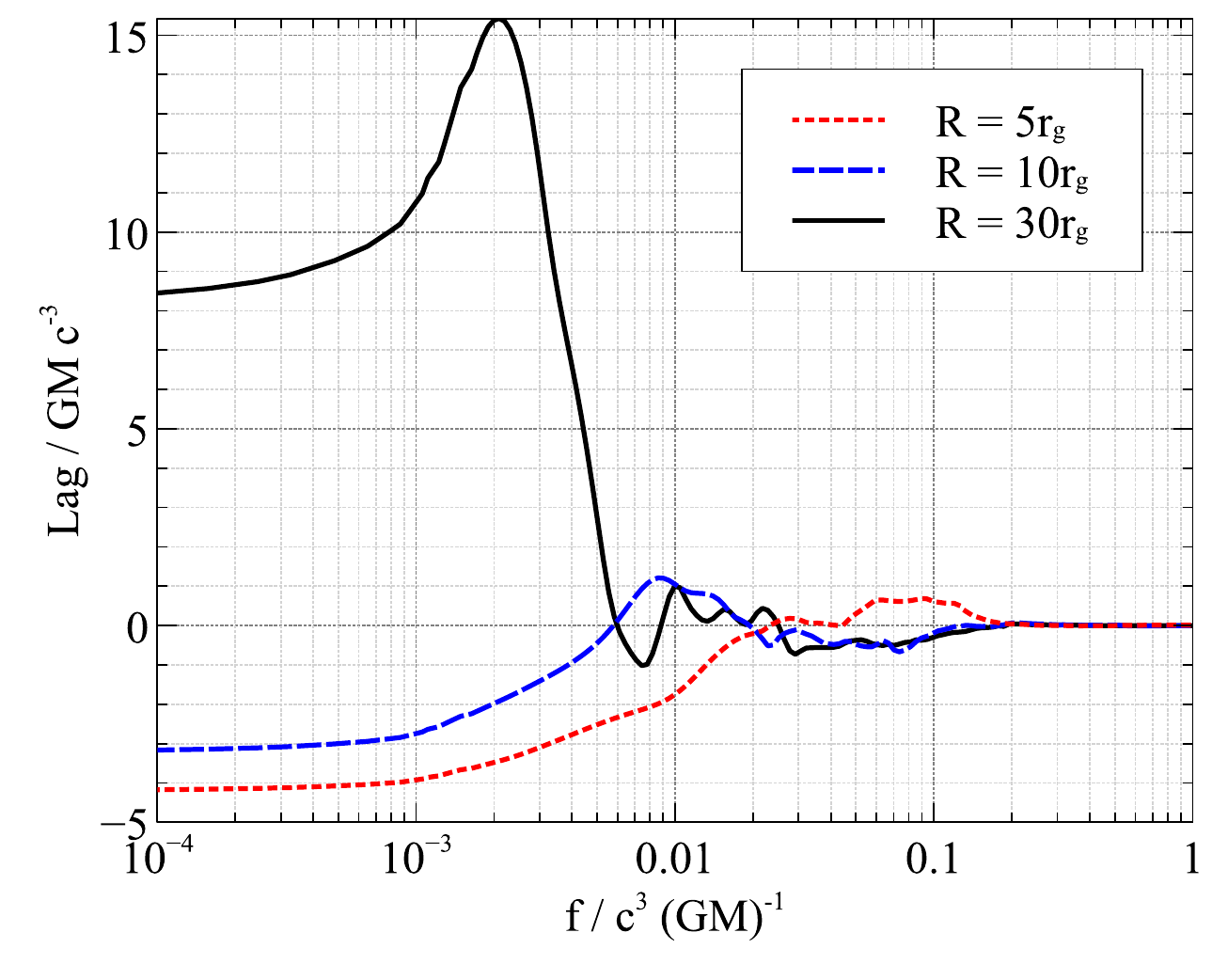}
}
\subfigure {
\includegraphics[width=56mm]{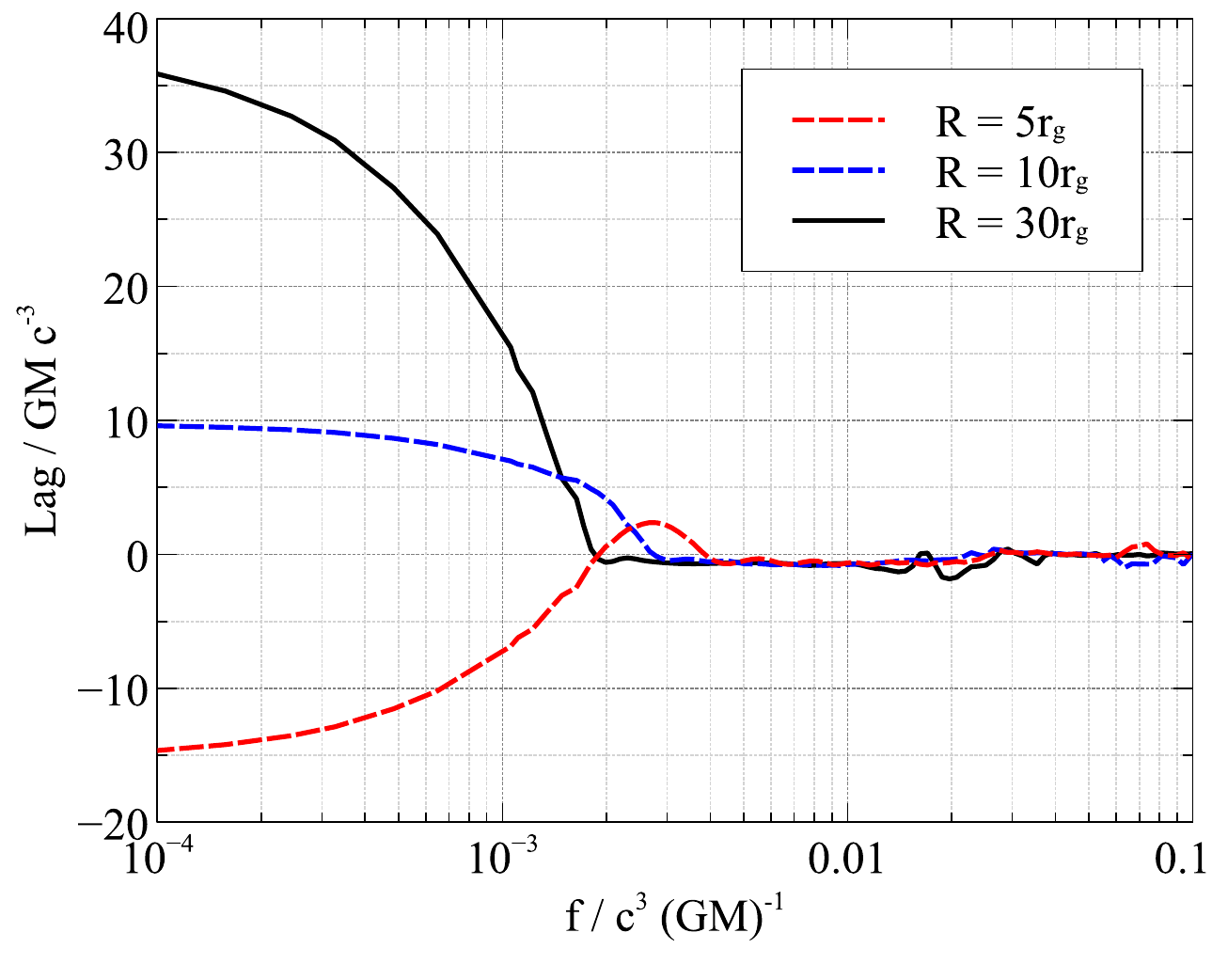}
}
\caption[]{\textit{Top:} The average arrival times of photons (both the directly observed continuum and reflection from the accretion disc) as a function of photon energy. \textit{Bottom:} the lag as a function of Fourier frequency between the 0.3-1\keV\ and 1-4\keV\ bands where fluctuations propagate inwards from the edge of radially extended coron\ae\ through which there is a linear gradient in photon index between the outer edge and centre to reproduce the hard lag.}
\label{lagspec_propin_gammagrad.fig}
\end{figure*}

The low frequency hard lag appears to be ubiquitous among accreting black holes. It was first discovered in the accreting X-ray binary Cygnus X-1 \citep{miyamoto+88} with the lag time found to have approximately a log-linear dependence on photon energy \citep{nowak+99}. Moreover, such a hard lag has been detected in the time variability of the accreting supermassive black hole in the AGN, NGC\,6814 in which  relativistically blurred reflection from an accretion disc was found not to make a significant contribution to the X-ray spectrum \citep{walton_hardlag}. These findings suggest that the hard lag is associated with the X-ray continuum itself, a position supported further by the lag-energy spectrum measured at these low frequencies in which the lag is seen to rise smoothly as a function of photon energy rather than changes in the lag coinciding with features characteristic of the reflection spectrum.

\citet{kotov+2001} propose that fluctuations in mass accretion rate propagate inwards through the accretion disc, energising successive sections of the corona as they do so. The inner parts of the corona are more energetic and produce a harder X-ray spectrum than the outer parts, producing the observed lag between the average arrival of softer and harder continuum photons. Implementing such a model, \citet{arevalo+2006} found that the observed power spectral density of the variability seen in the X-ray emission from X-ray binaries as well as the approximate log-linear dependence of the lag time on photon energy can be explained.

While \citet{arevalo+2006} modelled two continuum components, a soft and a harder power law spectrum, with varying contributions from radii in the corona according to power law emissivity profiles, we here model a continuous, linear variation in the continuum spectrum photon index with radius. The innermost parts of the corona are the most energetic, with $\Gamma=2.5$, and the spectrum softens to $\Gamma=3$ on the outer edge as illustrated in Fig.~\ref{propin_gammagrad.fig}. The average photon arrival times and lag spectra for fluctuations propagating inwards through such a corona are shown in Fig.~\ref{lagspec_propin_gammagrad.fig}. It should be stressed that we do not, in this work, aim to construct a detailed model of continuum lags arising in the corona but rather to determine if such a scenario can be reconciled with the detection of X-ray reverberation from the accretion disc and to discover under what conditions it is possible to detect both the low frequency continuum lags and high frequency reverberation lags that are seen side by side in AGN.

The average arrival time of photons as a function of X-ray energy (\textit{i.e.} the lag-energy spectrum that would be observed in the lowest frequency components of the variability) when the accretion disc is illuminated by a corona in which fluctuations propagate from a region producing a softer power law continuum to a harder one is shown in Fig.~\ref{lagspec_propin_gammagrad.fig} along with the broadband lag-frequency spectrum between the reflection-dominated 0.3-1\keV\ band and continuum-dominated 1-4\keV\ band. Coron\ae\ are again considered extending radially over the plane of the accretion disc to 5, 10 and 30\rg\ and in each case, the corona extends vertically between 1.5 and 2\rg\ above the plane of the disc. Fluctuations are considered propagating inward through the corona at speeds of $c$, $0.1c$ and $0.01c$.

By construction, the propagation through a gradient in the photon index leads to a steady increase in the lag time for increasing photon energies. The dependence of the lag in just the continuum component is approximately log-linear, as observed. Where the continuum reflects off the accretion disc, imprinted upon the log-linear rise in lag are the signatures of the reflection spectrum.

The `hard lag' introduced acts against the `soft lag' between the continuum and reverberating soft excess. The propagation of the fluctuation through the coronal gradient means that 1-4\keV\ photons in the continuum respond to the fluctuation, on average, later than 0.3-1\keV\ photons in the continuum since a greater fraction of the 1-4\keV\ come from the harder power law contributed by the inner regions of the corona. The effect of this is to delay 1-4\keV\ band with respect to the 0.3-1\keV\ band. The effect is slight when luminosity fluctuations propagate at light speed through the corona with the measured reverberation lag decreasing for the larger corona (where, for a given propagation speed, it takes longer for the fluctuation to travel from the outer to the inner regions), although the dominant process defining the lag-energy profile is still reverberation from the disc. For more slowly propagating fluctuations, however, the dominant process becomes the propagation of the fluctuation from softer to harder photons in the continuum. The lag-energy profile changes shape, now showing the increase in arrival time to higher photon energies but with signatures of the reflection spectrum clearly imprinted. Although, on average, continuum photons at higher energies arrive later, the softer continuum produced in the outer parts still contains sufficient photons above the iron K absorption edge at 7.1\keV\ and, hence, can excite the strong iron K$\alpha$ line at 6.4\keV\ from parts of the disc in close proximity to the early coronal emission. This results in the comparatively early arrival of line photons at energies corresponding to the approaching and receding sides of the disc beneath the outer edge of the corona, hence the double-peaked profile around 6.4\keV\ and the early arrival of photons in the soft excess as seen in the case of inward propagation with no gradient in continuum photon index. For the slowest propagating (constant velocity) fluctuations, the broadband lag-frequency spectrum turns over and shows only the harder energy bands lagging behind the soft with no measurement of a negative reverberation lag.

It is clear that where a `hard lag' is produced by propagation from less to more energetic regions of the corona, this process cannot trivially be separated from the measurement of the reverberation lag. When the fluctuation propagates sufficiently quickly, the negative reverberation lag is observed but the lag time is shortened by the opposing time lags in the continuum emission. Using the reverberation time lag as a measure of the (vertical) extent of the corona therefore requires simultaneous fitting of both the frequency and energy dependence of the propagation lag in the continuum. Care must be taken when fitting a phenomenological function, for instance a power law \citep[as in][]{emmanoul+2011,chainakun+2015}, to the low frequency portion of the lag-frequency spectrum without accounting for the energy dependence of this process as the hard lag in the continuum can be strongly degenerate with the reverberation lag. Where the fluctuation propagates at constant velocity through the corona it is not possible to simultaneously observe the hard, continuum lag at low frequency and reverberation from the disc at high frequency. Propagation as slow as $0.01c$ affects the lag spectrum at all frequencies either reducing the measured reverberation lag or cancelling it out entirely, damping the lag to zero across higher frequencies.

\subsection{Viscous Propagation}
\label{viscous.sec}
It might be expected that fluctuations in the luminosity of a corona extended over the surface of the accretion disc are associated with fluctuations propagating through the underlying disc. This might describe a situation in which the X-ray emitting corona is a hot atmosphere directly associated with the disc or is composed of energetic particles that are accelerated magnetically from the disc surface, for instance through the reconnection of magnetic flux loops arising from the disc. The luminosity of the corona would be expected to vary with the mass accretion rate and fluctuations therein, with overdense regions of the accretion disc releasing more gravitational potential energy as they pass inwards, potentially both directly into the energetic corona and through the production of more thermal seed photons that are up-scattered to become the X-ray continuum. If the corona were to be magnetically accelerated, the luminosity would also vary with the local magnetic flux density, increasing both the probability of reconnection events and the energy associated with each. Magnetic fields are expected to be accreted inwards along with the ionised material in the accretion disc \citep[\textit{e.g.}][]{mckinney+2012,sikora_begelman}. Predictions can be made for the simple case of coronal fluctuations associated with fluctuations in a simple, classical accretion flow driven by the viscous transfer of angular momentum \citep{shaksun}.

\begin{figure}
\centering
\includegraphics[width=85mm]{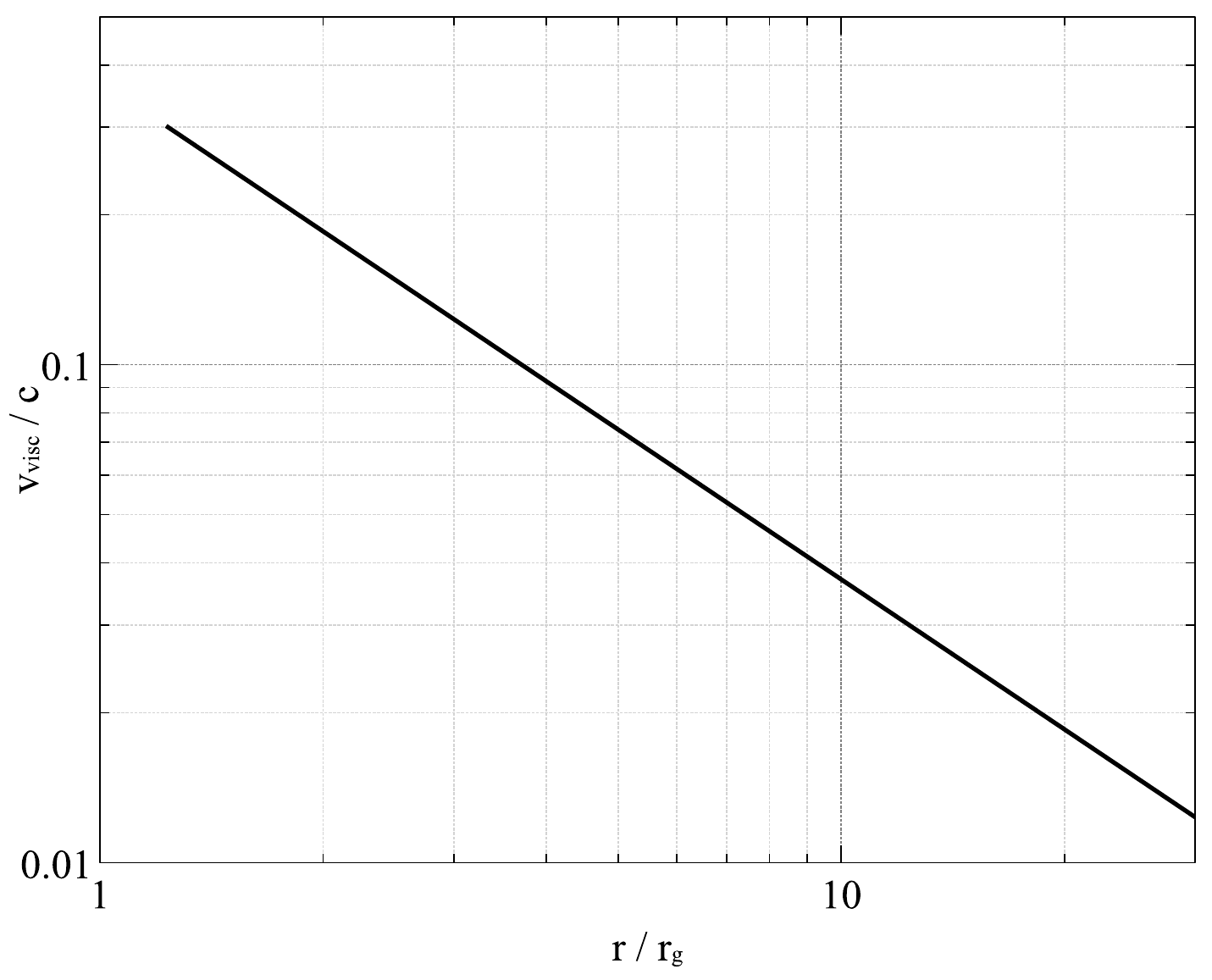}
\caption[]{The velocity as a function of radius of fluctuations propagating inwards through a Shakura-Sunyaev accretion disc where $\alpha(h/r)^2 = Cr^{-1}$ and $C$ is set to reproduce a thick disc with $\alpha = 0.3$ and $(h/r)=1$ on the inner edge. Viscous propagation of material through the disc is used as the basis for luminosity fluctuations through the overlying corona.}
\label{viscprop.fig}
\end{figure}

In a standard Shakura-Sunyaev accretion disc, inward accretion of material occurs through the transfer of angular momentum outwards in the disc through viscous friction between neighbouring radii orbiting at their respective Keplerian velocities (see, \textit{e.g.}, \citealt{czerny+06} for a review). The local viscous velocity at which material travels radially inward is
\begin{equation}
v(r) = r^{-\frac{1}{2}} \frac{\alpha}{\sqrt{GM}}\left(\frac{h}{r}\right)^2
\end{equation}
This can be integrated to give the travel time of an overdensity between two radii, describing the propagation through a corona from which reverberation is to be simulated. The values of the $\alpha$ viscosity parameter and the disc aspect ratio $(h/r)$ are often taken to follow a power law with radius, $\alpha(h/r)^2 = Cr^{-\beta}$ with the constant $C$ set to reproduce a thick disc with $\alpha = 0.3$ and $(h/r)=1$ on the inner edge \citep{arevalo+2006}. We here set $\beta=1$ which (should $\alpha$ be constant over the disc) represents a constant disc aspect ratio and, most likely, an upper limit on the propagation velocity, hence a lower limit on the propagation effect (given that propagation through the corona has a more pronounced effect on the low frequency reverberation response at lower velocities). The viscous propagation velocity as a function of radius through this model disc is shown in Fig.~\ref{viscprop.fig}. For a flat disc surface (parallel to the equatorial plane), expected in a radiation dominated accretion disc, $\beta=2$ producing slower propagation through the corona with a more rapid acceleration as the fluctuation reaches the inner regions.

\begin{figure*}
\centering
\subfigure[Lag-frequency spectrum] {
\includegraphics[width=56mm]{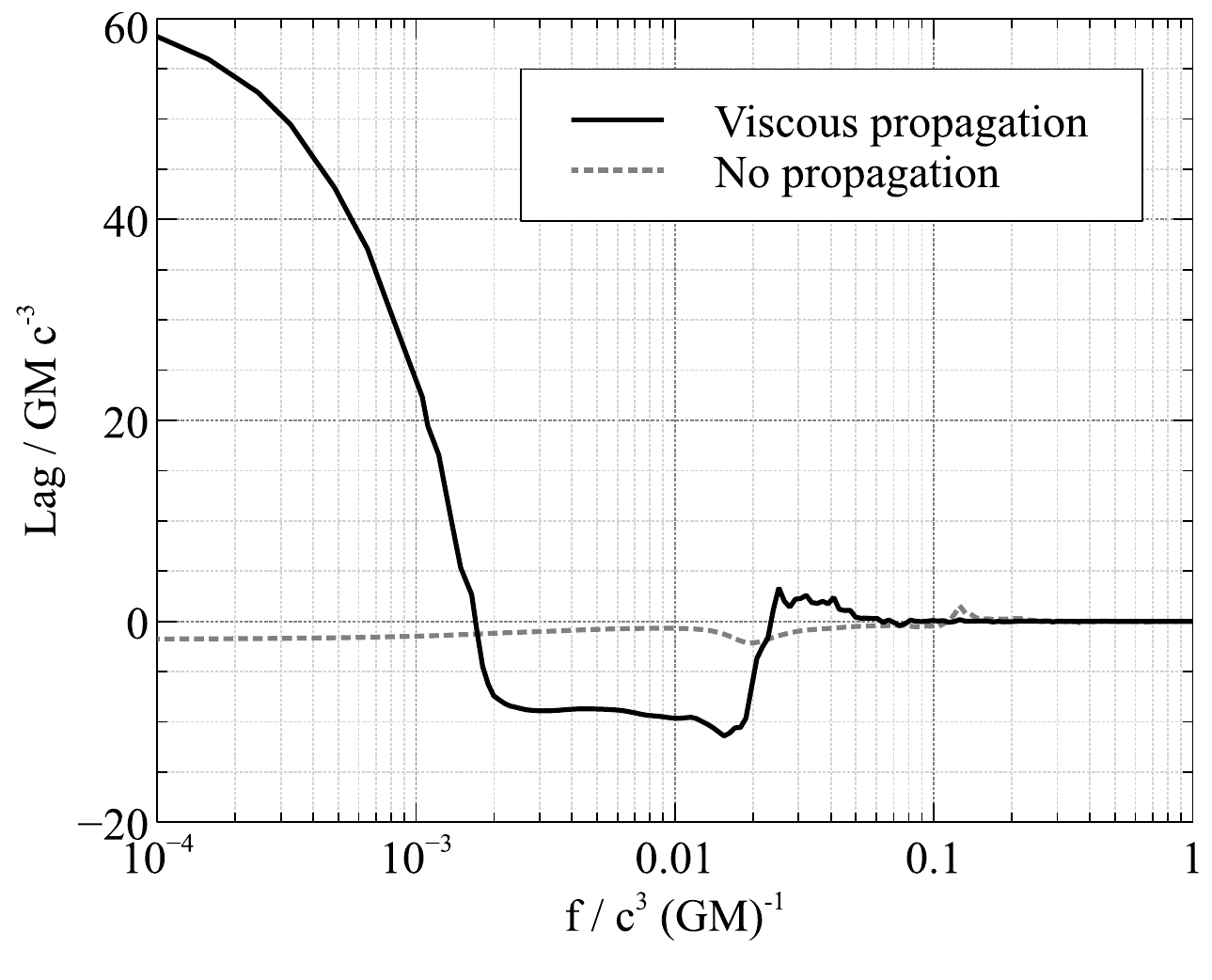}
\label{lagspec_viscous.fig:lagfreq}
}
\subfigure[Lag-energy, low frequency] {
\includegraphics[width=56mm]{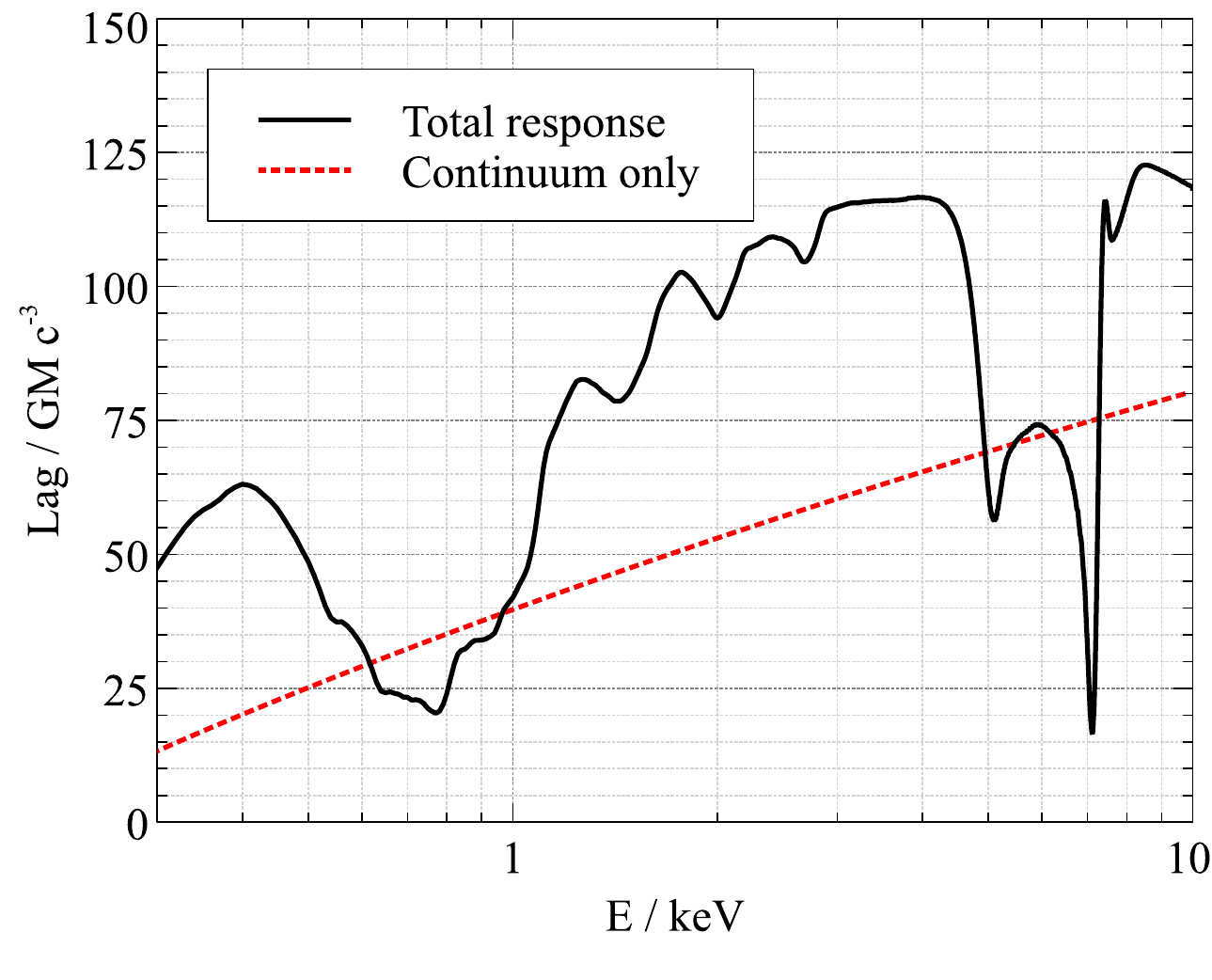}
\label{lagspec_viscous.fig:lagen_low}
}
\subfigure[Lag-energy, high frequency] {
\includegraphics[width=56mm]{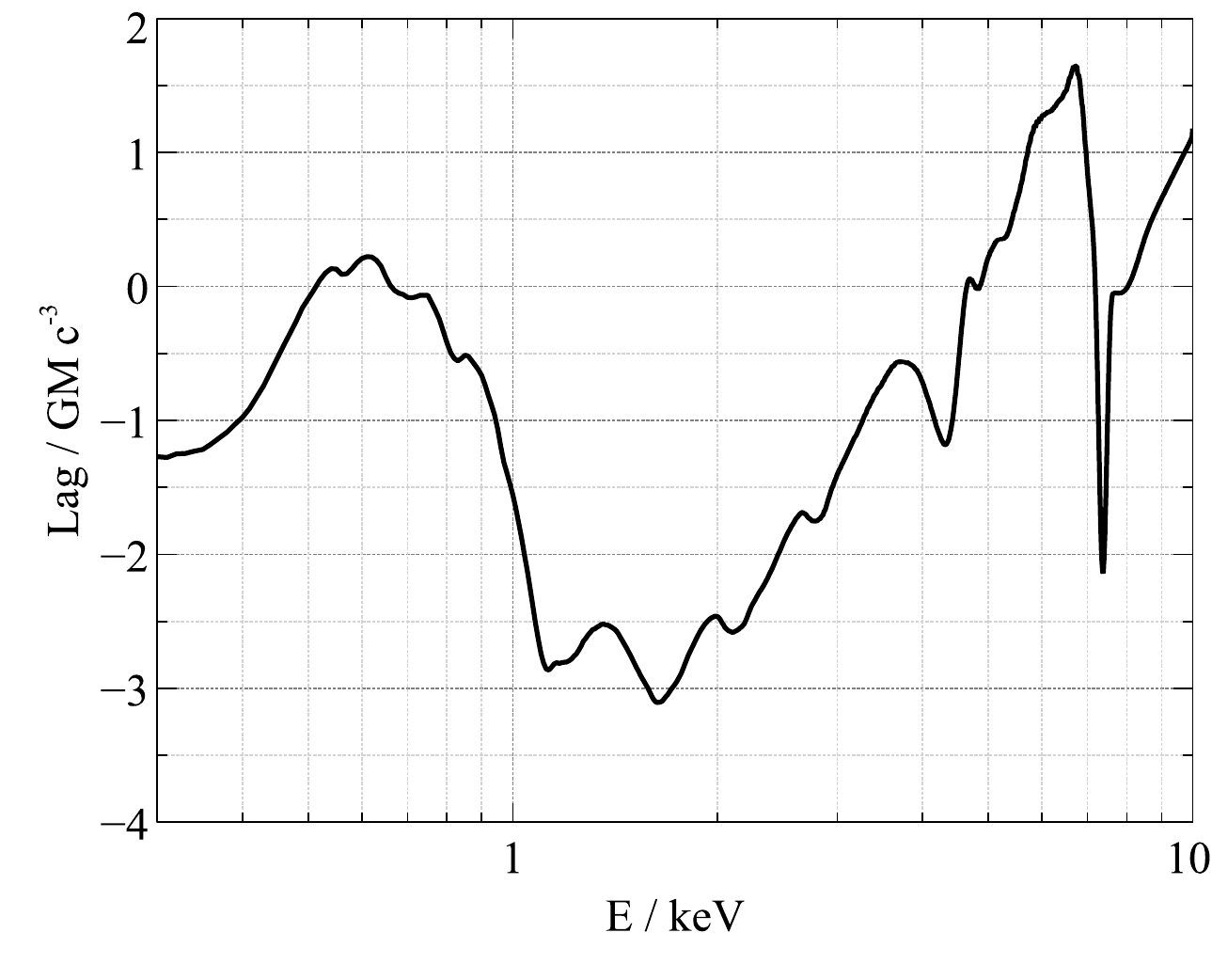}
\label{lagspec_viscous.fig:lagen_high}
}
\caption[]{\subref{lagspec_viscous.fig:lagfreq} The broadband lag-frequency spectrum between the 0.3-1\keV\ and 1-4\keV\ bands where fluctuations propagate inwards from the edge of a radially extended corona at the viscous timescale on the underlying Shakura-Sunyaev accretion disc where $\alpha(h/r)^2 = Cr^{-1}$. There is a linear gradient in photon index between the outer edge and centre. \subref{lagspec_viscous.fig:lagen_low} and \subref{lagspec_viscous.fig:lagen_high} The lag-energy spectra averaged over low frequencies, $10^{-4}$-$10^{-3}\,c^3(GM)^{-1}$ and high frequencies, $2\times 10^{-3}$-$10^{-2}\,c^3(GM)^{-1}$, respectively. These frequency ranges represent the regions of positive and negative lag in \subref{lagspec_viscous.fig:lagfreq}.The low frequency lag-energy spectrum also shows the lag spectrum if only the continuum emission were present, showing the approximate log-linear dependence of the lag on photon energy where there is a linear decrease in photon index from the edge to the centre of the corona.}
\label{lagspec_viscous.fig}
\end{figure*}

The case of a corona extended radially over the surface of the disc to 30\rg\ is considered, as suggested by measurement of the accretion disc emissivity profile in 1H\,0707$-$495. The photon index of the continuum produced at each location in the corona varies linearly with radius from $\Gamma=2.5$ in the centre to $\Gamma=3$ on the outer edge. Fluctuations in luminosity propagate inward through the corona at the local viscous velocity of material flowing through the underlying accretion disc. 

Fig.~\ref{lagspec_viscous.fig:lagfreq} shows the lag as a function of frequency between the reflection-dominated 0.3-1\keV\ energy band and the 1-4\keV\ energy band as well as the lag as a function of energy over both the low and high frequency ranges, identified as the frequency ranges in which the lag is found to be positive and negative. We find that viscous propagation is readily able to reproduce the observed combination of the hard lag, where harder X-rays in the continuum respond, on average, later than softer X-rays, below $10^{-3}\,c^3(GM)^{-1}$, and reverberation is seen between $2\times 10^{-3}$ and $2\times 10^{-2}\,c^3(GM)^{-1}$.

These two modes are distinctly seen in the energy dependence of the lag over the low and high frequency ranges. The lag-energy spectra are computed from the response functions as they would be from real observations \citep{reverb_review}. The cross spectrum is computed between each narrow energy band of the response and the reference band; the sum of the response across the full 0.1-10\keV\ range minus the energy band of interest. This cross spectrum is then multiplied by $f^{-2}$ to act as a high frequency filter in the same manner as the observed power spectral density of the variability of a real AGN, as in Equation~\ref{crossspec.equ}. Although unimportant when considering narrow frequency ranges, when the cross spectrum is averaged over broad frequency ranges, this enhances the contribution of the lower frequency components within the frequency range of interest in determining the lag-energy spectrum. The final, filtered cross spectrum is then averaged over the frequency range of interest and the lag of the present energy band behind the reference is, once again, given by the argument of the (average) cross spectrum. The low and high frequency lag-energy spectra, in which frequency ranges were selected to coincide with the regions of the lag-frequency spectrum where the hard lag and reverberation lag are respectively seen, are shown in Figs.~\ref{lagspec_viscous.fig:lagen_low} and \subref{lagspec_viscous.fig:lagen_high}.

The low frequency lag-energy spectrum shows the propagation of the fluctuation from the outer to inner parts of the corona, producing an increasing fraction of the harder X-rays as it does so. Imprinted on this are the signatures of the early arriving reflection; the soft excess leading to a notch at low frequencies and the double-trough profile of the iron K$\alpha$ line from the approaching and receding sides of the disc at, from the parts of the accretion disc beneath the outer edge of the corona where the luminosity fluctuation begins, as in the case of slow, constant velocity inward propagation. At high frequencies, the lag-energy spectrum shows the classic form expected for reflection from the accretion disc; the arrival of photons comprising the soft excess and the iron K$\alpha$ line later than continuum which dominates the 1-4\keV\ energy band.

When only the high frequency components of the variability are considered, the inner parts of the accretion disc are seen to respond to fluctuations in the continuum more rapidly than the outer part. The redshifted wing of the iron K$\alpha$ line responds sooner than the 5-7\keV\ core of the line that is dominated by reflection from large radii in the disc. In these high frequency components, the slow propagation of the fluctuation from the outer to the inner parts of the corona, which previously lead to the double-trough profile. The propagation and hence the time scale of this sharp feature in the lag-energy spectrum is longer than the frequency range considered, hence is averaged to zero while just the more rapid reverberation, of order the 2\rg\ distance between the corona and the disc is seen, leading to the simple lag-energy spectrum. The inner disc that is closer to the coronal emission is seen to respond before the outer disc.  

While both the hard lag and reverberation could not be seen simultaneously when the fluctuation propagates at constant velocity, in the case of viscous propagation, the propagation over the outer part of the disc is sufficiently slow that it influences the variability only at the lowest frequencies. For the case considered here of a corona extending radially out to 30\rg, the average velocity of the fluctuation is $0.005c$, limiting the influence of the propagation on the measured lag to frequencies below $8\times 10^{-5}\,c^3 / GM$. For constant velocity propagation, the hard lag is still able to cancel the effect of reverberation at higher frequencies. The inner parts of the corona still produce a hard lag over the higher frequency ranges, even though the outer parts of the corona are no longer contributing as their contributed lags wrap in phase and are too long and averaged to zero. In the case of viscous propagation, the velocity increases as the fluctuation propagates inward according to a power law (Fig.~\ref{viscprop.fig}). Once the propagation reaches the inner regions, it is travelling sufficiently quickly that the propagation lag is less than the reverberation lag and the negative reverberation lag can once again be measured over the higher frequency components. In the case considered here, the propagation is travelling faster than $0.1c$ over the innermost 4\rg. At this velocity, we see in Fig.~\ref{lagspec_propin_gammagrad.fig} that the reverberation lag dominates over the hard lag. We note that the high frequency reverberation is, therefore, dominated by X-rays that originate from the innermost regions of the corona with the hard lag dominated by propagation on larger scales throughout the extended corona.

Comparing the broadband lag-frequency spectrum for the case of viscous propagation through an extended corona to that for light-speed propagation through a corona of the same radial extent over the disc, we find that the inward viscous propagation causes a significant increase in the lag time from the $\sim 2\,GM/c^3$ `pure reverberation' time scale; \textit{i.e.} that determined by the average vertical distance between the corona and reflector, increased by the delay of photons travelling close to the black hole and diluted by the contributions of the reflected and continuum components to the others' dominant band. Photons that originate from the inner parts of the corona experience strong gravitational light bending close to the black hole and, hence, tend to be focused towards the black hole and onto the inner regions of the disc rather than being able to escape to be observed as part of the continuum. This means that the majority of directly observed continuum photons are emitted from the outer parts of the corona, while a greater portion of the reflected photons come from the inner regions. The measured lag is increased by a factor of around 5, from $2\,GM/c^3$ to $10\,GM/c^3$.

The slow propagation of the luminosity fluctuation from the outer to inner parts of the corona therefore increases the average time lag between continuum and reflected photons. Under these circumstances, the measured lag between the continuum- and reflection-dominated energy bands therefore gives only an upper limit to the reverberation time scale indicative of the vertical extent of the corona and its distance from the accretion disc. We note, however, that the measured reverberation time lag in 1H\,0707$-$495 is only around $2\,GM/c^3$, and similarly short in other Seyfert galaxies. We find that in a corona extending 30\rg\ over the disc, the excess lag introduced by propagation is approximately $8\,GM/c^3$ and still nearly $6\,GM/c^3$ if the corona were to only extend 5\rg\ over the disc. These additional lags due to the inward propagation alone are much longer than the total measured lag time, hence we can conclude that the propagation through the corona cannot be having a significant effect in the reverberation-dominated high frequency lag spectra of Seyfert galaxies.

\subsection{Propagating Fluctuations}
\label{propfluc.sec}
So far, X-ray reverberation has been considered from extended coron\ae\ through which a single fluctuation in luminosity propagates from a set location (either the centre or outermost edge of the corona). This is as if the underlying accretion disc is connected to the corona in just one location or if the accretion disc undergoes fluctuations in mass accretion rate or other properties that determine the radiative output of the corona in its outer regions and these fluctuations then propagate through the disc to the inner regions where they energise the corona. It is likely, however, that a corona extending over the surface of the disc is connected to the underlying accretion flow throughout its extent. For example particles making up the corona are accelerated locally by the underlying accretion flow, perhaps by the reconnection of magnetic flux loops arising from the disc. Fluctuations in the radiative output might therefore be expected to arise due to stochastic processes locally on the disc as well as variations (\textit{e.g.} in mass accretion rate) propagating in from large radii.

\begin{figure}
\centering
\includegraphics[width=85mm]{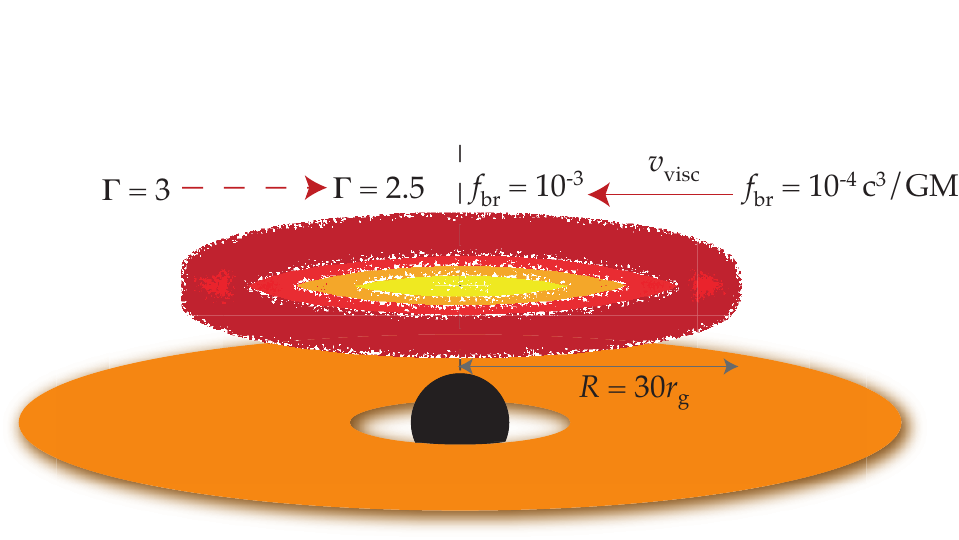}
\caption[]{An extended corona that is seeded by stochastic variability in the underlying accretion disc at all radii by summing the responses from extended coron\ae\ with different radial extents. Fluctuations in the emission from each radial component of the corona propagate inwards at the viscous velocity from their respective outer edges. Each component of the corona varies according to a randomly generated light curve with a broken power-law power spectral density (PSD), breaking from $f^{-1}$ at low frequencies to $f^{-2}$ at high frequencies at a break frequency, $f_\mathrm{br}$ that corresponds to the orbital frequency on the disc beneath the outer edge. As such, the larger outer regions of the corona vary more slowly while the inner regions dominate the short timescale variability.}
\label{radial_prop_model.fig}
\end{figure}

A model of a radially extended corona in which stochastic variations arise at all radii and then propagate inwards was constructed, akin to the \textsc{propfluc} model of \citet{ingram+2011,ingram+2013} in which variability at progressively higher frequencies is introduced at smaller and smaller radii. The response of both the continuum and reflected spectral components was calculated by summing the response functions due to  viscous propagation of fluctuations through coron\ae\ of different radii. At each radius, random fluctuations were generated according to a broken power law PSD. In models of propagating fluctuations, the time series describing the variability that propagates in from larger radii is typically multiplied by the time series describing the stochastic variability introduced locally (\textit{i.e.} the variability amplitude at any given radius is modulated by the underlying variation in the accretion disc) in order to reproduce the lognormal flux distribution that is widely observed. To more simply combine the ray-traced reverberation responses from different radii within the corona, the responses due to uncorrelated fluctuations propagating inwards from each radius are summed (whereas in a full, multiplicative propagating fluctuations model, the responses would be summed for fluctuations propagating from each radius where the time series arising at each radius is modulated by that propagating inwards from larger radii). We are just concerned with the effect on the measured X-ray reverberation here, however, and the important feature is the introduction of uncorrelated variability at higher frequencies closer to the black hole, thus the simple addition of uncorrelated responses will preserve the key features of the lag spectra.

The PSD of the variability seen in AGN is commonly found to follow a broken power law, falling off as $f^{-1}$ up to a break frequency $f_\mathrm{br}$, after which it falls off as $f^{-2}$ to high frequencies \citep{mchardy+2004,uttley_mchardy-2005}. The break frequency is found to scale inversely with black hole mass \citep{mchardy+2006} and might therefore be attributed to the characteristic size of the emitting region, if the variability is generated within a similar range of gravitational radii in all sources (with the gravitational radius scaling with mass). For the variability introduced at each radius in the corona, we set the break frequency to the Keplerian orbital frequency at that radius. One might also consider damping of variability on the time scale corresponding to the viscous inflow of material through each radius, which is of a similar order to the orbital time scale \citep[\textit{e.g.}][]{czerny+06}. This allows each radius to generate red noise fluctuations with equal energy per unit frequency up to the orbital frequency, which will smooth out any fluctuations and limit the ability of the large, slowly-moving outer parts of the corona to generate high frequency variability.  The model is outlined in Fig.~\ref{radial_prop_model.fig}.

For each radial component, a random light curve was generated with the desired power spectrum using the method of \citet{timmer_konig} where the Fourier transform of the time series is created. The amplitude of each frequency component is set to follow the broken power law and each component is assigned a random phase. The light curve for each component of the corona was convolved with the impulse response function (including both continuum and reflected emission with the photon index decreasing linearly inwards through the corona) for a fluctuation that propagates inwards on the viscous time scale from that radius. The reflection fraction for each radial component of the corona was set to the value expected from ray tracing calculations that compute the number of photons that hit the disc relative to that able to escape to be observed as part of the continuum (as in \citealt{1h0707_jan11}).

This prescription is found to produce realistic light curves, reproducing the amplitude and timescales of the variability typically seen in AGN with the power spectral density following $f^{-2}$ up to frequencies of 1\,$c^3 (GM)^{-1}$. While rapid variability arising from the outer regions of the disc is smoothed out as it propagates inwards, the inner regions of the disc are able to produce the rapid components of the variability that are observed. The lag spectrum was computed from the generated light curves following the standard procedure.

\begin{figure}
\centering
\includegraphics[width=85mm]{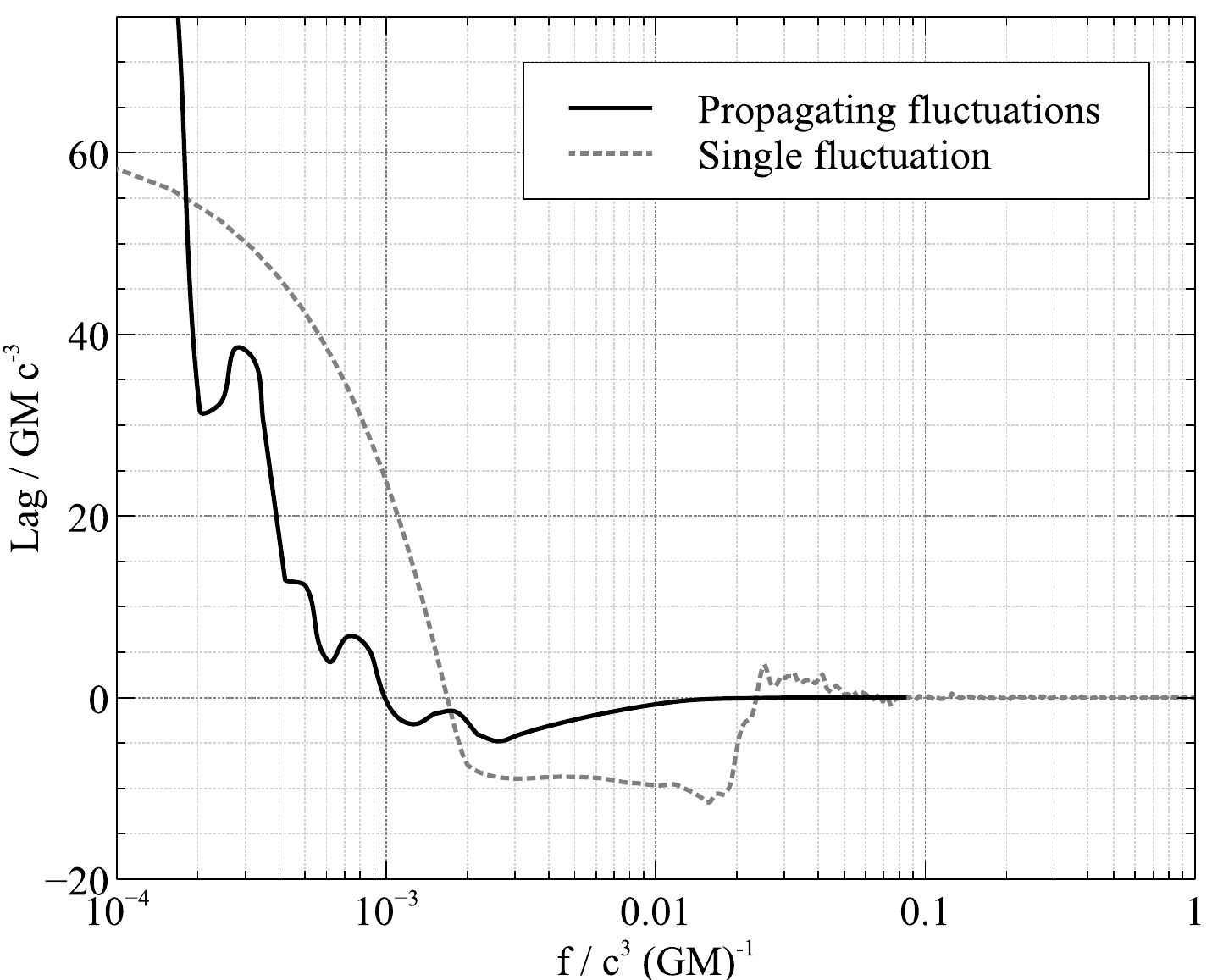}
\caption[]{The lag-frequency spectrum that arises from an accretion disc illuminated by an extended corona that is seeded by stochastic variability in the underlying accretion disc at all radii, compared to that considered previously where a single luminosity fluctuation propagates in from the outer edge.}
\label{lagfreq_radialprop.fig}
\end{figure}

The lag-frequency spectrum that arises in this scenario is shown in Fig.~\ref{lagfreq_radialprop.fig}. The features of the single fluctuation propagating through this corona are retained; the hard lag due to the propagation of the fluctuation from the less energetic outer to more energetic inner parts of the corona at low frequencies, then transitioning to the soft, reverberation lag at higher frequencies where the reflection-dominated 0.3-1\keV\ band lags behind the continuum-dominated 1-4\keV\ band. Previously, the measured reverberation lag time was increased from that in the case of a simple corona with either no or light-speed propagation. The slow, viscous propagation of the luminosity fluctuation from the outer part of the corona, which contributes a greater fraction of the directly detected continuum emission, to the inner parts from which more emission is focused onto the disc to be reflected, lengthening the measured lag. However, (incoherent) stochastic variability is now introduced into the corona at all radii, with the inner parts dominating the high frequency variations. The propagation effect, delaying the bulk of the reflection with regard to the bulk of the continuum emission, is greatly reduced now that uncorrelated fluctuations are introduced throughout the corona. The key factor in determining the broad band reverberation time lag is, once again, the vertical distance between the site of the coronal emission and the reflector. The reverberation feature in the broad band lag-frequency spectrum is smoothed out to a shape similar to that seen in the observed spectra.

\subsection{Propagation up Vertically Extended Coronae}
\label{propup.sec}
\begin{figure}
\centering
\includegraphics[width=85mm]{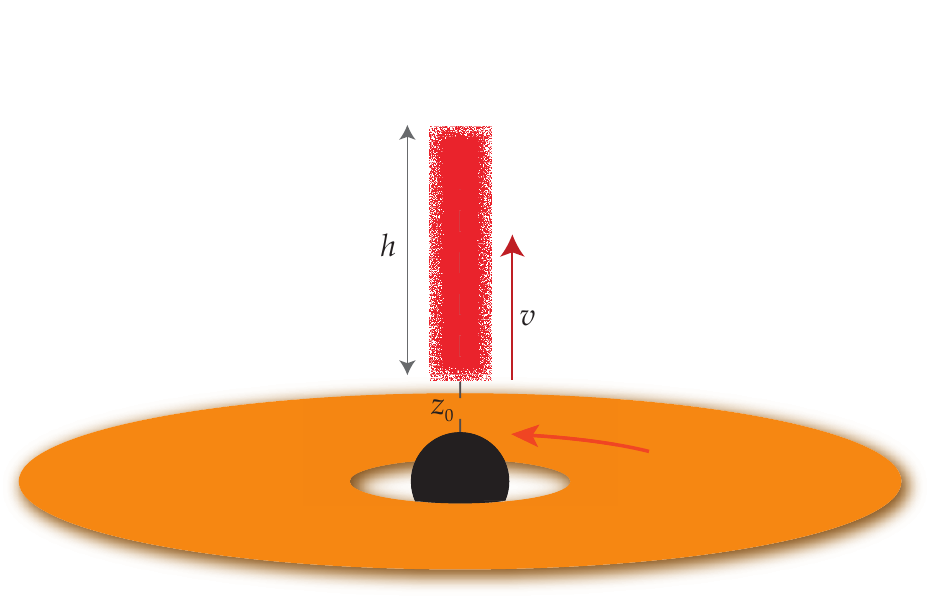}
\caption[]{The propagation of luminosity fluctuations up a collimated corona that is extended vertically above the plane of the accretion disc. The corona begins 1.5\rg\ from the black hole and extends to varying height $h$ from the singularity with fluctuations propagating up at constant velocity $v$. The photon index of the continuum produced throughout this model corona is constant at all heights.}
\label{vertical_corona.fig}
\end{figure}

In addition to coron\ae\ that are radially extended over the surface of the accretion disc, one might also consider the case of a collimated corona that extends vertically above the disc plane in a jet-like configuration (Fig.~\ref{vertical_corona.fig}). In this scenario, one might imagine the injection of energy from the innermost regions on the accretion disc (or, for example, magnetic fields anchored to these regions) into the base of the structure from which point fluctuations in the luminosity will propagate upward.

\begin{figure*}
\centering
\subfigure[$v = c$] {
\includegraphics[width=85mm]{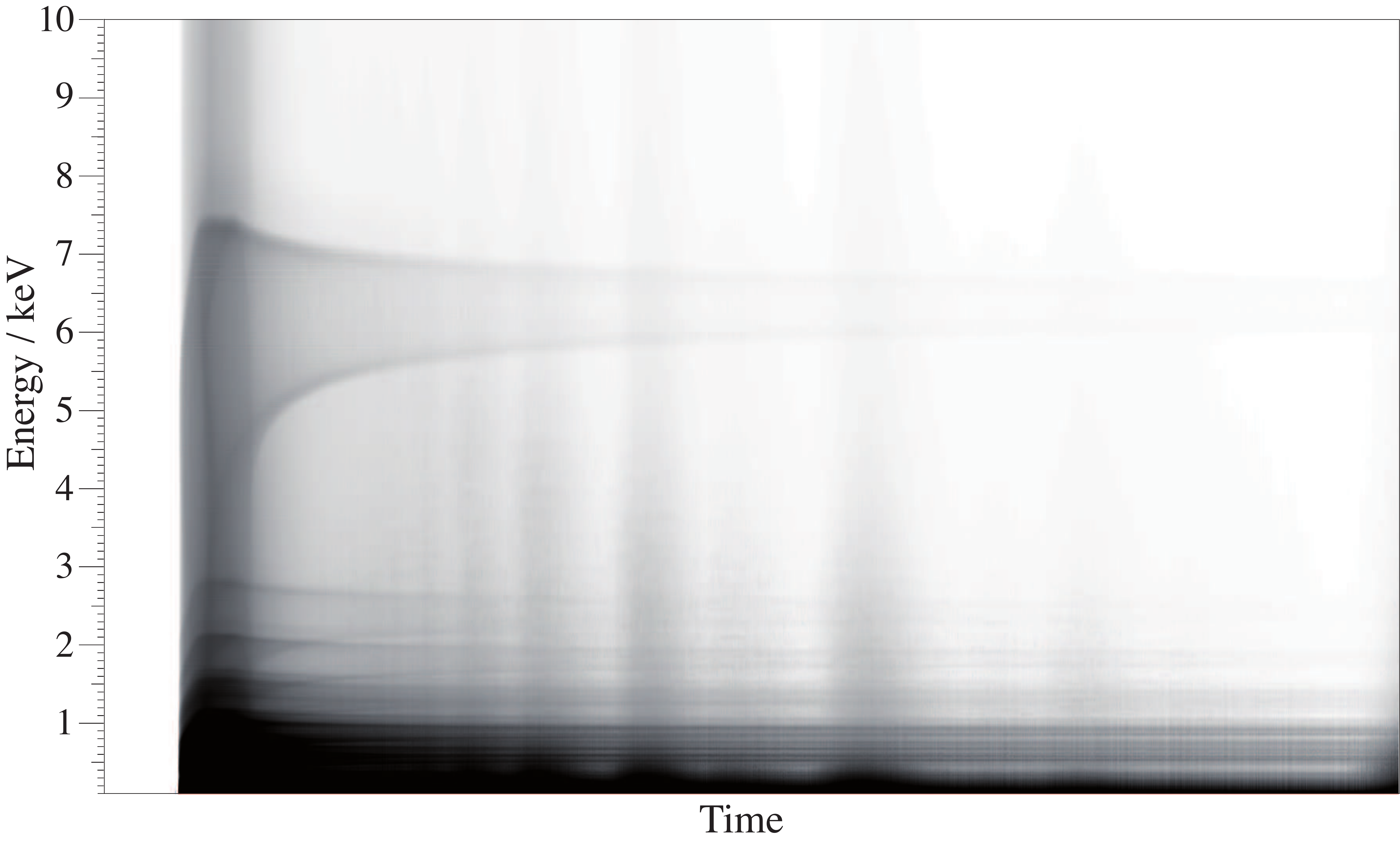}
\label{ent_propup.fig:c}
}
\subfigure[$v = 0.1c$] {
\includegraphics[width=85mm]{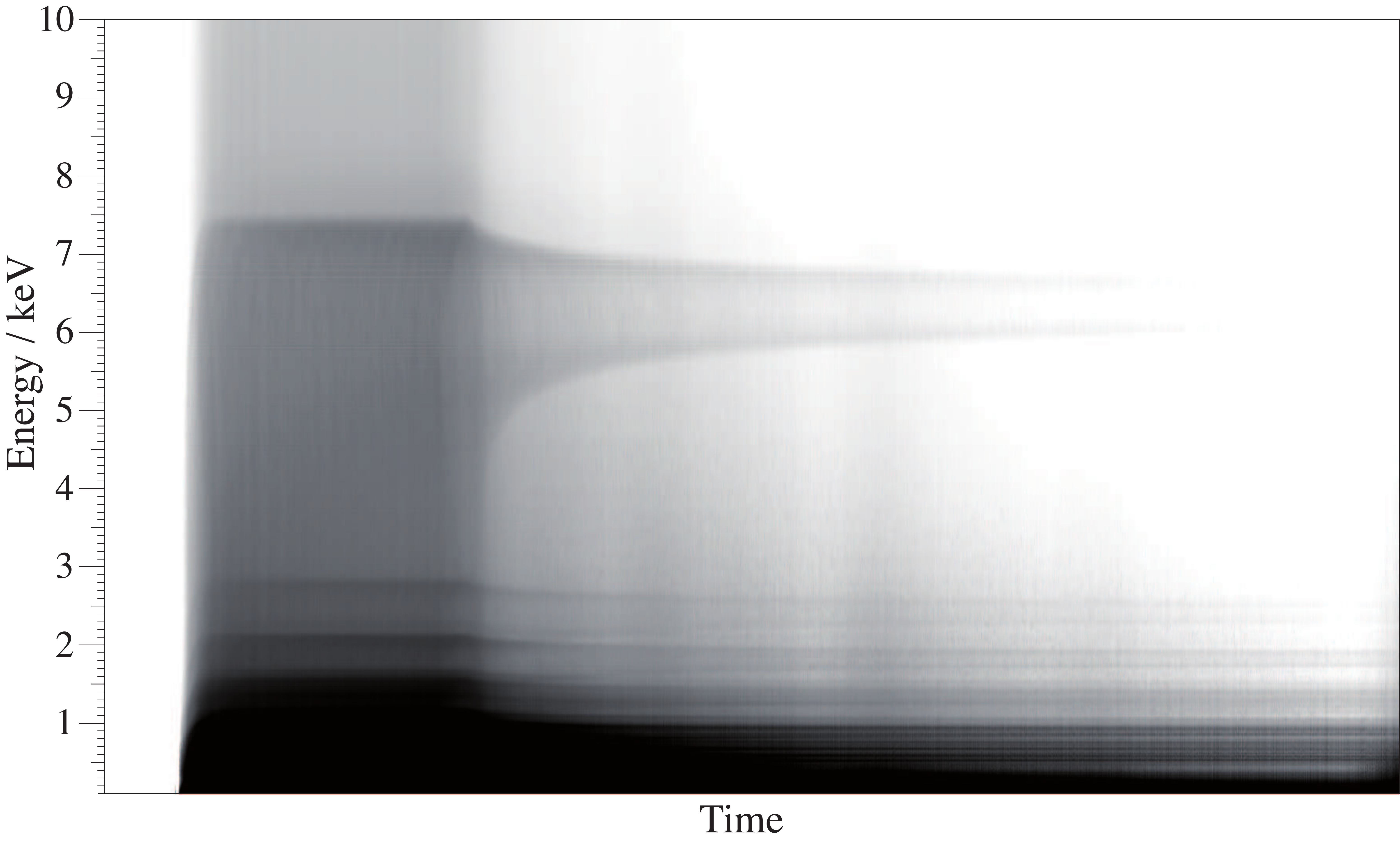}
\label{ent_propup.fig:0.1c}
}
\caption[]{Time and energy resolved response functions for reverberation from an accretion disc illuminated by a single flash propagating up a corona extended vertically to 10\rg. }
\label{ent_propup.fig}
\end{figure*}

Time- and energy-resolved response functions of the reverberation off an accretion disc illuminated by such a corona are shown in Fig.~\ref{ent_propup.fig}. The form of the response in this instance can be understood as the superposition of the responses to illumination by successive point sources at increasing height above the disc. The redshifted tails from the inner disc as well as the loop-back and the delayed re-emergence of photons from the back side of the inner disc stack with the response from higher portions of the corona shifted to later times (both due to the extended light travel time from the higher source and the time taken for the fluctuation to propagate up the corona). This leads to smearing of the inner disc response over time. Finally, the late response is seen from outer disc, illuminated predominantly by the uppermost parts of the corona, giving the narrow line response.

\begin{figure*}
\centering
\subfigure[] {
\includegraphics[width=85mm]{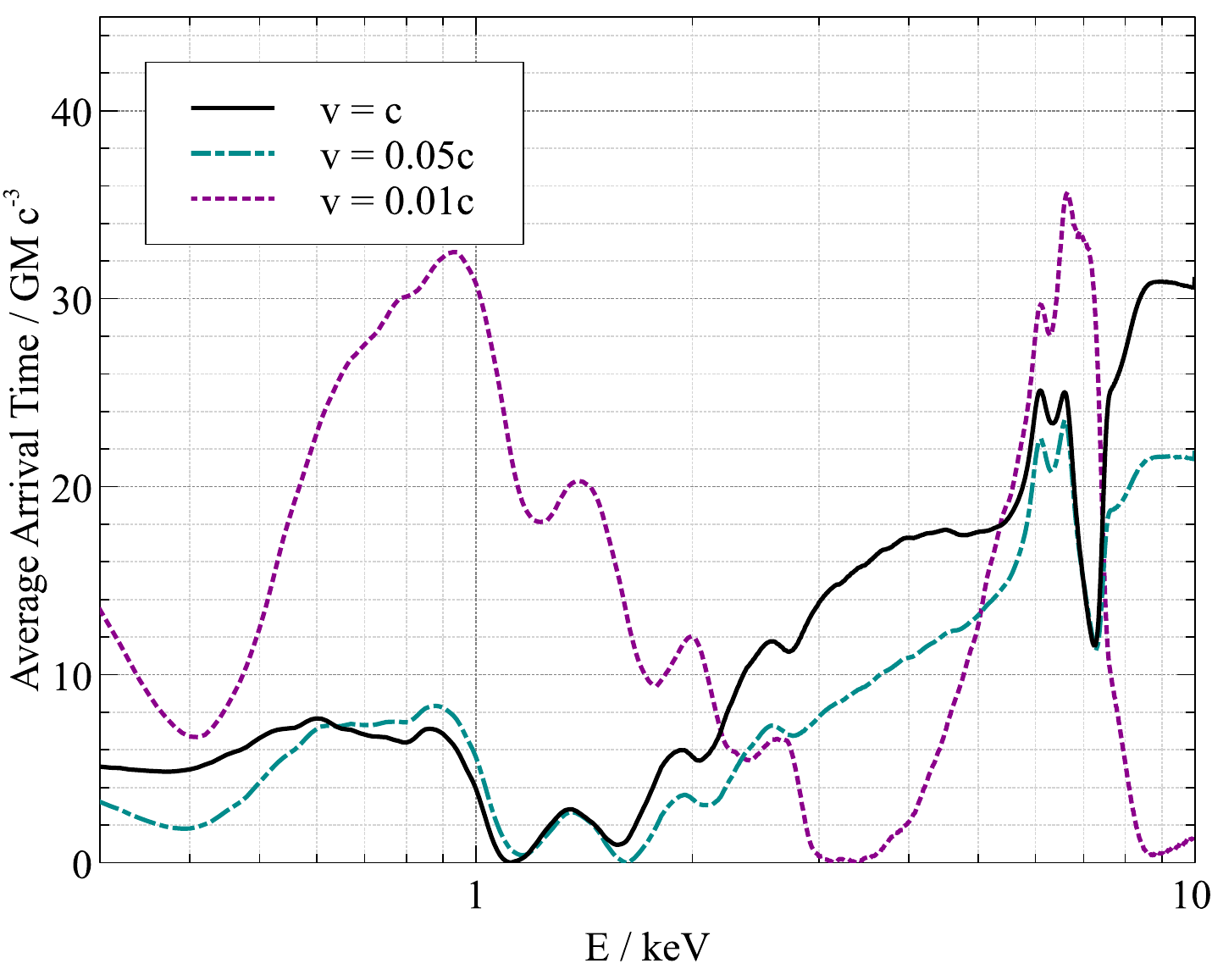}
\label{avg_arrival_propup.fig:vel}
}
\subfigure[] {
\includegraphics[width=85mm]{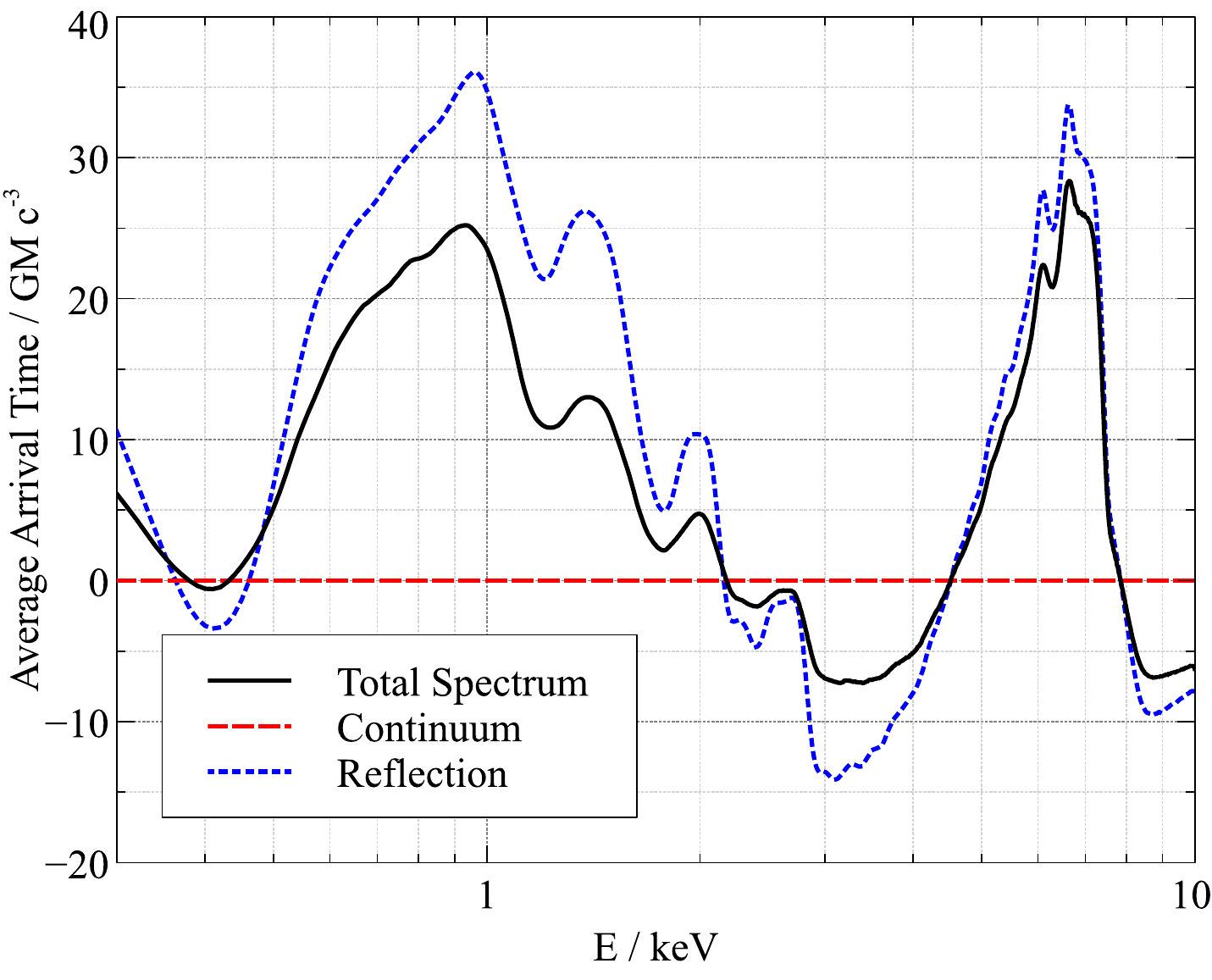}
\label{avg_arrival_propup.fig:components}
}
\caption[]{\subref{avg_arrival_propup.fig:vel} The average arrival times of photons as a function of energy where the accretion disc is illuminated by a vertically collimated corona extending between 1.5 and 10\rg\ above the singularity. The overall arrival time including both continuum and reflected photons is shown for fluctuations propagating at varying speed. \subref{avg_arrival_propup.fig:components} The average arrival times of photons originating from a vertically collimated corona extending between 1.5 and 10\rg\ from a luminosity fluctuation propagating upwards at $0.01c$ separated into the directly observed continuum and reflected components.}
\label{avg_arrival_propup.fig}
\end{figure*}

Fig.~\ref{avg_arrival_propup.fig:vel} shows the average arrival time of (both continuum and reflected) photons as a function of energy (\textit{i.e.} the proxy to the lag-energy spectrum) for fluctuations propagating at varying velocity up a collimated corona extending from 1.5 to 10\rg\ vertically above the disc plane. For more slowly propagating fluctuations, it can be seen that the dip in the lag-energy spectrum corresponding to the earliest arriving emission shifts from the continuum-dominated 1-2\keV\ band to around 3\keV, with the the lag time rising again either side of this dip towards both 2 and 4\keV, as is seen in the observed lag-energy spectra of Seyfert galaxies.

The origin of this 3\keV\ dip can be understood in the context of the average arrival time of photons that constitute the directly observed continuum and reflected spectral components. These are shown in Fig.~\ref{avg_arrival_propup.fig:components} and can be interpreted from the forms of the energy- and time-resolved response function for photons originating from a point source just 1.5\rg\ from the black hole shown in Fig.~\ref{ent_point.fig:h2}, \textit{i.e.} the lower part of the vertically extended corona that is illuminated first. 

The dip arises naturally in the average arrival time of photons in just the reflected component. It corresponds to the intersection between the outward-propagating reverberation response from successively larger radii in the disc towards the 6.4\keV\ rest-frame energy of the line and the inward-propagating response over the innermost few gravitational radii seen through the most redshifted photons. The passage of photons reflected from these radii (as well as the redshifted re-emerging reflection from the back side of the accretion disc) is delayed by the strong gravitational field through which they must pass, thus a distant observer will see reflection from closer to the black hole responding later. This point of intersection appears at 3\keV\ in the broad reverberation response for a point source located just 1.5\rg\ from the black hole which corresponds to the base of the vertically extended corona; the first part to be illuminated. Higher parts of the corona, which provide a greater fraction of the reflection from the outer disc (whereas the photons from lower parts of the corona are focused more strongly towards the black hole and hence onto the inner parts of the disc) are illuminated later as the flash, or luminosity fluctuation, propagates up. This steepens the profile of the lag-energy spectrum as the 6.4\keV\ rest frame emission in the iron K$\alpha$ line is delayed with respect to the redshifted emission arising from the inner disc to a greater extent in the case of vertical propagation.

Normally the 3\keV\ dip would not be directly observable in the lag-energy spectrum once the arrival times of the directly observed continuum photons are accounted for. For any given point within a corona, the path travelled by the directly observed continuum photons to the observer is always shorter than that travelled by reflected photons passing via the accretion disc. In this case, however, the luminosity fluctuation is propagating up from close to the black hole where the reflection fraction is high (the majority of photons are focused towards the black hole and onto the inner regions of the accretion disc rather than being able to escape to be detected as part of the continuum, \textit{e.g.} \citealt{1h0707_jan11,mrk335_corona_paper}) to greater heights where the reflection fraction falls closer to unity (in the absence of gravitational light bending, 50 per cent of photons from an isotropic point source travel upwards to be detected in the continuum and 50 per cent downward to be reflected from a disc that spans to infinity). When the fluctuation travels up the vertically extended corona sufficiently slowly (we find velocities slower than $0.02c$ to be required), the \textit{average} arrival time of all continuum photons is delayed sufficiently with respect to the reflection, because the bulk of the continuum is not produced until the fluctuation reaches the upper parts of the corona. The reflected component then dominates the earliest arriving photons and the 3\keV\ dip survives in the resultant lag-energy spectrum, as can be seen in Fig.~\ref{avg_arrival_propup.fig:components}. The precise energy at which the dip is found can vary slightly within the 3-4\keV\ range depending on the height of the base of the vertically collimated corona above the disc, determining the radius to which the light travel time is the least, and the reflection fraction which determines the energies at which the average arrival time of the continuum is later than the earliest arriving reflection.

\begin{figure}
\centering
\includegraphics[width=85mm]{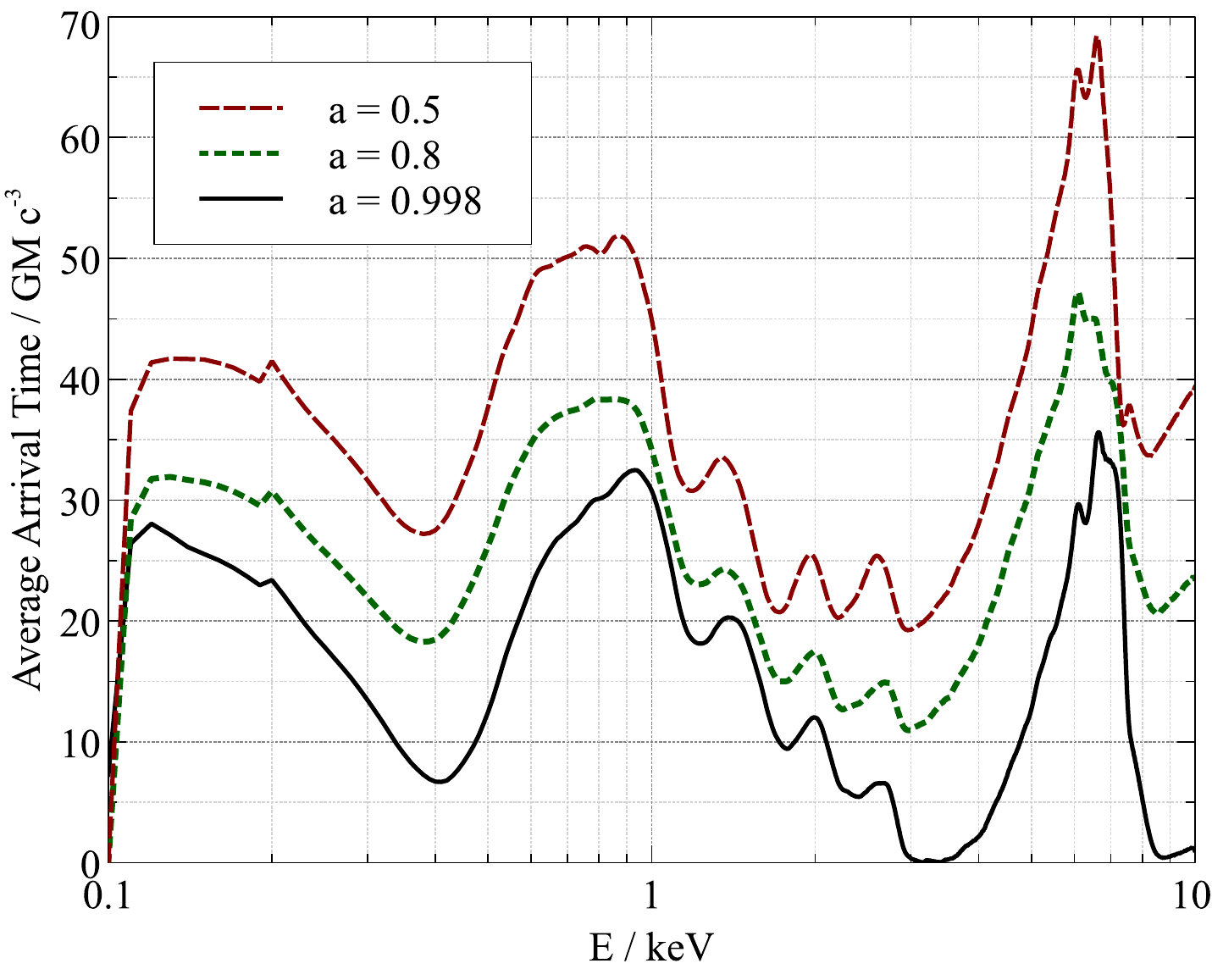}
\caption[]{The variation in the lag-energy spectrum, represented by the average arrival time of photons as a function of energy, as the black holes spin parameter is varied. Luminosity fluctuations propagate up a vertically collimated corona extending 10\rg\ above the plane of the disc at $0.01c$.}
\label{propup_spin.fig}
\end{figure}

The detection of the 3\keV\ dip in the lag-energy spectrum represents a direct detection of the effects of strong gravity in the immediate vicinity of the black hole. The dip arises at the intersection of the outward-propagating reverberation response over the outer regions of the accretion disc (which are expected classically) and the inward-propagating response due to the extreme redshift and time delay of photons reflected in the innermost few gravitational radii of the accretion disc, as well as the re-emergence of photons lensed into the line of sight from the back side of the disc. The 3\keV\ dip is expected only in the presence of strong gravity and when the accretion disc extends inwards within around 5\rg\ of the black hole such that a significant contribution can be seen from the inward-travelling response over the inner disc. The feature becomes weaker for more slowly spinning black holes or where the accretion disc is truncated and is not seen for spin parameter $a < 0.5\,GMc^{-2}$ (Fig.~\ref{propup_spin.fig}).

\section{Discussion}
By considering the lag spectra (both as a function of photon energy and Fourier frequency) that would be expected due to the reverberation from the accretion disc of continuum X-rays, we have found that the energy dependence of the reverberation lag can be a powerful discriminator of the means by which luminosity fluctuations propagate through extended coron\ae. This will give important clues as to the mechanism by which the corona is energised by the accretion flow. The finite speed of propagation through an extended corona (as is required to maintain causality if luminosity fluctuations are generated either within the corona or accretion disc) imprints its signature on the arrival times of photons in relativistically broadened emission lines.

Where fluctuations propagate outwards through a radially extended corona, the inner parts of the disc are illuminated initially resulting in a prompt response from the redshifted photons comprising the wing of iron K$\alpha$ line, while photons around the 6.4\keV\ core of the line, predominantly arising from the outer disc, do not respond until later. This flattens the redshifted wing of the line and giving the core of the line a distinctive square shape. On the other hand, inward propagation results in an early response from a double-troughed structure, corresponding to the Doppler shifts from the approaching and receding sides of the disc from beneath the outer edge of the corona, with a later response from the core of the line and the redshifted wing. Upward propagation through a collimated corona produces a distinctive V-shaped dip around 3\keV\ due to the early response from the inner parts of the disc illuminated by the base of the corona close to the black hole.

The limited resolution of lag measurements that is achievable by the present generation of X-ray observatories means that these subtle features are not readily distinguished. Future observatories, however, with large collecting area around 6\keV\ including the Athena X-ray observatory (see \citealt{reverb_review} for a discussion of lag measurements with future observatories) promise enhanced detail in the lag spectra allowing features to be distinguished in narrower energy bins and potentially direct measurement of the impulse response functions if the cross-spectra can be measured to sufficiently high resolution in both energy and frequency. This will not only allow for the unambiguous detection of the signatures of propagation through the corona and inner regions of the accretion flow but will allow for the precise measurement of its parameters, such as the propagation speed and the gradients in the coronal energetics and continuum photon index. This requires detailed modelling of the passage of X-rays around the black hole through ray tracing calculation and full characterisation of the process by which the measurements are made, most notably through the frequency dependence of the lags.

Consideration must be given to the statistical techniques that are to be employed to fit models to X-ray timing data in order to robustly measure parameters related to the coronal geometry and the propagation of fluctuations (or indeed other models). At the simplest level, measurements of the lag-frequency spectrum between two given energy bands or the lag-energy spectrum averaged over a given frequency range can be fit with corresponding models, computed for the energy bands and frequency ranges in question as presented in this work, using a maximum likelihood method ($\chi^2$ if the errors in each data point can be assumed to be Gaussian), as employed by \citet{emman+2014}. Each of these spectra, however, contains only a subset of the available information, being averaged over frequency and X-ray energy and with the quality of data available with present observatories, the number of data points that can be reasonably measured may be small compared to the number of model parameters leading to significant degeneracy. To maximise use of the available data, lags should be fit as a function of both energy and frequency using as smaller bin in each to give robust signal-to-noise. Timing measurements can be combined with the X-ray spectrum which, for instance, also contains information about the coronal geometry through the profile of broadened emission lines \citep{understanding_emis_paper,dauser+13} but care must be taken in combining the statistics from the two data sets as there will typically be far more data points in the X-ray spectrum than in either the lag-frequency or lag-energy spectrum. There are typically up to a few tens of data points in timing data products but hundreds of energy bins in spectra meaning that if the statistics from each are simply added to compute the likelihood or $\chi^2$ statistic, a poor fit to the timing data can be permitted by the statistical techniques employed. Both the spectral and timing data are encoded in the cross spectrum (Equation \ref{crossspec.equ}). If the data quality from future generation observatories is sufficient, fitting the cross spectrum, which is a complex number with both magnitude and phase as a function of both energy and frequency, will in principle maximally exploit the available data.

Before models can be rigorously fit to X-ray timing data and values as well as errors associated with model parameters can be measured, not only should the best of these techniques be determined but the degeneracies between the parameters of interest (given the expected data quality) should be explored. This is, however, beyond the scope of the current work and will be considered in detail in a future publication. Rather than directly fitting models to the available data and measuring parameters of the corona and the propagation of fluctuation, we here explore the general features of these models and determine the regions of the parameter space that reproduce the observed features. This yields a broad-brush picture of the features the corona must have to explain the observed lag-frequency and lag-energy spectra.

\subsection{Confronting the theory with observations}

While it is difficult to reconcile the full details of X-ray reverberation measurements if the continuum arises from a compact, point-like corona, we find it to be possible to explain the different features commonly observed in the timing analysis of Seyfert galaxies if the corona is extended. Although the resolution of the lag-energy spectra measured to date is limited, important constraints can be placed on the geometry of the corona and propagation mechanisms in order to self-consistently reproduce the full range of observed behaviour.

The broad band lag-frequency spectrum between the reflection-dominated 0.3-1\keV\ and the continuum dominated 1-4\keV\ bands shows a distinctive transition from a `hard lag' at low frequencies to a `soft lag' characteristic of reverberation from the disc at higher frequencies. At low frequencies, the average lag between the variability in higher energy X-ray bands behind correlated variations in a reference band is seen to increase steadily as a function of X-ray energy. At high frequencies, energies dominated by the soft excess reflected from the disc (0.3-1\keV) and the iron K$\alpha$ line are seen to lag behind energy bands dominated by directly-observed continuum emission. The redshifted wing of the iron line, below 5\keV\ is seen to respond earlier than the core of the line between 5 and 7\keV. The earliest response is seen in photons around 3\keV, producing a distinctive dip in the high frequency lag-energy spectrum that is not trivially explained by models of X-ray reflection.

The interpretation of the X-ray timing data are to be reconciled with measurements of the corona made from the time-averaged X-ray spectrum. The illumination patterns of accretion discs (emissivity profiles) suggest that the X-rays that are reflected in emission lines, broadened by Doppler shifts and gravitational redshifts originate from coron\ae\ that extend radially over the surface disc. The short reverberation lags that are measured show that while extending radially to tens of gravitational radii, these coron\ae\ extend just a few gravitational radii vertically above the plane of the disc.

\subsection{The propagation of luminosity fluctuations and the origin of the hard lag}

The low frequency hard lag is commonplace among accreting black holes and is commonly attributed to the propagation of luminosity fluctuations through an extended corona. The variation begins in a less energetic  region of the corona, producing a softer continuum spectrum, propagating to a more energetic region where a harder continuum is produced, hence delaying the average arrival time of the more energetic photons with respect to those less energetic. The gradient in energetics across the corona might be associated with the outer and inner parts of the accretion disc with fluctuations in luminosity connected to fluctuations in the mass accretion rate through the underlying disc, hence producing the hard lag as fluctuations propagate inwards through the accretion disc \citep{kotov+2001}. Such a model can readily reproduce the approximate log-linear dependence of the lag on photon energy that is observed as well as the PSDs of the X-ray variability \citep{arevalo+2006}. It is not immediately obvious, however, that this prescription can be combined with reverberation from the accretion disc to enable the hard lag to be observed in low frequency components of the variability at the same time as the soft (reverberation) lag is observed at higher frequencies, since the effects of the two processes counteract one another.

We find that it is possible to self-consistently combine a hard lag in the X-ray continuum with detectable reverberation from the accretion disc in the case of a corona extending radially at a low height over the surface of the accretion disc in which fluctuations in luminosity originate at the outer edge and propagate inwards. This is consistent with the geometry of the extended coron\ae\ inferred from the measurement of the accretion disc emissivity profile (the pattern of illumination of the disc by the coronal X-ray source) from the time-averaged X-ray spectra of the NLS1 galaxies in which reverberation has been detected; 1H\,0707$-$495 \citep{1h0707_emis_paper,understanding_emis_paper} and IRAS\,13224$-$3809 \citep{iras_fix}, and during the 2006 high flux observation of Markarian 335 \citep{mrk335_corona_paper} during which both low frequency hard lags and reverberation lags were detected \citep{kara+13}.

In these models, the hard lag is produced by a gradient in the photon index of the spectrum generated as a function of radius within the corona. The photon index decreases (\textit{i.e.} the spectrum hardens) linearly from the outer to the inner radii and it was found that this can produce the approximate lognormal dependence of the lag on photon energy that is observed. A linear variation in photon index was adopted for simplicity. This variation could, however, take a number of forms which can, in principle, be constrained given a sufficiently high resolution measurement of both the energy and frequency dependence of the lag as well as the X-ray spectrum. A measurement of the variation in photon index produced across the corona would constrain the energetics of the corona; its characteristic temperature and optical depth as a function of position that will, in turn, place important constraints on the mechanisms by which the corona is produced and energised and how it is related to the underlying accretion flow.

Where there is a gradient in the photon index produced as a function of radius in the corona, we find that the propagation of luminosity fluctuations must obey certain conditions if the hard and soft lags are to be detectable simultaneously. If luminosity fluctuations propagate at constant velocity, the positive hard lag between 1-4\keV\ energy band and the 0.3-1\keV\ band acts against the negative reverberation lag at all frequencies. This means that only a negative reverberation-dominated lag is seen for the fastest moving propagations, the magnitude of which decreases as the propagation is slowed until only a positive continuum-dominated lag is detected. It is necessary that the propagation of the luminosity fluctuation accelerates as it reaches the inner parts of the corona such that over the short distances travelled over the inner regions, the hard lag induced by the gradient in photon index over these parts does not counteract the time lag due to reverberation. The low frequency lag spectrum is dominated by the outer corona where the fluctuation propagates slowly.

Although these criteria place stringent constraints on the means by which luminosity fluctuations can propagate through an extended corona in the observed Seyfert galaxies, they are naturally met by propagation inwards from the outer edge of the corona on the viscous time scale within the underlying accretion disc. This represents a scenario in which the luminosity of a given patch of the corona is dictated by fluctuations locally in the disc. This could either be directly through variations in the mass accretion rate as over- and underdense regions accrete inwards, varying the thermal seed photon flux radiated from the disc to be up-scattered by the corona, or indirectly, injecting more energy into the corona, for instance through magnetic flux loops that undergo reconnection events to accelerate particles, forming the corona.

The propagation of a single luminosity fluctuation from the outer to the inner parts of a radially extended corona was, however, found to overestimate the measured reverberation lag. Due to their close passage by the black hole, photons from the inner part of the corona tend to be focused onto the accretion disc rather than being able to escape to be detected as part of the continuum, hence the inner parts of the corona contribute the greatest fraction of the reflected X-rays, while the outer parts of the corona contribute more to the detected continuum. The slow propagation of luminosity fluctuations from the outer to the inner part of the corona thus increases the time lag that would be measured from the `pure reverberation' time scale if there were no propagation (or propagation at light speed). Hence without knowing the details of the propagation, only an upper limit on the reverberation time scale is obtained and hence an upper limit on the vertical distance between the corona and the accretion disc. We find, however, that when the corona is seeded with stochastic luminosity fluctuations at all radii on the disc with faster fluctuations originating at smaller radii, this effect is substantially reduced and propagation effects at high frequencies become negligible. Measured reverberation time lags in Seyfert galaxies are as short as $2\,GM / c^3$, thus either the coron\ae\ are constrained to a very small region vertically above the disc (the reverberation lag is artificially increased by at least a factor of 5 due to viscous propagation), or luminosity fluctuations are introduced into the corona throughout its extent.

The outward propagation of luminosity fluctuations through radially extended coron\ae\ was also considered and found to reproduce the overall features of the observed high frequency lag-energy spectrum and even the hard lag at low frequencies under specific combinations of coronal extent and propagation velocity. This is due to the delay in the fluctuation reaching the outer parts of the corona that dominate the observed continuum emission. Being an artefact of disc reverberation, however, this hard lag follows the same energy dependence as the soft reverberation lag, failing to produce the increasing time lag with increasing photon energy that is observed. This effect leads to the earliest emission (in the case of the slowest propagation) from photons between 3 and 5\keV, though appearing as a broad trough in the lag-energy spectrum rather than the sharper 3\keV\ dip that is observed, that we find cannot be reproduced in the case of outward or inward propagating fluctuations from a corona extending radially over the accretion disc.

A hard lag can be introduced in the continuum in a model of outward propagating luminosity fluctuations that could be tuned to reproduce the low frequency hard lag that is observed. It would be necessary in this case for the fluctuations to propagate from an inner, less energetic region of the corona that produces a softer X-ray continuum than the more energetic outer parts of the corona. This could be envisaged if, for example, a greater flux of low energy seed photons were scattered by particles in the inner regions of the corona, since the inner radii of the disc will produce a greater thermal flux. This would also require that the energy injected from the accretion flow into the corona does not increase at smaller radii and that, even though cooling be allowed by seed photons, the inner corona cannot be reheated by flux from the outer parts. While viscous propagation inward through the accretion disc and the associated corona naturally reproduces the time scales required to detect both hard and soft lags, there is no such natural explanation for the equivalent outward propagation. The variability of X-ray emission from quasars during microlensing events suggests that harder X-rays originate from a more compact region than the softer X-rays \citep{dai+2010,chen+2012,macleod+2015}, suggesting that the gradient in coronal energetics producing the hard lag accompanies inward rather than outward propagation.

\subsection{The 3\,keV dip}
The dip in the high frequency lag-energy spectrum, with the earliest response at 3\keV, is seen almost universally in Seyfert galaxies \citep{kara+13}. The shape of the lag-energy spectrum (once scaled to the absolute value of the lag across objects, determined by the black hole mass) between around 2 and 8\keV\ is almost identical, highly suggestive of a common configuration of the primary X-ray source and reprocessor to form the final spectrum. This common feature is, however, not trivially predicted in simple models of X-ray reverberation from the accretion disc. In reflection and reverberation models, photons detected at 3\keV\ are predominantly reflected from the inner regions of the accretion disc. For a primary X-ray source above the disc (whether a point source or an extended corona) through which there is no significant propagation delay, there will always be a shorter path directly to the observer from the corona than travelling via the disc, hence the continuum-dominated 1-2\keV\ energy band should, in principle, always lead emission at 3\keV.

It is possible to reproduce the observed dip in the high frequency lag-energy spectrum (that is, in the reverberation from the accretion disc) in the case the luminosity fluctuations propagate upwards through a collimated, vertically extended corona. In this case, the 3\keV\ dip arises naturally in the arrival times of photons that have been reflected from the disc forming the iron K$\alpha$ fluorescence line. Photons above 3\keV\ are reflected further out in the disc and hence further from the compact footprint of the vertically extended corona, so are seen to respond later owing to the longer light travel time. Line photons that are extremely redshifted and seen below 3\keV\ arise from the very inner regions of the disc. The strong gravitational field in such close proximity to the black hole substantially slows the propagation of these photons so they too are seen to respond later, producing the dip at 3\keV. The dip is detectable only when propagation up the vertically extended corona is sufficiently slow, delaying the average arrival time of the continuum photons until the fluctuation propagates upwards to a region of the corona where a substantial fraction of the photons can escape to be detected in the continuum without being focused towards the black hole and onto the disc.

Where the observed X-ray emission is produced by the reflection of an X-ray continuum arising from an energetic corona from the surface of the accretion disc, the detection of the 3\keV\ dip in the high frequency lag-energy spectrum represents a direct detection of the effects of strong gravity in close proximity to the black hole, as predicted by General Relativity. It signals the delay of the passage of photons reflected from the innermost parts of the disc, the extreme redshifting of photons reflected from these regions and the bending of a substantial portion of continuum rays emitted from lower parts of the extended corona towards the black hole. The detection of the 3\keV\ dip requires that the accretion disc extends sufficiently close to the black hole that a substantial fraction of photons are reflected within 5\rg. Within this radius, the delay and extreme redshift experienced by reflected photons produces the upturn in the lag-energy profile. A strong dip at 3\keV\ in a high frequency lag-energy spectrum showing reverberation from the accretion disc is a further indication of a rapidly spinning black hole.

\subsection{Piecing together the corona}

The results of these simulations of simplified models of the coronal geometry and the propagation of luminosity fluctuations suggest that it is possible to reproduce the features of the observed lag-energy and lag-frequency spectra of Seyfert galaxies through the combination of two scenarios. The increasing time lag as a function of photon energy in the continuum at low frequencies can be produced self-consistently with the reverberation of this continuum off the accretion disc at higher frequencies if luminosity fluctuations propagate inwards through a corona extended radially over the accretion disc on the viscous time scale within the underlying disc. On the other hand, the 3\keV\ dip in the high frequency (reverberation) lag-energy spectrum is reproduced by the upward propagation of luminosity fluctuations through a vertically extended corona.

\begin{figure}
\centering
\includegraphics[width=85mm]{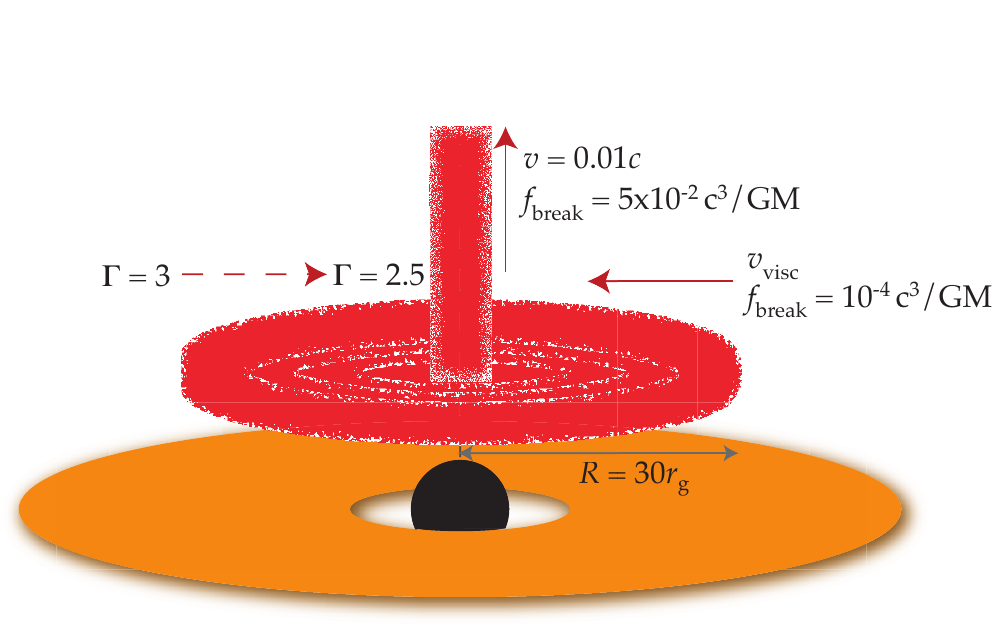}
\caption[]{The combined model in which the low frequency variability is dominated by fluctuations propagating through a corona extending vertically over the plane of the accretion disc. The corona is seeded with variations at all radii with the break frequency of the power spectral density at each radius corresponding to the orbital frequency. Fluctuations propagate inwards on the viscous timescale through the underlying disc. The corona has a bright central region driven through which fluctuations propagate upwards at constant velocity.}
\label{combined_corona.fig}
\end{figure}

\begin{figure*}
\centering
\subfigure[Lag-frequency spectrum] {
\includegraphics[width=55mm]{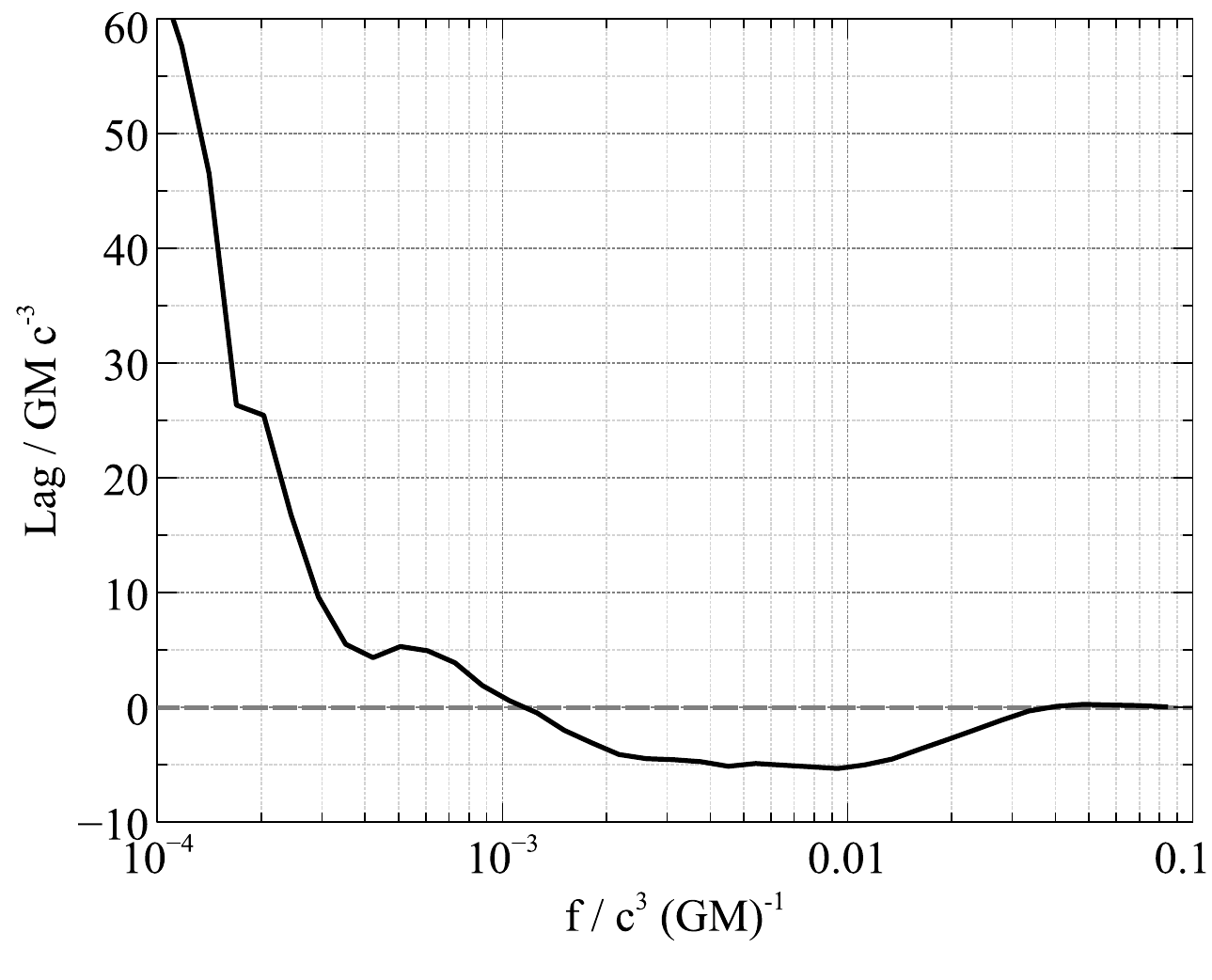}
\label{lagspec_combined.fig:lagfreq}
}
\subfigure[Lag-energy, low frequency] {
\includegraphics[width=55mm]{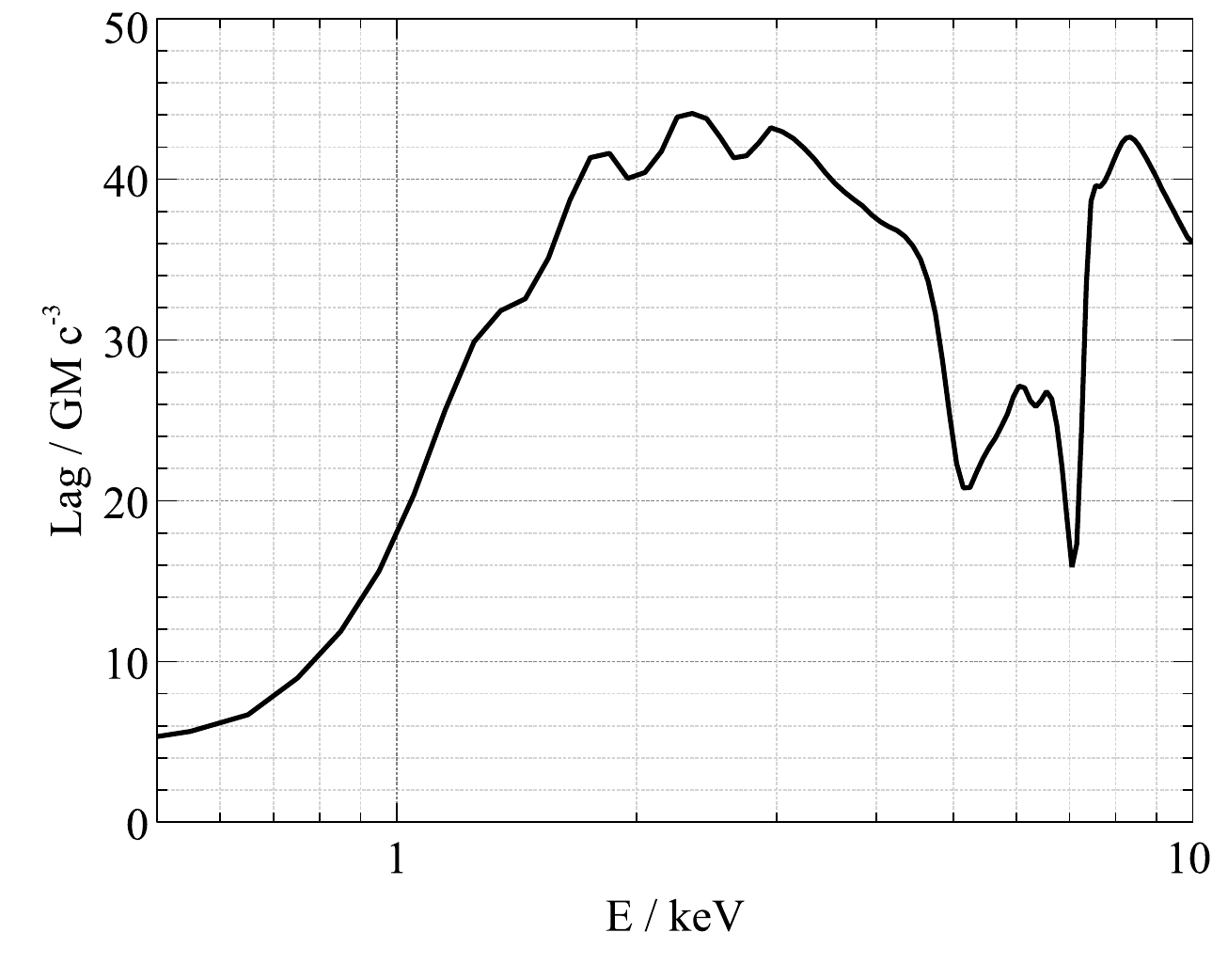}
\label{lagspec_combined.fig:lagen_low}
}
\subfigure[Lag-energy, high frequency] {
\includegraphics[width=55mm]{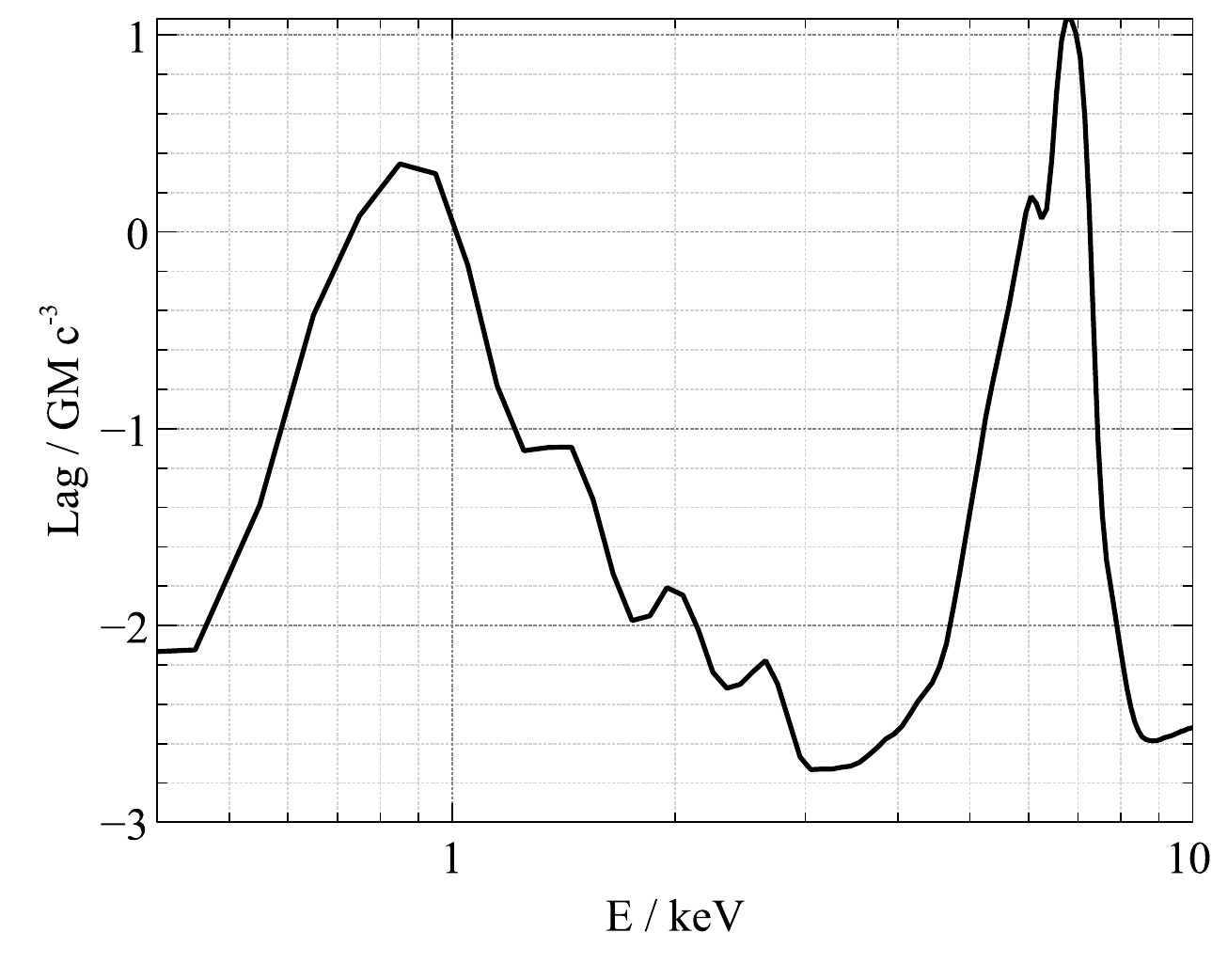}
\label{lagspec_combined.fig:lagen_high}
}
\caption[]{\subref{lagspec_combined.fig:lagfreq} The broadband lag-frequency spectrum between the 0.3-1\keV\ and 1-4\keV\ bands for the hybrid corona model. A corona extending radially over the surface of the disc to 30\rg\ is seeded with stochastic luminosity fluctuations throughout its extent that then propagate inwards on the viscous timescale. There is a linear gradient in photon index between the outer edge and centre. In addition, there is a collimated core to the corona, producing one third the X-ray count rate as the entire extended region, seeded by luminosity fluctuations from the innermost stable orbit on the disc. These fluctuations then propagate upwards at $0.01c$. \subref{lagspec_combined.fig:lagen_low} and \subref{lagspec_combined.fig:lagen_high} The lag-energy spectra averaged over low frequencies, $10^{-4}$ to $5\times 10^{-4}\,c^3(GM)^{-1}$ and high frequencies, $2\times 10^{-3}$ to $10^{-2}\,c^3(GM)^{-1}$, respectively.}
\label{lagspec_combined.fig}
\end{figure*}

The simultaneous detection of the hard lag and the 3\keV\ dip suggests that these two modes of propagation operate simultaneously within the corona. On the longest time scales, the variability is dominated by luminosity fluctuations that arise in the outer parts of a corona (or the corresponding radii on the accretion disc) extending radially at a low height over the disc surface and then propagate inwards on the viscous time scale. In addition to the slow variability that is generated at large radii in the disc (limited in frequency by either the orbital or viscous time scale at the radius at which it arises), more rapid variability may be introduced into the corona at successively smaller radii. This would reflect stochastic processes occurring locally; turbulence within the disc or the generation and reconnection of magnetic flux loops, for instance. In the innermost regions, the corona becomes collimated vertically and in these regions, fluctuations in the luminosity propagate upwards (Fig.~\ref{combined_corona.fig}). This collimated inner part of the corona is driven by processes arising on the inner regions of the accretion flow where the orbital and viscous time scale are much shorter.

The lag-energy and lag-frequency spectra expected from such a hybrid model is shown in Fig.~\ref{lagspec_combined.fig}. The emission from a radially extended corona is seeded with stochastic fluctuations throughout its extent as in Section~\ref{propfluc.sec} (with PSDs governed by the orbital frequency at each radius). This is combined with emission from a collimated core as in Section~\ref{propup.sec} driven by stochastic variability with a PSD appropriate to the orbital velocity on the innermost stable orbit of the accretion disc. The relative proportion of X-ray counts from the extended and collimated portions of the corona is a free parameter and we find that the collimated part producing around one third the total count rate of the entire extended portion approximately reproduces the observed behaviour.

At low frequencies, the measured time lags are dominated by propagation through the radially extended corona which is slowly varying in time and hence dominates the time-averaged X-ray spectrum of the AGN (providing the flattened accretion disc emissivity profiles that are observed). The low frequency lag-energy spectrum approximates the observed hard lag. An approximate log-linear increase in time lag as a function of photon energy is produced with a slight flattening seen between 0.5 and 1\keV\ (as seen in 1H\,0707$-$495) which arises due to the early response of reflected photons in the soft excess from regions of the disc below the outer edge of the corona. No evidence is seen in the observations for the double-trough structure associated with the iron K$\alpha$ line where fluctuations propagate inwards slowly. Despite the predicted structure being finer than the achievable energy binning (a drop in the average lag over these bins should still be seen). In this model, we find that the double-trough structure becomes smoothed out and its lead time over the continuum at those energies (\textit{i.e.} the depth of the troughs) reduced by a combination of factors which may account for its non-detection. The seeding of the corona with luminosity fluctuations throughout its extent means that the early response to any given fluctuation is not dominated by a single annulus on the disc, which is required to produce the double-trough structure, spreading it over a broader range in energy. The reduced reflection fraction associated with fluctuations propagating through the outer regions of the corona compared to those arising closer to the black hole where more photons are lensed onto the inner disc, reduces the time difference associated with these energies and, finally, the feature becomes smoothed out and hence less detectable in higher frequency components. Sharp troughs associated with a long time delay are only detectable in the lowest frequency components of the variability, reducing this feature over a low frequency range selected to coincide with that selected in observations.

The slowly varying extended corona makes a lesser contribution to the high frequency lag measurements which are dominated by the rapidly varying core. This vertically extended structure produces the observed lag-energy spectrum with the characteristic dip at 3\keV\ that appears in many sources. It is important to emphasise that these are simplified models of the corona and the propagation of luminosity fluctuations through its extent. The combination of more detailed modelling with high resolution measurements, exploiting bright sources observed with large collecting areas to obtain accurate lag measurements in narrow energy bins will enable detailed modelling of the structure and processes occurring within the corona.

There is a small number of Seyfert galaxies in which no dip is seen at 3\keV\ in the high frequency lag-energy spectrum, despite there being strong evidence for the reflection and reverberation of continuum X-rays off the accretion disc. These include NCG\,4151 \citep{cackett_ngc4151} and PG\,1244+026 \citep{kara_pg1244}. These non-detections suggest that while common, the collimated central part of the corona that produces the dip is not ubiquitous. It is possible that while the radially extended portion of the corona exists in these objects, the central portion of the corona with upward-propagating fluctuations does not form. \citet{mrk335_flare_paper} find that the corona in Mrk\,335 can become collimated and be ejected, suggesting that it is plausible that the collimation of the corona is a transient process. In many objects it is sustained for prolonged periods (the extreme of this, where it extends well beyond the inner regions of the accretion flow would be a jet, though on smaller scales it co-exists with the radially extended portion of the corona), while in these few it is not found. It is not clear what drives the collimation. it is clearly not just the spin of the black hole since NGC\,4151 and PG\,1244+026 appear to be rapidly spinning black holes, though it may be due to the collimation of magnetic fields near the black hole event horizon such as those theorised to power radio jets.

This vastly simplified model of a multicomponent corona does not perfectly reproduce the observed lag spectra. It includes just the reverberation of the X-ray continuum from the accretion disc and uses a simplified prescription for the coronal geometry and gradient in energetics that produces the hard lag. It predicts features in the low frequency lag-energy spectrum around the iron K$\alpha$ spectrum that have not been observed and does not include X-ray time lags that might arise from the thermal reprocessing of X-rays incident on the disc or the `soft Comptonisation' of seed photons by a hot disc atmosphere before they reach the corona which might account for the mismatch in relative arrival times of the soft excess and iron K$\alpha$ photons between the model and observations. This simplified model does, however, reveal how measurements of X-ray reverberation are starting to reveal structures that must be present within the corona.

\section{Conclusions}

Simplified models of X-ray reverberation from the accretion disc demonstrate how measurement of the both the energy and frequency dependence of lags enables not only the time averaged geometry of the extended X-ray emitting corona to be measured, but also how fluctuations in luminosity propagate through its extent. This potentially gives clues to the mechanism by which the corona is formed and  energy is injected from the accretion flow. Measurements of X-ray reverberation lags to sufficiently high resolution and their dependence on both photon energy and Fourier frequency reveal the vertical extent of an extended corona above the plane of the accretion disc as well as the direction and velocity of propagation through its extent.

The energy and frequency dependence of reverberation lags that has been observed to date across a growing sample Seyfert galaxies is already placing important constraints on the nature of the corona if they are to be explained through X-ray reverberation.

It is possible to simultaneously observe X-ray reverberation at high frequencies with the hard lag at low in the longest timescale variability, produced by the propagation of fluctuations radially inward from less to more energetic regions of the corona as had been previously suggested in self-consistent models, only if the propagation of fluctuations through the corona obeys particular criteria. The simultaneous detection of the hard lag in the continuum at low frequencies and the reverberation lag at high frequencies requires sufficiently slow propagation through the outer regions of the corona, accelerating over the inner regions.

Propagation through an underlying Shakura-Sunyaev accretion disc on the viscous time scale naturally meets these criteria while faster propagation through the corona itself does not work. It is plausible that coron\ae\ extend over the inner regions of the accretion disc (extending to a few tens of gravitational radii, consistent with previous measurements of the corona) and are coupled to the underlying disc. The corona is driven by stochastic variability arising across its entire extent, consistent with propagating fluctuation models of accretion flows, to reproduce the observed frequency dependence of the reverberation lag.

The dip in the high frequency (reverberation) lag-energy spectrum in which the earliest response is seen in photons around 3\keV, commonly observed in Seyfert galaxies, is not trivially explained by simple reverberation models. Detection of 3\keV\ dip in the lag-energy spectrum from the reverberation of X-rays from the inner accretion disc requires relatively slow (slower than $0.01c$) propagation up a vertically collimated corona in the innermost regions. The dip represents a direct detection of the effect of strong gravity on X-ray propagation from the innermost few gravitational radii around the black hole.

The simultaneous detection of a hard lag, increasing smoothly with photon energy at low frequencies, X-ray reverberation at high frequencies and a dip at 3\keV\ in the reverberation-dominated high frequency lag-energy spectrum suggests two modes of propagation operating simultaneously within the corona. On the longest time scales, the variability is dominated by stochastic luminosity fluctuations that arise across the accretion disc and propagate inwards viscously through the disc, energising the corona as they pass through. In the innermost regions, a bright core of the corona becomes collimated vertically and in these regions, fluctuations in the luminosity propagate upwards though relatively slowly (slower than $0.01c$). This collimated inner part of the corona is driven by more rapid processes arising on the inner regions of the accretion flow, hence is found to dominate the high frequency lag spectrum.

\section*{Acknowledgements}
DRW was supported by a CITA National Fellowship for the duration of this work. EMC gratefully acknowledges support from the National Science Foundation through CAREER award number 1351222. ACF thanks the ERC for support from Advanced Grant
Feedback 340442.

%\vspace{-0.6cm}
\bibliographystyle{mnras2}
\bibliography{agn}

\begin{thebibliography}{}

\bibitem[\protect\citeauthoryear{{Alston}, {Vaughan} \& {Uttley}}{{Alston}
  et~al.}{2013}]{alston+2013}
{Alston} W.~N.,  {Vaughan} S.,    {Uttley} P.,  2013, \mnras, 429, 75

\bibitem[\protect\citeauthoryear{{Ar{\'e}valo} \& {Uttley}}{{Ar{\'e}valo} \&
  {Uttley}}{2006}]{arevalo+2006}
{Ar{\'e}valo} P.,  {Uttley} P.,  2006, \mnras, 367, 801

\bibitem[\protect\citeauthoryear{{Cackett}, {Fabian}, {Zogbhi}, {Kara},
  {Reynolds} \& {Uttley}}{{Cackett} et~al.}{2013}]{cackett+2013}
{Cackett} E.~M.,  {Fabian} A.~C.,  {Zogbhi} A.,  {Kara} E.,  {Reynolds} C.,
  {Uttley} P.,  2013, \apjl, 764, L9

\bibitem[\protect\citeauthoryear{{Cackett}, {Zoghbi}, {Reynolds}, {Fabian},
  {Kara}, {Uttley} \& {Wilkins}}{{Cackett} et~al.}{2014}]{cackett_ngc4151}
{Cackett} E.~M.,  {Zoghbi} A.,  {Reynolds} C.,  {Fabian} A.~C.,  {Kara} E.,
  {Uttley} P.,    {Wilkins} D.~R.,  2014, \mnras, 438, 2980

\bibitem[\protect\citeauthoryear{{Chainakun} \& {Young}}{{Chainakun} \&
  {Young}}{2012}]{chainakun+2012}
{Chainakun} P.,  {Young} A.~J.,  2012, \mnras, 420, 1145

\bibitem[\protect\citeauthoryear{{Chainakun} \& {Young}}{{Chainakun} \&
  {Young}}{2015}]{chainakun+2015}
{Chainakun} P.,  {Young} A.~J.,  2015, \mnras, 452, 333

\bibitem[\protect\citeauthoryear{{Chen}, {Dai}, {Kochanek}, {Chartas},
  {Blackburne} \& {Morgan}}{{Chen} et~al.}{2012}]{chen+2012}
{Chen} B.,  {Dai} X.,  {Kochanek} C.~S.,  {Chartas} G.,  {Blackburne} J.~A.,
  {Morgan} C.~W.,  2012, \apj, 755, 24

\bibitem[\protect\citeauthoryear{{Czerny}}{{Czerny}}{2006}]{czerny+06}
{Czerny} B.,  2006, in {Gaskell} C.~M.,  {McHardy} I.~M.,  {Peterson} B.~M.,
  {Sergeev} S.~G.,  eds, Astronomical Society of the Pacific Conference Series
  Vol.~360 of Astronomical Society of the Pacific Conference Series, {The Role
  of the Accretion Disk in AGN Variability}.
p.~265

\bibitem[\protect\citeauthoryear{{Dai}, {Kochanek}, {Chartas}, {Koz{\l}owski},
  {Morgan}, {Garmire} \& {Agol}}{{Dai} et~al.}{2010}]{dai+2010}
{Dai} X.,  {Kochanek} C.~S.,  {Chartas} G.,  {Koz{\l}owski} S.,  {Morgan}
  C.~W.,  {Garmire} G.,    {Agol} E.,  2010, \apj, 709, 278

\bibitem[\protect\citeauthoryear{{Dauser}, {Garcia}, {Wilms}, {B{\"o}ck},
  {Brenneman}, {Falanga}, {Fukumura} \& {Reynolds}}{{Dauser}
  et~al.}{2013}]{dauser+13}
{Dauser} T.,  {Garcia} J.,  {Wilms} J.,  {B{\"o}ck} M.,  {Brenneman} L.~W.,
  {Falanga} M.,  {Fukumura} K.,    {Reynolds} C.~S.,  2013, \mnras, 430, 1694

\bibitem[\protect\citeauthoryear{{De Marco}, {Ponti}, {Cappi}, {Dadina},
  {Uttley}, {Cackett}, {Fabian} \& {Miniutti}}{{De Marco}
  et~al.}{2013}]{demarco+2012}
{De Marco} B.,  {Ponti} G.,  {Cappi} M.,  {Dadina} M.,  {Uttley} P.,  {Cackett}
  E.~M.,  {Fabian} A.~C.,    {Miniutti} G.,  2013, \mnras, 431, 2441

\bibitem[\protect\citeauthoryear{{De Marco}, {Ponti}, {Uttley}, {Cappi},
  {Dadina}, {Fabian} \& {Miniutti}}{{De Marco} et~al.}{2011}]{demarco+2011}
{De Marco} B.,  {Ponti} G.,  {Uttley} P.,  {Cappi} M.,  {Dadina} M.,  {Fabian}
  A.~C.,    {Miniutti} G.,  2011, \mnras, 417, L98

\bibitem[\protect\citeauthoryear{{Emmanoulopoulos}, {McHardy} \&
  {Papadakis}}{{Emmanoulopoulos} et~al.}{2011}]{emmanoul+2011}
{Emmanoulopoulos} D.,  {McHardy} I.~M.,    {Papadakis} I.~E.,  2011, \mnras,
  416, L94

\bibitem[\protect\citeauthoryear{{Emmanoulopoulos}, {Papadakis}, {Dov{\v c}iak}
  \& {McHardy}}{{Emmanoulopoulos} et~al.}{2014}]{emman+2014}
{Emmanoulopoulos} D.,  {Papadakis} I.~E.,  {Dov{\v c}iak} M.,    {McHardy}
  I.~M.,  2014, \mnras, 439, 3931

\bibitem[\protect\citeauthoryear{{Fabian}, {Kara}, {Walton}, {Wilkins} \&
  {Ross}}{{Fabian} et~al.}{2013}]{iras_fix}
{Fabian} A.~C.,  {Kara} E.,  {Walton} D.~J.,  {Wilkins} D.~R.,    {Ross} R.~R.,
   2013, \mnras, 429, 2917

\bibitem[\protect\citeauthoryear{{Fabian}, {Rees}, {Stella} \&
  {White}}{{Fabian} et~al.}{1989}]{fabian+89}
{Fabian} A.~C.,  {Rees} M.~J.,  {Stella} L.,    {White} N.~E.,  1989, \mnras,
  238, 729

\bibitem[\protect\citeauthoryear{{Fabian}, {Zoghbi}, {Ross}, {Uttley}, {Gallo},
  {Brandt}, {Blustin}, {Boller}, {Caballero-Garcia}, {Larsson}, {Miller},
  {Miniutti}, {Ponti}, {Reis}, {Reynolds}, {Tanaka} \& {Young}}{{Fabian}
  et~al.}{2009}]{fabian+09}
{Fabian} A.~C.,  {Zoghbi} A.,  {Ross} R.~R.,  {Uttley} P.,  {Gallo} L.~C.,
  {Brandt} W.~N.,  {Blustin} A.~J.,  {Boller} T.,  {Caballero-Garcia} M.~D.,
  {Larsson} J.,  {Miller} J.~M.,  {Miniutti} G.,  {Ponti} G.,  {Reis} R.~C.,
  {Reynolds} C.~S.,  {Tanaka} Y.,    {Young} A.~J.,  2009, \nat, 459, 540

\bibitem[\protect\citeauthoryear{{Fabian}, {Zoghbi}, {Wilkins}, {Dwelly},
  {Uttley}, {Schartel}, {Miniutti}, {Gallo}, {Grupe}, {Komossa} \&
  {Santos-Lle{\'o}}}{{Fabian} et~al.}{2012}]{1h0707_jan11}
{Fabian} A.~C.,  {Zoghbi} A.,  {Wilkins} D.,  {Dwelly} T.,  {Uttley} P.,
  {Schartel} N.,  {Miniutti} G.,  {Gallo} L.,  {Grupe} D.,  {Komossa} S.,
  {Santos-Lle{\'o}} M.,  2012, \mnras, 419, 116

\bibitem[\protect\citeauthoryear{{Galeev}, {Rosner} \& {Vaiana}}{{Galeev}
  et~al.}{1979}]{galeev+79}
{Galeev} A.~A.,  {Rosner} R.,    {Vaiana} G.~S.,  1979, \apj, 229, 318

\bibitem[\protect\citeauthoryear{{Haardt} \& {Maraschi}}{{Haardt} \&
  {Maraschi}}{1991}]{haardt+91}
{Haardt} F.,  {Maraschi} L.,  1991, \apjl, 380, L51

\bibitem[\protect\citeauthoryear{{Ingram} \& {Done}}{{Ingram} \&
  {Done}}{2011}]{ingram+2011}
{Ingram} A.,  {Done} C.,  2011, \mnras, 415, 2323

\bibitem[\protect\citeauthoryear{{Ingram} \& {van der Klis}}{{Ingram} \& {van
  der Klis}}{2013}]{ingram+2013}
{Ingram} A.,  {van der Klis} M.,  2013, \mnras, 434, 1476

\bibitem[\protect\citeauthoryear{{Kara}, {Cackett}, {Fabian}, {Reynolds} \&
  {Uttley}}{{Kara} et~al.}{2014}]{kara_pg1244}
{Kara} E.,  {Cackett} E.~M.,  {Fabian} A.~C.,  {Reynolds} C.,    {Uttley} P.,
  2014, \mnras, 439, L26

\bibitem[\protect\citeauthoryear{{Kara}, {Fabian}, {Cackett}, {Steiner},
  {Uttley}, {Wilkins} \& {Zoghbi}}{{Kara} et~al.}{2013}]{kara_1h0707}
{Kara} E.,  {Fabian} A.~C.,  {Cackett} E.~M.,  {Steiner} J.~F.,  {Uttley} P.,
  {Wilkins} D.~R.,    {Zoghbi} A.,  2013, \mnras, 428, 2795

\bibitem[\protect\citeauthoryear{{Kara}, {Fabian}, {Cackett}, {Uttley},
  {Wilkins} \& {Zoghbi}}{{Kara} et~al.}{2013}]{kara+13}
{Kara} E.,  {Fabian} A.~C.,  {Cackett} E.~M.,  {Uttley} P.,  {Wilkins} D.~R.,
   {Zoghbi} A.,  2013, \mnras, 434, 1129

\bibitem[\protect\citeauthoryear{{Kara}, {Zoghbi}, {Marinucci}, {Walton},
  {Fabian}, {Risaliti}, {Boggs}, {Christensen}, {Fuerst}, {Hailey}, {Harrison},
  {Matt}, {Parker}, {Reynolds}, {Stern} \& {Zhang}}{{Kara}
  et~al.}{2015}]{kara+2015}
{Kara} E.,  {Zoghbi} A.,  {Marinucci} A.,  {Walton} D.~J.,  {Fabian} A.~C.,
  {Risaliti} G.,  {Boggs} S.~E.,  {Christensen} F.~E.,  {Fuerst} F.,  {Hailey}
  C.~J.,  {Harrison} F.~A.,  {Matt} G.,  {Parker} M.~L.,  {Reynolds} C.~S.,
  {Stern} D.,    {Zhang} W.~W.,  2015, \mnras, 446, 737

\bibitem[\protect\citeauthoryear{{Kotov}, {Churazov} \& {Gilfanov}}{{Kotov}
  et~al.}{2001}]{kotov+2001}
{Kotov} O.,  {Churazov} E.,    {Gilfanov} M.,  2001, \mnras, 327, 799

\bibitem[\protect\citeauthoryear{{Laor}}{{Laor}}{1991}]{laor-91}
{Laor} A.,  1991, \apj, 376, 90

\bibitem[\protect\citeauthoryear{{Leighly}}{{Leighly}}{1999}]{leighly-99_2}
{Leighly} K.~M.,  1999, \apjs, 125, 317

\bibitem[\protect\citeauthoryear{{Liu}, {Mineshige} \& {Ohsuga}}{{Liu}
  et~al.}{2003}]{liu+03}
{Liu} B.~F.,  {Mineshige} S.,    {Ohsuga} K.,  2003, \apj, 587, 571

\bibitem[\protect\citeauthoryear{{MacLeod}, {Morgan}, {Mosquera}, {Kochanek},
  {Tewes}, {Courbin}, {Meylan}, {Chen}, {Dai} \& {Chartas}}{{MacLeod}
  et~al.}{2015}]{macleod+2015}
{MacLeod} C.~L.,  {Morgan} C.~W.,  {Mosquera} A.,  {Kochanek} C.~S.,  {Tewes}
  M.,  {Courbin} F.,  {Meylan} G.,  {Chen} B.,  {Dai} X.,    {Chartas} G.,
  2015, \apj, 806, 258

\bibitem[\protect\citeauthoryear{{McHardy}, {Koerding}, {Knigge}, {Uttley} \&
  {Fender}}{{McHardy} et~al.}{2006}]{mchardy+2006}
{McHardy} I.~M.,  {Koerding} E.,  {Knigge} C.,  {Uttley} P.,    {Fender} R.~P.,
   2006, \nat, 444, 730

\bibitem[\protect\citeauthoryear{{McHardy}, {Papadakis}, {Uttley}, {Page} \&
  {Mason}}{{McHardy} et~al.}{2004}]{mchardy+2004}
{McHardy} I.~M.,  {Papadakis} I.~E.,  {Uttley} P.,  {Page} M.~J.,    {Mason}
  K.~O.,  2004, \mnras, 348, 783

\bibitem[\protect\citeauthoryear{{McKinney}, {Tchekhovskoy} \&
  {Blandford}}{{McKinney} et~al.}{2012}]{mckinney+2012}
{McKinney} J.~C.,  {Tchekhovskoy} A.,    {Blandford} R.~D.,  2012, \mnras, 423,
  3083

\bibitem[\protect\citeauthoryear{{Merloni} \& {Fabian}}{{Merloni} \&
  {Fabian}}{2001}]{merloni_fabian}
{Merloni} A.,  {Fabian} A.~C.,  2001, \mnras, 328, 958

\bibitem[\protect\citeauthoryear{{Miyamoto} \& {Kitamoto}}{{Miyamoto} \&
  {Kitamoto}}{1989}]{miyamoto+89}
{Miyamoto} S.,  {Kitamoto} S.,  1989, \nat, 342, 773

\bibitem[\protect\citeauthoryear{{Miyamoto}, {Kitamoto}, {Mitsuda} \&
  {Dotani}}{{Miyamoto} et~al.}{1988}]{miyamoto+88}
{Miyamoto} S.,  {Kitamoto} S.,  {Mitsuda} K.,    {Dotani} T.,  1988, \nat, 336,
  450

\bibitem[\protect\citeauthoryear{{Nowak}, {Vaughan}, {Wilms}, {Dove} \&
  {Begelman}}{{Nowak} et~al.}{1999}]{nowak+99}
{Nowak} M.~A.,  {Vaughan} B.~A.,  {Wilms} J.,  {Dove} J.~B.,    {Begelman}
  M.~C.,  1999, \apj, 510, 874

\bibitem[\protect\citeauthoryear{{Ponti}, {Papadakis}, {Bianchi}, {Guainazzi},
  {Matt}, {Uttley} \& {Bonilla}}{{Ponti} et~al.}{2012}]{ponti+2012}
{Ponti} G.,  {Papadakis} I.,  {Bianchi} S.,  {Guainazzi} M.,  {Matt} G.,
  {Uttley} P.,    {Bonilla} N.~F.,  2012, \aap, 542, A83

\bibitem[\protect\citeauthoryear{{Reynolds}, {Young}, {Begelman} \&
  {Fabian}}{{Reynolds} et~al.}{1999}]{reynolds+99}
{Reynolds} C.~S.,  {Young} A.~J.,  {Begelman} M.~C.,    {Fabian} A.~C.,  1999,
  \apj, 514, 164

\bibitem[\protect\citeauthoryear{{Ross} \& {Fabian}}{{Ross} \&
  {Fabian}}{2005}]{ross_fabian}
{Ross} R.~R.,  {Fabian} A.~C.,  2005, \mnras, 358, 211

\bibitem[\protect\citeauthoryear{{Shakura} \& {Sunyaev}}{{Shakura} \&
  {Sunyaev}}{1973}]{shaksun}
{Shakura} N.~I.,  {Sunyaev} R.~A.,  1973, \aap, 24, 337

\bibitem[\protect\citeauthoryear{Shapiro}{Shapiro}{1964}]{shapiro}
Shapiro I.~I.,  1964, Phys. Rev. Lett., 13, 789

\bibitem[\protect\citeauthoryear{{Sikora} \& {Begelman}}{{Sikora} \&
  {Begelman}}{2013}]{sikora_begelman}
{Sikora} M.,  {Begelman} M.~C.,  2013, \apjl, 764, L24

\bibitem[\protect\citeauthoryear{{Sunyaev} \& {Tr{\"u}mper}}{{Sunyaev} \&
  {Tr{\"u}mper}}{1979}]{sunyaev_trumper}
{Sunyaev} R.~A.,  {Tr{\"u}mper} J.,  1979, \nat, 279, 506

\bibitem[\protect\citeauthoryear{{Timmer} \& {Koenig}}{{Timmer} \&
  {Koenig}}{1995}]{timmer_konig}
{Timmer} J.,  {Koenig} M.,  1995, \aap, 300, 707

\bibitem[\protect\citeauthoryear{{Turner}, {George}, {Nandra} \&
  {Turcan}}{{Turner} et~al.}{1999}]{turner+99}
{Turner} T.~J.,  {George} I.~M.,  {Nandra} K.,    {Turcan} D.,  1999, \apj,
  524, 667

\bibitem[\protect\citeauthoryear{{Uttley}, {Cackett}, {Fabian}, {Kara} \&
  {Wilkins}}{{Uttley} et~al.}{2014}]{reverb_review}
{Uttley} P.,  {Cackett} E.~M.,  {Fabian} A.~C.,  {Kara} E.,    {Wilkins} D.~R.,
   2014, \aapr

\bibitem[\protect\citeauthoryear{{Uttley} \& {McHardy}}{{Uttley} \&
  {McHardy}}{2005}]{uttley_mchardy-2005}
{Uttley} P.,  {McHardy} I.~M.,  2005, \mnras, 363, 586

\bibitem[\protect\citeauthoryear{{Uttley}, {Wilkinson}, {Cassatella}, {Wilms},
  {Pottschmidt}, {Hanke} \& {B{\"o}ck}}{{Uttley} et~al.}{2011}]{uttley+2011}
{Uttley} P.,  {Wilkinson} T.,  {Cassatella} P.,  {Wilms} J.,  {Pottschmidt} K.,
   {Hanke} M.,    {B{\"o}ck} M.,  2011, \mnras, 414, L60

\bibitem[\protect\citeauthoryear{{Walton}, {Zoghbi}, {Cackett}, {Uttley},
  {Harrison}, {Fabian}, {Kara}, {Miller}, {Reis} \& {Reynolds}}{{Walton}
  et~al.}{2013}]{walton_hardlag}
{Walton} D.~J.,  {Zoghbi} A.,  {Cackett} E.~M.,  {Uttley} P.,  {Harrison}
  F.~A.,  {Fabian} A.~C.,  {Kara} E.,  {Miller} J.~M.,  {Reis} R.~C.,
  {Reynolds} C.~S.,  2013, \apjl, 777, L23

\bibitem[\protect\citeauthoryear{{Wilkins} \& {Fabian}}{{Wilkins} \&
  {Fabian}}{2011}]{1h0707_emis_paper}
{Wilkins} D.~R.,  {Fabian} A.~C.,  2011, \mnras, 414, 1269

\bibitem[\protect\citeauthoryear{{Wilkins} \& {Fabian}}{{Wilkins} \&
  {Fabian}}{2012}]{understanding_emis_paper}
{Wilkins} D.~R.,  {Fabian} A.~C.,  2012, \mnras, 424, 1284

\bibitem[\protect\citeauthoryear{{Wilkins} \& {Fabian}}{{Wilkins} \&
  {Fabian}}{2013}]{lag_spectra_paper}
{Wilkins} D.~R.,  {Fabian} A.~C.,  2013, \mnras, 430, 247

\bibitem[\protect\citeauthoryear{{Wilkins} \& {Gallo}}{{Wilkins} \&
  {Gallo}}{2015}]{mrk335_corona_paper}
{Wilkins} D.~R.,  {Gallo} L.~C.,  2015, \mnras, 449, 129

\bibitem[\protect\citeauthoryear{{Wilkins}, {Gallo}, {Grupe}, {Bonson},
  {Komossa} \& {Fabian}}{{Wilkins} et~al.}{2015}]{mrk335_flare_paper}
{Wilkins} D.~R.,  {Gallo} L.~C.,  {Grupe} D.,  {Bonson} K.,  {Komossa} S.,
  {Fabian} A.~C.,  2015, \mnras, 454, 4440

\bibitem[\protect\citeauthoryear{{Young} \& {Reynolds}}{{Young} \&
  {Reynolds}}{2000}]{young_reynolds}
{Young} A.~J.,  {Reynolds} C.~S.,  2000, \apj, 529, 101

\bibitem[\protect\citeauthoryear{{Zhou} \& {Wang}}{{Zhou} \&
  {Wang}}{2005}]{zhou_wang}
{Zhou} X.-L.,  {Wang} J.-M.,  2005, \apjl, 618, L83

\bibitem[\protect\citeauthoryear{{Zoghbi}, {Cackett}, {Reynolds}, {Kara},
  {Harrison}, {Fabian}, {Lohfink}, {Matt}, {Balokovic}, {Boggs}, {Christensen},
  {Craig}, {Hailey}, {Stern} \& {Zhang}}{{Zoghbi} et~al.}{2014}]{zoghbi+2014}
{Zoghbi} A.,  {Cackett} E.~M.,  {Reynolds} C.,  {Kara} E.,  {Harrison} F.~A.,
  {Fabian} A.~C.,  {Lohfink} A.,  {Matt} G.,  {Balokovic} M.,  {Boggs} S.~E.,
  {Christensen} F.~E.,  {Craig} W.,  {Hailey} C.~J.,  {Stern} D.,    {Zhang}
  W.~W.,  2014, \apj, 789, 56

\bibitem[\protect\citeauthoryear{{Zoghbi} \& {Fabian}}{{Zoghbi} \&
  {Fabian}}{2011}]{zoghbi+2011}
{Zoghbi} A.,  {Fabian} A.~C.,  2011, \mnras, 418, 2642

\bibitem[\protect\citeauthoryear{{Zoghbi}, {Fabian}, {Reynolds} \&
  {Cackett}}{{Zoghbi} et~al.}{2012}]{zoghbi+2012}
{Zoghbi} A.,  {Fabian} A.~C.,  {Reynolds} C.~S.,    {Cackett} E.~M.,  2012,
  \mnras, 422, 129

\bibitem[\protect\citeauthoryear{{Zoghbi}, {Fabian}, {Uttley}, {Miniutti},
  {Gallo}, {Reynolds}, {Miller} \& {Ponti}}{{Zoghbi} et~al.}{2010}]{zoghbi+09}
{Zoghbi} A.,  {Fabian} A.~C.,  {Uttley} P.,  {Miniutti} G.,  {Gallo} L.~C.,
  {Reynolds} C.~S.,  {Miller} J.~M.,    {Ponti} G.,  2010, \mnras, 401, 2419

\bibitem[\protect\citeauthoryear{{Zoghbi}, {Reynolds}, {Cackett}, {Miniutti},
  {Kara} \& {Fabian}}{{Zoghbi} et~al.}{2013}]{zoghbi+2013}
{Zoghbi} A.,  {Reynolds} C.,  {Cackett} E.~M.,  {Miniutti} G.,  {Kara} E.,
  {Fabian} A.~C.,  2013, \apj, 767, 121

\end{thebibliography}

\label{lastpage}

\end{document}